\documentclass[11pt]{article}
\usepackage[utf8]{inputenc}
\usepackage{jheppub}
\usepackage{graphicx}
\usepackage{caption}
\usepackage{subcaption}
\usepackage{hyperref}
\usepackage[T1]{fontenc}
\usepackage{float}

\newcommand{\be}{\begin{equation}}
\newcommand{\ee}{\end{equation}}
\newcommand{\bea}{\begin{eqnarray}}
\newcommand{\eea}{\end{eqnarray}}

\newcommand\underrel[3][]{\mathrel{\mathop{#3}\limits_{%
      \ifx c#1\relax\mathclap{#2}\else#2\fi}}}
\def\CC{\mathcal{C}}
\def\CD{\mathcal{D}}

\def\CI{\mathcal{I}}

\def\CO{\mathcal{O}}
\def\CP{\mathcal{P}}
\def\CR{\mathcal{R}}

\title{Mixed five-point correlators in the 3d Ising model}

\author{David Poland,$^a$ Valentina Prilepina, Petar Tadi\' c$^{b,c}$}

\affiliation{$^{a}$ Department of Physics, Yale University, New Haven, CT 06520, USA}
\affiliation{$^{b}$ Maxwell Institute for Mathematical Sciences, Department of Mathematics, Heriot-Watt University, Edinburgh EH14, UK}
\affiliation{$^{c}$ Institute for Interdisciplinary and Multidisciplinary Studies, University of Montenegro, Podgorica, Montenegro}

\emailAdd{david.poland@yale.edu, valentina.prilepina.1@ulaval.ca, p.tadic@hw.ac.uk}

\abstract{We study five-point correlators of the $\sigma$, $\epsilon$, and $\epsilon'$ operators in the critical 3d Ising model. We consider the $\sigma \times \sigma$ and $\sigma \times \epsilon$ operator product expansions (OPEs) and truncate them by including a finite set of exchanged operators. We approximate the truncated operators by the corresponding contributions in appropriate disconnected five-point correlators. We compute a number of OPE coefficients that were previously unknown and show that these are consistent with predictions obtained using the fuzzy sphere regularization of the critical 3d Ising model.}  

\preprint{OUTP-21-11P}

\begin{document}
\maketitle
\flushbottom

\newpage

\section{Introduction}

The conformal bootstrap program, developed following \cite{Ferrara:1973yt, Polyakov:1974gs, Rattazzi:2008pe, El-Showk:2012cjh, El-Showk:2014dwa}, is a powerful non-perturbative approach to studying conformal field theories (CFTs) which utilizes the consequences of conformal and internal symmetries to impose consistency conditions on a theory. Solving these conditions puts nontrivial bounds on the dynamics of the theory, and in some cases, completely solves it.

In principle, it is possible to obtain the complete set of conformal bootstrap constraints by studying four-point correlation functions of all primary operators in the spectrum of the theory in question. In practice, however, doing so for operators with nonzero spin becomes challenging very quickly, and this is why most current conformal bootstrap results are obtained using four-point functions of external scalar operators. Some results for external spinning operators in four-point correlators were obtained in~\cite{Iliesiu:2015qra, Iliesiu:2017nrv, Dymarsky:2017xzb, Dymarsky:2017yzx, Karateev:2019pvw, Reehorst:2019pzi, Erramilli:2020rlr, Erramilli:2022kgp, He:2023ewx, Bartlett-Tisdall:2023ghh, Mitchell:2024hix, Bartlett-Tisdall:2024mbx}. Recently, the most precise results for the conformal dimensions of the leading scalars in the critical 3d Ising model were computed in~\cite{Chang:2024whx} by bootstrapping the system of correlators containing the $\sigma$, $\epsilon$, and $T_{\mu\nu}$ operators. 

Instead of considering four-point correlators of external spinning states, one can study higher-point correlators of external scalars, and, in principle, access the same data as with the spinning four-point correlators. Over the last few years there has been considerable progress in bootstrapping higher-point correlators using both numerical \cite{Poland:2023vpn, Poland:2023bny, Antunes:2023kyz} and analytical \cite{Bercini:2020msp, Antunes:2021kmm, Anous:2021caj, Kaviraj:2022wbw, Goncalves:2023oyx, Costa:2023wfz, Bercini:2024pya, Harris:2024nmr, Bargheer:2024hfx, Artico:2024wut} techniques. 

In \cite{Li:2023tic, Poland:2023bny} it was observed that one can improve the accuracy of the truncation method (originally developed in \cite{Gliozzi:2013ysa, Gliozzi:2014jsa}), for the $\sigma \times \sigma$ operator product expansion (OPE) in the critical 3d Ising model, by approximating the truncated contributions with their counterparts from mean field theory (MFT). In \cite{Poland:2023bny} we also demonstrated that the truncated contributions in the $\sigma \times \sigma$ OPE in the five-point correlator $\langle\sigma\sigma\epsilon\sigma\sigma\rangle$ can be accurately approximated by the corresponding contributions in a disconnected five-point correlator. This allowed us to accurately compute some of the OPE coefficients involving two spinning states and the scalar $\epsilon$ state. 

In this paper we show that this same technique can be applied to other five-point correlators containing $\sigma$, $\epsilon$, and $\epsilon'$ states. We will evaluate these correlators using the $\sigma \times \sigma$ and $\sigma \times \epsilon$ OPEs. In all cases involving the irrelevant scalar $\epsilon'$, $\epsilon'$ would appear as the middle operator in the correlator. Since we do not consider the OPE with the middle operator, we therefore expect that its large conformal dimension will not dramatically slow down the convergence of the OPE expansions. Using this approach, we compute several previously unknown OPE coefficients. The OPE coefficients that we obtain here can be also used as inputs for the Hamiltonian truncation technique (see eg.~\cite{Fitzpatrick:2022dwq}), which is a non-perturbative method for studying quantum field theories along the RG flow induced by deformations of a conformal field theory.

In \cite{Hu:2023xak} it was established that one can also accurately determine OPE coefficients in the critical 3d Ising model by regularizing it on the fuzzy sphere \cite{Zhu:2022gjc}. In this work we will compare our results with those of \cite{Hu:2023xak} and also use the code developed in \cite{Zhou:2025liv} to compute fuzzy-sphere predictions for some additional OPE coefficients that we determined by means of the five-point bootstrap. We observe that the results are nontrivially consistent between the two approaches.

This paper is organized as follows. In section~\ref{sectwo} we review the conventions for general five-point correlators and elucidate the numerical method we use to extract unknown OPE coefficients. In section~\ref{secthree} we study five-point correlators that exclusively feature the $\sigma \times \sigma$ OPE, namely $\langle\sigma\sigma\epsilon\sigma\sigma\rangle$ and $\langle\sigma\sigma\epsilon'\sigma\sigma\rangle$. In section~\ref{secfour} we consider the truncation applied to the $\sigma \times \epsilon$ OPE. We first study the four-point function $\langle\sigma\epsilon\sigma\epsilon\rangle$ and compute the OPE coefficients $\lambda_{\sigma\epsilon \Sigma_j}$, where $\Sigma_j$ are spin-$j$ operators on the leading Regge trajectory in $\sigma \times \epsilon$. Here we show that the results we extract approximately agree with the ones already known from the four-point bootstrap~\cite{Simmons-Duffin:2016wlq}. Then we proceed to study five-point correlators that can be decomposed in the $\sigma \times \epsilon$ OPE, namely $\langle\sigma\epsilon\epsilon\epsilon\sigma\rangle$,
$\langle\sigma\epsilon\epsilon'\epsilon\sigma\rangle$, and $\langle\sigma\sigma\sigma\sigma\epsilon\rangle$, computing several new OPE coefficients.
In section~\ref{fuzzysphere} we employ the fuzzy sphere regularization to extract some of the OPE coefficients previously computed in section~\ref{secfour} (containing at least one $\sigma$ or $\epsilon$ operator). In section~\ref{secfive} we discuss our results. In a number of appendices, we state our parameterization of the five-point conformal cross-ratios, present explicit correlation matrices for OPE coefficients with different tensor structure labels, furnish numerical CFT data that we fix in our analysis, and include details on the computations of  OPE coefficients on the fuzzy sphere.

\section{Five-point correlators and the numerical bootstrap}\label{sectwo}

In this section we set up the conventions that we use for five-point correlators, review the crossing relation, explain how we compute constraints from the crossing relation, and define the cost function that we minimize to compute unknown OPE coefficients.

Let us consider decomposing the following five-point correlator into five-point conformal blocks in the $\phi_1(x_1)\times \phi_2(x_2)$ and $\phi_4(x_4)\times \phi_5(x_5)$ operator product expansions (following the conventions of \cite{Rosenhaus:2018zqn, Poland:2023vpn, Poland:2023bny}):
\begin{equation}\label{t-channel}
\begin{split}
\langle \phi_1(x_1)\phi_2(x_2)&\phi_3(x_3)\phi_4(x_4)\phi_5(x_5) \rangle = \\
&P(x_i)\sum_{\mathcal{O},\mathcal{O'}}\sum_{n_{IJ}=0}^{{\rm min}(\ell_{\mathcal{O}},\ell_{\mathcal{O'}})} \lambda_{\phi_1\phi_2 \mathcal{O}}\lambda_{\phi_4\phi_5 \mathcal{O}'}\lambda_{\mathcal{O}\phi_3 \mathcal{O'}}^{n_{IJ}} G_{(\mathcal{O},\mathcal{O'})}^{(n_{IJ})}(u_1', v_1', u_2', v_2', w'),
\end{split}
\end{equation}
where
\begin{equation}\label{crosratiodef}
\begin{split}
&P(x_i)=\frac{1}{x_{12}^{\Delta_1+\Delta_2}x_{34}^{\Delta_3}x_{45}^{\Delta_4+\Delta_5}}\left( \frac{x_{23}}{x_{13}}\right)^{\Delta_1-\Delta_2}\left( \frac{x_{24}}{x_{23}}\right)^{\Delta_3}
\left( \frac{x_{35}}{x_{34}}\right)^{\Delta_4-\Delta_5}, \qquad x_{ij}=x_i-x_j,\\
&u_1'=\frac{x_{12}^{2}x_{34}^{2}}{x_{13}^{2}x_{24}^{2}},\quad v_1'=\frac{x_{14}^{2}x_{23}^{2}}{x_{13}^{2}x_{24}^{2}},\quad u_2'=\frac{x_{23}^{2}x_{45}^{2}}{x_{24}^{2}x_{35}^{2}},\quad v_2'=\frac{x_{25}^{2}x_{34}^{2}}{x_{24}^{2}x_{35}^{2}},\quad w'=\frac{x_{15}^{2}x_{23}^{2}x_{34}^{2}}{x_{24}^{2}x_{13}^{2}x_{35}^{2}},
\end{split}
\end{equation}
and $\ell_{\CO}$ denotes the spin of the operator $\CO$. Here $P(x_i)$ labels the external dimension-dependent prefactor (or leg factor), while $G_{(\mathcal{O},\mathcal{O'})}^{(n_{IJ})}(u_1', v_1', u_2', v_2', w')$ represents the five-point conformal block that encodes the contribution of the exchanged operators $\mathcal{O}$ and $\mathcal{O'}$ in the OPE. Equivalently, the five-point correlator can be expanded using the $\phi_1(x_1)\times \phi_4(x_4)$ and $\phi_2(x_2)\times \phi_5(x_5)$ OPEs, giving
\begin{equation}\label{s-channel}
\begin{split}
\langle \phi_1(x_1)\phi_2(x_2)&\phi_3(x_3)\phi_4(x_4)\phi_5(x_5) \rangle = \\
&\tilde{P}(x_i)\sum_{\mathcal{O},\mathcal{O'}}\sum_{n_{IJ}=0}^{{\rm min}(\ell_{\mathcal{O}},\ell_{\mathcal{O'}})} \lambda_{\phi_1\phi_4 \mathcal{O}}\lambda_{\phi_2\phi_5 \mathcal{O}'}\lambda_{\mathcal{O}\phi_3 \mathcal{O'}}^{n_{IJ}} G_{(\mathcal{O},\mathcal{O'})}^{(n_{IJ})}(v_1', u_1', v_2', u_2', w'),
\end{split}
\end{equation}
where $\tilde{P}$ is obtained from $P$ by $\phi_2(x_2) \leftrightarrow \phi_4(x_4)$ exchange. The associativity of the operator product expansion implies that the correlator is independent of the choice of OPEs used to expand it, so \eqref{t-channel} must be equal to \eqref{s-channel}. This requirement is referred to as the crossing relation and can be expressed as a sum rule
\begin{equation}\label{crossing-relation}
\sum_{(\mathcal{O},\mathcal{O'})}\sum_{n_{IJ}=0}^{{\rm min}(\ell_{\mathcal{O}},\ell_{\mathcal{O'}})}\lambda_{\mathcal{O}\phi_3 \mathcal{O'}}^{n_{IJ}} F_{(\mathcal{O},\mathcal{O'})}^{(n_{IJ})}(u_1', v_1', u_2', v_2', w') = 0,
\end{equation}
where
\begin{equation}
\begin{split}
F_{(\mathcal{O},\mathcal{O'})}^{(n_{IJ})}(u_1', v_1', u_2', v_2', w') =\,& \lambda_{\phi_1\phi_2 \mathcal{O}}  \lambda_{\phi_4 \phi_5 \mathcal{O}'}\, {v_1'}^{\frac{\Delta_1+\Delta_4}{2}} {v_2'}^{\frac{\Delta_2+\Delta_5}{2}} G_{(\mathcal{O},\mathcal{O'})}^{(n_{IJ})}(u_1', v_1', u_2', v_2', w') \\
&- \lambda_{\phi_1\phi_4 \mathcal{O}}  \lambda_{\phi_2 \phi_5 \mathcal{O}'}\, {u_1'}^{\frac{\Delta_1+\Delta_2}{2}} {u_2'}^{\frac{\Delta_4+\Delta_5}{2}}  G_{(\mathcal{O},\mathcal{O'})}^{(n_{IJ})}(v_1', u_1', v_2', u_2', w').
\end{split}
\end{equation}
Here, we use the standard box basis \cite{Costa:2011mg} to describe the conformal blocks corresponding to different conformally-invariant tensor structures in the $\langle\CO \phi_3 \CO' \rangle$ correlator labeled by $n_{IJ}$. Our five-point conformal block normalization is given by eq.~(2.7) of~\cite{Poland:2023bny}. This normalization is consistent with the normalization of the four-point block given by eq.~(52) of~\cite{Poland:2018epd}.

We will study the crossing relation \eqref{crossing-relation} for various five-point correlators in the critical 3d Ising model. We shall calculate a number of previously unknown OPE coefficients by truncating the sum over exchanged primary operators and only including a small subset of pairs of primary operators $\mathcal{S}_{\rm Ising}^{\langle\phi_1\phi_2\phi_3\phi_4\phi_5 \rangle}$ that we explicitly specify for each correlator we consider. The truncated contributions will be approximated using the corresponding contributions in a suitable disconnected five-point correlator, defined by
\begin{equation}
\langle \phi_1(x_1)\phi_2(x_2)\phi_3(x_3)\phi_4(x_4)\phi_5(x_5) \rangle_{\rm d} \equiv \langle \phi_1(x_1)\phi_2(x_2) \rangle \langle \phi_3(x_3)\phi_4(x_4)\phi_5(x_5) \rangle + {\rm perm.}
\end{equation}
In order to roughly approximate the infinite number of truncated contributions in the critical Ising correlators using the disconnected contributions, we subtract the contributions from the disconnected correlator that are analogs of those included in the Ising case, and we denote these pairs of operators  by $\mathcal{S}_{\rm disc.}^{\langle\phi_1\phi_2\phi_3\phi_4\phi_5 \rangle}$. In this way, we make explicit use of the fact that the disconnected correlator is crossing symmetric, so that we only need to treat a finite number of contributions in both the critical Ising and disconnected correlators. 

Upon approximating the truncated states by their corresponding contributions in the disconnected correlator, the crossing relation assumes the form
\begin{equation}\label{truncated-crossing}
\begin{split}
&\sum_{(\mathcal{O},\mathcal{O'})\in \mathcal{S}_{\rm Ising}^{\langle\phi_1\phi_2\phi_3\phi_4\phi_5 \rangle}}\sum_{n_{IJ}=0}^{{\rm min}(\ell_{\mathcal{O}},\ell_{\mathcal{O'}})} \lambda_{\mathcal{O}\phi_3 \mathcal{O'}}^{n_{IJ}} F_{(\mathcal{O},\mathcal{O'})}^{(n_{IJ})}(u_1', v_1', u_2', v_2', w') \\
&-\sum_{(\mathcal{O},\mathcal{O'})\in \mathcal{S}_{\rm disc.}^{\langle\phi_1\phi_2\phi_3\phi_4\phi_5 \rangle}}\sum_{n_{IJ}=0}^{{\rm min}(\ell_{\mathcal{O}},\ell_{\mathcal{O'}})} \Lambda_{\mathcal{O}\phi_3 \mathcal{O'}}^{n_{IJ}} F_{(\mathcal{O},\mathcal{O'})}^{(n_{IJ})}(u_1', v_1', u_2', v_2', w') \simeq 0,
\end{split}
\end{equation}
where $\Lambda_{\mathcal{O}\phi_3 \mathcal{O'}}^{n_{IJ}}$ are the OPE coefficients in the disconnected correlator. The unknown parameters in \eqref{truncated-crossing} are a subset of the OPE coefficients in the 3d critical Ising model $\lambda_{\mathcal{O}\phi_3 \mathcal{O'}}^{n_{IJ}}$. We will be fixing all the conformal dimensions of the external and exchanged operators, as well as previously-computed OPE coefficients, to their best known values from~\cite{Simmons-Duffin:2016wlq, Chang:2024whx, Su:unpublised}. We explicitly list all the data we use in Appendix~\ref{a:data}.

Along with the truncated crossing relation, we obtain additional constraints for the unknown OPE coefficients \eqref{truncated-crossing} by taking derivatives of the crossing relation with respect to the cross-ratios. We then evaluate all resulting constraints at the crossing symmetric point 
\begin{equation}\label{symmetric-point}
u_1'=v_1'=u_2'=v_2'=1, \quad w'=\frac{3}{2}.
\end{equation}
The constraints can be schematically represented by
\begin{equation}\label{constraints}
\begin{split}
e_i(\lambda)=\CD_{i}\partial_{a^+}\Bigg(&\sum_{(\mathcal{O},\mathcal{O'})\in \mathcal{S}_{\rm Ising}^{\langle\phi_1\phi_2\phi_3\phi_4\phi_5 \rangle}}\sum_{n_{IJ}=0}^{{\rm min}(\ell_{\mathcal{O}},\ell_{\mathcal{O'}})} \lambda_{\mathcal{O}\phi_3 \mathcal{O'}}^{n_{IJ}} F_{(\mathcal{O},\mathcal{O'})}^{(n_{IJ})}(a^+, b^+, a^-, b^-, w) \\
&-\sum_{(\mathcal{O},\mathcal{O'})\in \mathcal{S}_{\rm disc.}^{\langle\phi_1\phi_2\phi_3\phi_4\phi_5 \rangle}}\sum_{n_{IJ}=0}^{{\rm min}(\ell_{\mathcal{O}},\ell_{\mathcal{O'}})} \Lambda_{\mathcal{O}\phi_3 \mathcal{O'}}^{n_{IJ}} F_{(\mathcal{O},\mathcal{O'})}^{(n_{IJ})}(a^+, b^+, a^-, b^-, w)\Bigg)\Bigg|_{\eqref{configurationab}},
\end{split}
\end{equation}
where $\lambda$ collectively denotes all unknown OPE coefficients and $\CD_i$ labels derivatives taken with respect to the cross-ratios $(a^+, b^+, a^-, b^-, w)$, whose relation to the cross-ratios defined in eq.~\eqref{crosratiodef} is given in Appendix~\ref{parametrization}. In this parametrization, the crossing symmetric point \eqref{symmetric-point} is given by
\begin{equation}\label{configurationab}
a^{+}=1\,,\qquad b^{+}=-3\,, \qquad a^{-}=b^{-}=0\,, \qquad w=\frac{1}{2}\,.
\end{equation}
We observe that one needs to take at least one derivative with respect to the $a^+$ cross-ratio in order to get a non-zero constraint at the crossing symmetric point. Hence, we explicitly separate $\partial_{a^+}$ from $\CD_i$. The numerical values of the five-point conformal blocks and their derivatives are computed using the method described in~\cite{Poland:2023bny}.

Following this approach, we generate more constraints than we have unknown OPE coefficients, so the question becomes how to choose the relevant constraints to solve for the OPE coefficients most accurately. We define a cost function
\begin{equation}\label{costf}
f(r,\lambda)=\left(\frac{1}{\sum_{i=1}^{\mathcal{C}}r_{i}}\right)\sum_{i=1}^{\mathcal{C}}r_{i}\left(\frac{e_{i}(\lambda)}{e_{i}(0)} \right)^2,
\end{equation}
where the $r_i$ are pseudo-randomly generated weights $0\leqslant r_i \leqslant 1$, collectively denoted by $r$. The sum ranges over the subset of constraints obtained by taking derivatives of the truncated crossing relation. We remark that the number of constraints included in the cost function $\CC$ must be larger than the number of unknown OPE coefficients in the system. 

Our approach in this work will be to include the constraints in the cost function $f(r,\lambda)$ that give the largest value of the quantity
\begin{equation}\label{dethessian}
\mathcal{I}=\frac{\det\left( \frac{\partial^2 f(1,L)}{\partial_{L_i}\partial_{L_j}}\Big|_{\rm min} \right)}{f(1,\lambda)|_{\rm min}},
\end{equation}
for each $\mathcal{C}$, where $L$ denotes the rescaled unknown OPE coefficients such that $L=1$ corresponds to the value of the OPE coefficient in the disconnected correlator. The idea is that we seek to select those constraints that yield the minima of the cost function with the fewest nearly-flat directions; hence, we require the determinant of the Hessian matrix at the minima to be as large as possible. We also aim for the crossing relation to be satisfied to the best possible extent at the minima, so we divide by the value of the cost function at the minima. Here, we use unit weights of the constraints in the cost function, so that each constraint is treated on the equal footing. But we also note that averaging over different weights would give the same results, since we are using multiple sets of constraints for each $\CC$. Rescaling the OPE coefficients ensures that at the minima of the cost function, the unknown $L$'s are $\CO(1)$, which eliminates the artificial flat directions that would otherwise be present and would correspond to OPE coefficients with small absolute values.

In practice, we calculate a large fixed set of constraints and subsequently scan through the subsets containing $\CC$ constraints to determine which ones give the largest $\CI$. For each $\CC$, we select the constraints with the largest $\CI$ and calculate the unknown OPE coefficients by randomly varying the weights of each constraint in the cost function. We expect that the number of constraints $\CC$ in the cost function needs to be large enough to lift potential flat directions, since in some correlators for low $\CC$ even the sets with the largest $\CI$ will come with almost flat directions, which lead to large standard deviations in the OPE coefficients. At the same time, $\CC$ cannot be too large so that we have too many constraints in the cost function, which make it very difficult to satisfy the truncated crossing relation. 

To reach optimal balance, we choose the bounds on the values of $\CC$ and the number of subsets for each $\CC$ to be included in the final averaging such that the results for the calculated OPE coefficients stabilize, meaning that both the mean values and their standard deviation calculated for each $\CC$ become roughly the same. Then the total average over all $\CC$'s included becomes nearly independent of the number of sets we are using for each $\CC$. Finally, we merge all results across the different $\CC$'s and compute the total mean and the standard deviation of each unknown OPE coefficient. The OPE coefficients that come with relatively small standard deviations are deemed to be accurate, since these are insensitive to the number of constraints $\CC$ in the cost function that we use or to the particular choice of constraints. Those that come with relatively large error bars are obviously sensitive to these choices. Our interpretation is that our method cannot accurately predict these coefficients.

\section{$\sigma \times \sigma$ OPE}\label{secthree}

In this section we use the truncation method to study the five-point correlators $\langle \sigma \sigma \epsilon \sigma \sigma \rangle$ and $\langle \sigma \sigma \epsilon' \sigma \sigma \rangle$. Here, we only consider taking the $\sigma \times \sigma$ OPE, where we expect that the truncation method with truncated operators approximated by their counterparts in the disconnected correlator should work well, as established in~\cite{Li:2023tic, Poland:2023bny}. We are at liberty to use one irrelevant operator as an external state, since we do not consider the OPE that involves this state, and therefore the convergence of the OPE is not significantly affected. With this said, we do observe somewhat larger error bars for the unknown OPE coefficients in this case, which is likely a consequence of the fact that the irrelevant external operator changes the dominant contributions in the OPEs, as we will see below.

\subsection{$\langle \sigma \sigma \epsilon \sigma \sigma \rangle$}

We begin by examining the $\langle \sigma \sigma \epsilon \sigma \sigma \rangle$ correlator, previously studied in \cite{Poland:2023bny, Poland:2023vpn}, but now using the method described in section~\ref{sectwo}. In particular, we will not treat $\Delta_\sigma$ as an unknown parameter to guide our selection of constraints to include in the cost function as in these previous works, but instead choose the constraints with the largest ratio of the determinant of the Hessian matrix at the minima of the cost function to the value of the cost function at the minima, given by \eqref{dethessian}. 

In the $\sigma \times \sigma$ OPEs we including the following contributions in the truncated crossing relation 
\begin{equation}\label{Sssess}
\begin{split}
\mathcal{S}_{\rm Ising}^{\langle\sigma \sigma \epsilon \sigma \sigma \rangle} = \{&\mathbf{1}, \epsilon, \epsilon', T_{\mu\nu}, C_{\mu\nu\rho\sigma}| \,\, {\rm all \,\, pairs},\\
& (\epsilon, S_6), (S_6, \epsilon),  (\epsilon, X_{\ell}), (X_{\ell}, \epsilon) \, | \, \ell=\{8, 10\},\\
&(T_{\mu\nu}, S_{6}), (S_6, T_{\mu\nu}) \}.
\end{split}
\end{equation}
The first line of this equation contains the states with conformal dimension below $\Delta \leqslant 5.03$. In the second line we include $\epsilon$ contributions paired with operators of leading twist and spin from 6 to 10. The reason we include these is that the double-trace operator $[\sigma, \sigma]_{0,0}$ in the disconnected correlator is not a very good approximation for $\epsilon$, since $\epsilon$ has a large anomalous dimension. In the third line we include the leading-twist spin-6  operator $S_6$ paired with the stress tensor. The stress tensor Ward identity relates the OPE coefficients of the three tensor structures that appear in the $\langle T_{\mu\nu} \epsilon S_6 \rangle$ correlator, leaving only one unknown (see eq.~(B.4) and (B.5) in~\cite{Meltzer:2018tnm}). Including the rest of the spin-6 contributions would introduce many more unknown OPE coefficients, which would consequently require many new constraints in order to lift the flat directions in the cost function. These new constraints may be even more sensitive to the further truncated contributions, so it is unclear how to consistently truncate the system and select the best constraints in this case. 

To avoid this issue, we approximate the rest of the contributions in the OPEs by the corresponding contributions in the disconnected correlator
\begin{equation}
\langle \sigma(x_1)\sigma(x_2)\epsilon(x_3)\sigma(x_4)\sigma(x_5) \rangle_{\rm d} = \langle \sigma(x_1)\sigma(x_2) \rangle \langle \epsilon(x_3)\sigma(x_4)\sigma(x_5) \rangle + {\rm perm.}
\end{equation}
The pairs of operators that contribute to the disconnected correlator in the operator product expansion are $(\mathbf{1}, \epsilon)$, $(\epsilon,\mathbf{1})$ and $([\sigma,\sigma]_{n,\ell},[\sigma,\sigma]_{n',\ell'})$. 
In order to approximate the truncated operators in the critical Ising $\langle\sigma\sigma\epsilon\sigma\sigma \rangle$ correlator, we need to subtract the following contributions from the disconnected correlator:
\begin{equation}\label{ssessSdisconnected}
\begin{split}
\mathcal{S}_{\rm disc.}^{\langle\sigma \sigma \epsilon \sigma \sigma \rangle} = \{
& (\mathbf{1}, \epsilon), (\epsilon, \mathbf{1}),\\
&[\sigma,\sigma]_{0,0}, [\sigma,\sigma]_{1,0}, [\sigma,\sigma]_{0,2}, [\sigma,\sigma]_{0,4} |\,\, {\rm all \,\, pairs},\\
& ([\sigma,\sigma]_{0,0}, [\sigma,\sigma]_{0,\ell}), ([\sigma,\sigma]_{0,\ell}, [\sigma,\sigma]_{0,0}) \, | \, \ell=\{6, 8, 10\},\\
&([\sigma,\sigma]_{0,2}, [\sigma,\sigma]_{0,6}), ([\sigma,\sigma]_{0,6}, [\sigma,\sigma]_{0,2}) \}.
\end{split}
\end{equation}
We rescale the unknown OPE coefficients 
\begin{equation}\label{rescaling}
\begin{split}
&\lambda_{T\epsilon T}^0 = \Lambda_{[\sigma, \sigma]_{0,2}\epsilon [\sigma, \sigma]_{0,2}}^0 L_{T\epsilon T}^0, \qquad \lambda_{T\epsilon C}^0 = \Lambda_{[\sigma, \sigma]_{0,2}\epsilon [\sigma, \sigma]_{0,4}}^{0} L_{T\epsilon C}^0,\\
&\lambda_{T\epsilon S_6}^0 = \Lambda_{[\sigma, \sigma]_{0,2}\epsilon [\sigma, \sigma]_{0,6}}^0 L_{T\epsilon S_6}^0, \qquad \lambda_{C\epsilon C}^{n_{IJ}} = \Lambda_{[\sigma, \sigma]_{0,4}\epsilon [\sigma, \sigma]_{0,4}}^{n_{IJ}}  L_{C\epsilon C}^{n_{IJ}}, \\
&\lambda_{\epsilon'\epsilon C} = \Lambda_{[\sigma, \sigma]_{1,0}\epsilon [\sigma, \sigma]_{0,4}}^{0} L_{\epsilon'\epsilon C},
\end{split}
\end{equation}
where $\Lambda$ denotes the corresponding OPE coefficient in the disconnected correlator. 

\begin{figure}[t]
\centering
\includegraphics[width=0.315\textwidth]{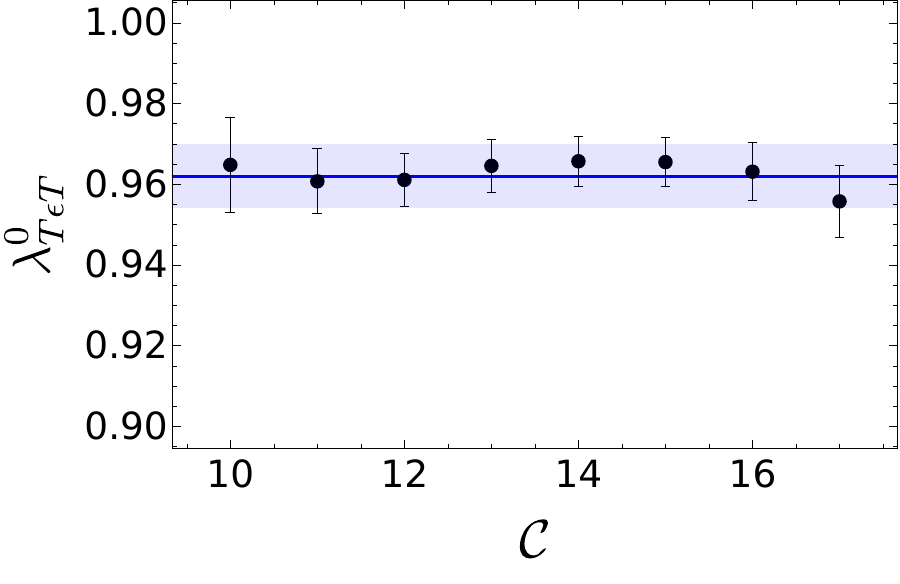}
\hspace{0.01\textwidth}
\includegraphics[width=0.315\textwidth]{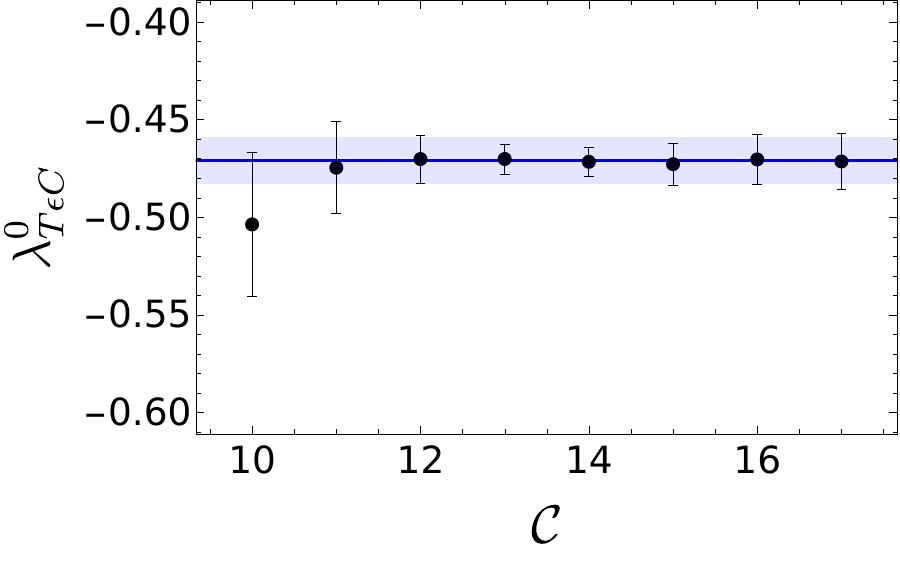}
\hspace{0.01\textwidth}
\includegraphics[width=0.315\textwidth]{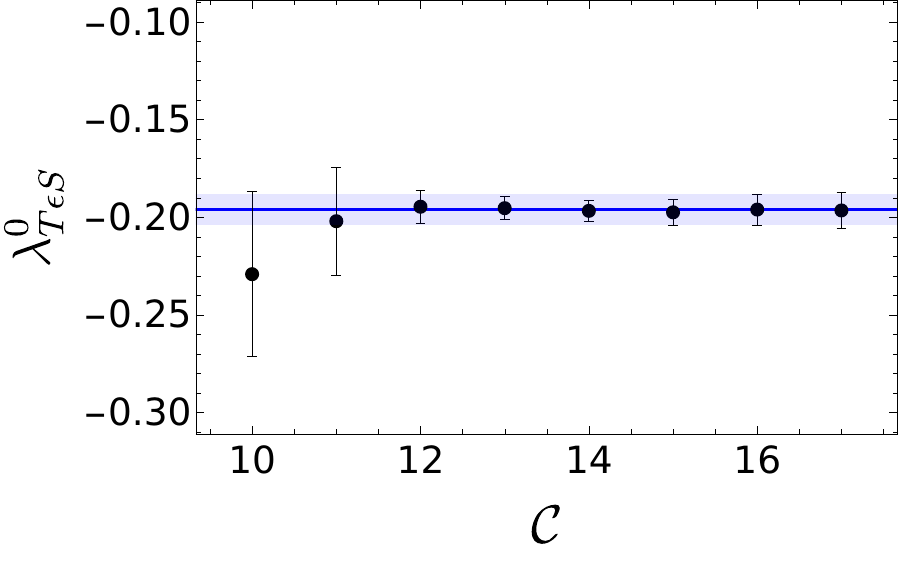}\\
\includegraphics[width=0.315\textwidth]{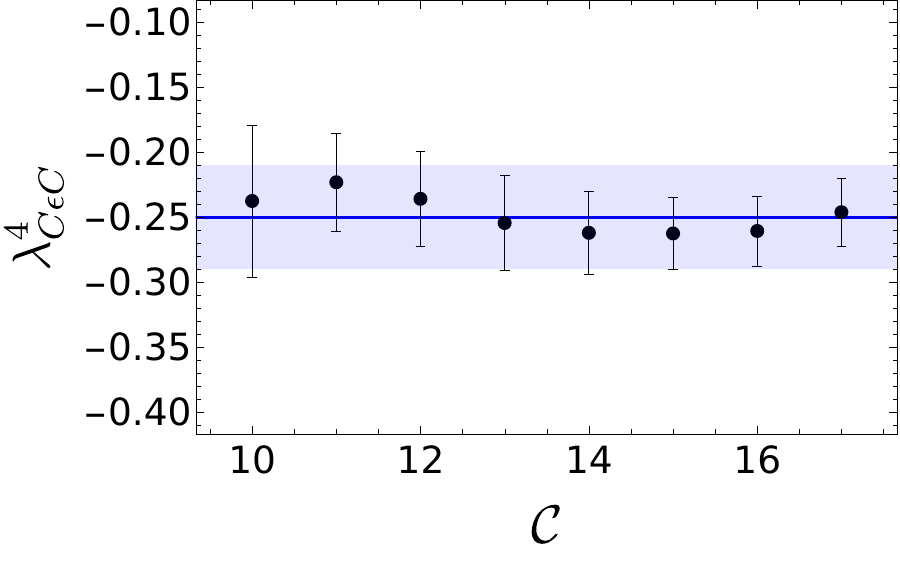}
\hspace{0.01\textwidth}
\includegraphics[width=0.315\textwidth]{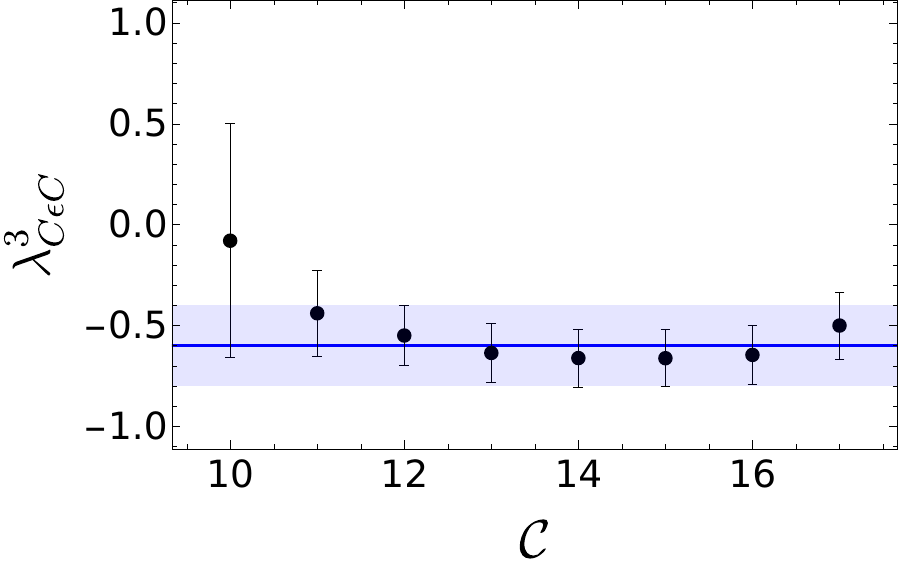}
\hspace{0.01\textwidth}
\includegraphics[width=0.315\textwidth]{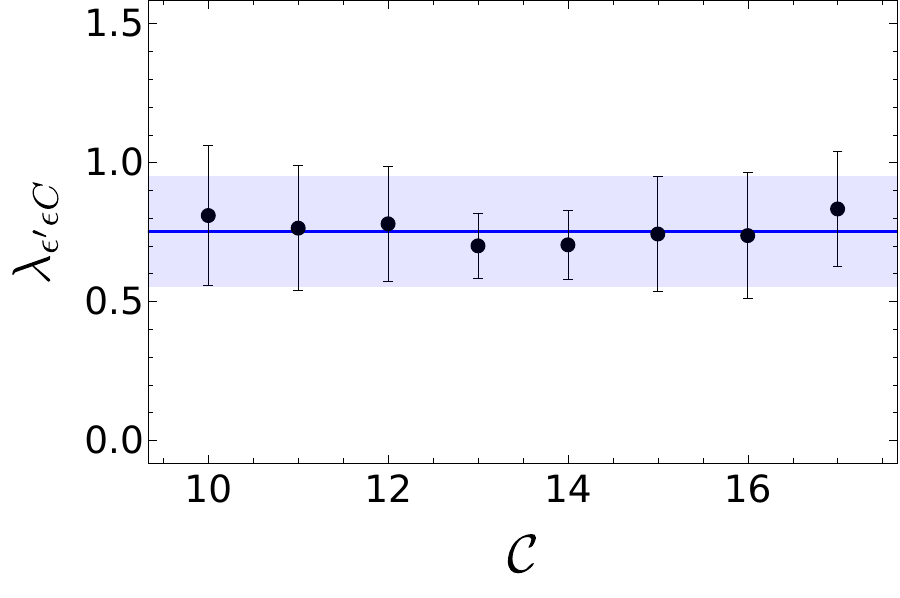}
\caption{{OPE coefficients computed from $\langle\sigma\sigma\epsilon\sigma\sigma\rangle$ by averaging over the minima of the cost function for the 20 sets of $\CC$ constraints with the largest $\CI$, where the weights of individual constraints in each set are pseudo-randomly varied. Here, $\lambda_{T\epsilon T}$ is treated as an unknown. The horizontal blue lines and shaded regions correspond to the results given in table~\ref{ssess}, left. These values are computed by averaging the results with $12\leqslant \CC \leqslant 17$. One can notice that the standard deviations for multiple OPE coefficients are large for low values of $\CC$, which is a consequence of the presence of flat directions of the cost function near the minima. Adding more constraints lifts these flat directions. Other OPE coefficients show similar behavior with larger standard deviations.}}
\label{fig:ssess-1}
\end{figure}

\begin{figure}[t]
\centering
\includegraphics[width=0.315\textwidth]{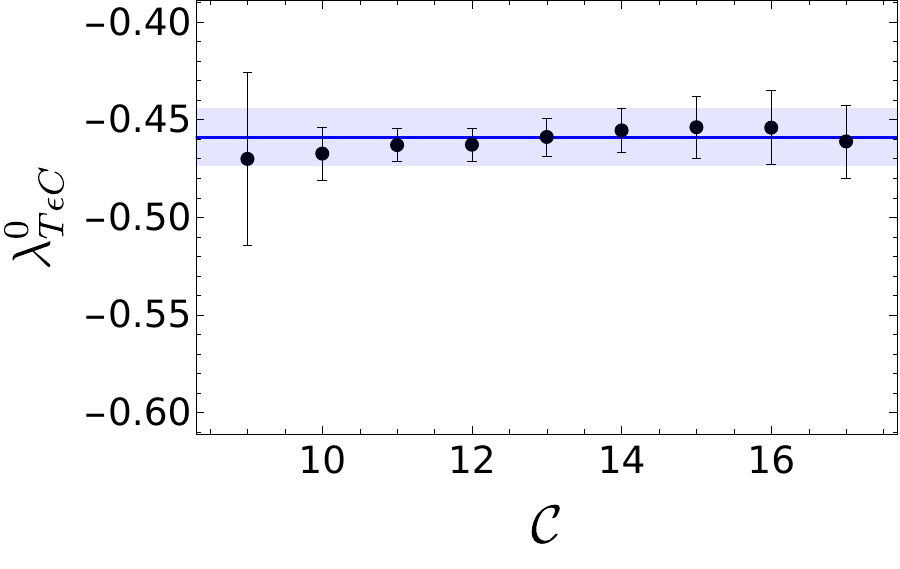}
\hspace{0.01\textwidth}
\includegraphics[width=0.315\textwidth]{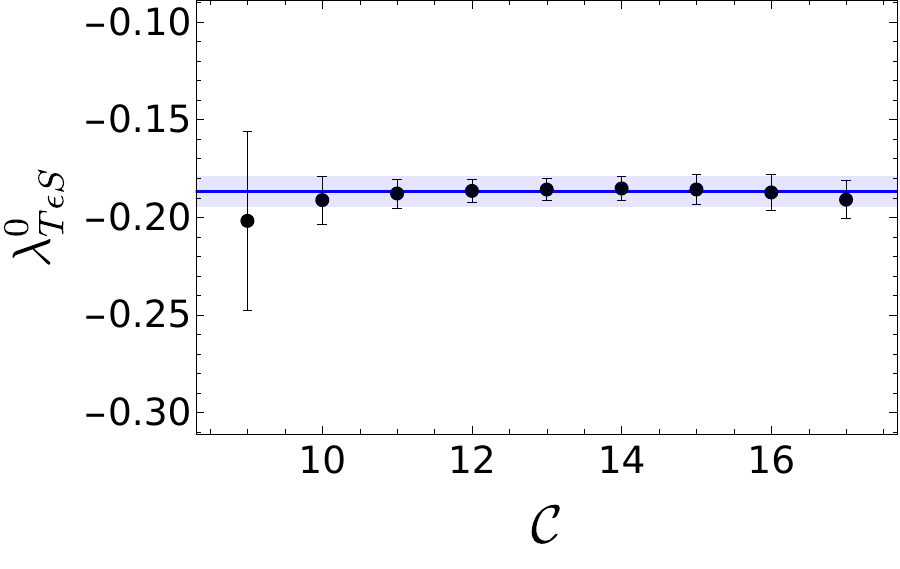}
\hspace{0.01\textwidth}
\includegraphics[width=0.315\textwidth]{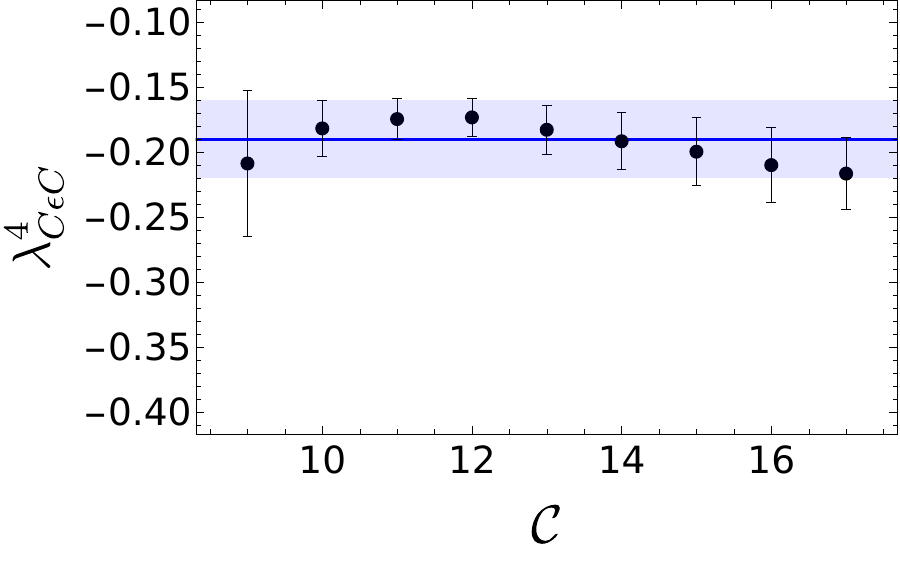}
\caption{{Similar to fig.~\ref{fig:ssess-1} but with $\lambda_{T\epsilon T}$ fixed to its best known value from \cite{Chang:2024whx}. The horizontal blue lines and shaded regions correspond to the results given in table~\ref{ssess}, right. These values are computed by averaging the results from 20 sets of constraints with $11\leqslant \CC \leqslant 17$. One can notice that the standard deviations for multiple OPE coefficients are large for low values of $\CC$, which is a consequence of the presence of flat directions of the cost function near the minima. Adding more constraints lift these flat directions. Other OPE coefficients show similar behavior with larger standard deviations.}}
\label{fig:ssess-2}
\end{figure}

We will consider the constraints obtained by taking up to three  derivatives in $\CD_i$. In this way, we obtain a total of 19 possible constraints. Then, for each $\mathcal{C}$, we identify the sets of constraints with the largest $\mathcal{I}$ and select 20 of them. We assign (pseudo-)randomly generated weights $r_i$ to the  constraints included in the cost function and subsequently  combine the results for all 20 sets and all $\mathcal{C}$'s such that $12\leqslant \mathcal{C} \leqslant 17$.  We remark that changing the number of sets only shifts the results for the unknown OPE coefficients within the estimated error bars. Hence, the results are not sensitive to this choice. We also repeat the calculation with $\lambda_{T\epsilon T}^0$ fixed to the central value found in \cite{Chang:2024whx}. In this case, we take $11\leqslant \mathcal{C} \leqslant 17$. The overall results are shown in table~\ref{ssess}, and the behavior of the coefficients with changing $\CC$ is shown in figs.~\ref{fig:ssess-1} and~\ref{fig:ssess-2}.

\begin{table}[t!]
\centering
\begin{minipage}{0.48\textwidth}
\centering
\begin{tabular}{|l|l|}
\hline
                                                   & est.\,values \\ \hline
$\lambda_{T\epsilon T}^{0}$      				   & 0.962(8)         \\
$\lambda_{T\epsilon C}^{0}$      				   & -0.471(12)         \\
$\lambda_{T\epsilon S}^{0}$      				   & -0.196(8)         \\
$\lambda_{C\epsilon C}^{4}$   					   & -0.25(4)         \\
$\lambda_{C\epsilon C}^{3}$   					   & -0.6(2)           \\
$\lambda_{C\epsilon C}^{2}$   					   & 1.8(8)           \\
$\lambda_{C\epsilon C}^{1}$  				       & -8(2)          \\
$\lambda_{C\epsilon C}^{0}$   					   & -2.9(9)          \\ 
$\lambda_{\epsilon' \epsilon C}$				   & 0.75(20)          \\ \hline
\end{tabular}
\end{minipage}
\hfill
\begin{minipage}{0.48\textwidth}
\centering
\begin{tabular}{|l|l|}
\hline
                                                   & est.\,values \\ \hline
$\lambda_{T\epsilon C}^{0}$      				   & -0.459(15)         \\
$\lambda_{T\epsilon S}^{0}$      				   & -0.187(8)         \\
$\lambda_{C\epsilon C}^{4}$   					   & -0.19(3)         \\
$\lambda_{C\epsilon C}^{3}$   					   & -0.5(2)           \\
$\lambda_{C\epsilon C}^{2}$   					   & 2(1)           \\
$\lambda_{C\epsilon C}^{1}$  				       & -7(2)          \\
$\lambda_{C\epsilon C}^{0}$   					   & -2(2)          \\ 
$\lambda_{\epsilon' \epsilon C}$				   & 0.79(22)          \\ \hline
\end{tabular}
\end{minipage}
\caption{{Left: Numerical data for unknown OPE coefficients in the truncated $\langle\sigma\sigma\epsilon\sigma\sigma \rangle$ correlator, obtained from the 20 sets of constraints with the largest $\mathcal{I}$ for $12\leqslant \mathcal{C}\leqslant 17$. The correlation matrix $\hat{\rho}_1$ for the $\lambda_{C\epsilon C}^{n_{IJ}}$ OPE coefficients is given in Appendix~\ref{correlationmatrix}. Right: Numerical data for unknown OPE coefficients in the truncated $\langle\sigma\sigma\epsilon\sigma\sigma \rangle$ correlator, with $\lambda_{T\epsilon T}^0$ fixed to the central value from~\cite{Chang:2024whx}. We use the 20 sets of constraints with the largest $\mathcal{I}$ for $11\leqslant \mathcal{C}\leqslant 17$. The correlation matrix $\hat{\rho}_2$ for the $\lambda_{C\epsilon C}^{n_{IJ}}$ OPE coefficients is given in Appendix~\ref{correlationmatrix}.}}\label{ssess}
\end{table}

We observe that the results given here agree with the results found in \citep{Poland:2023vpn}, confirming the validity of the method for selecting constraints in the cost function that we employ here. The minor shift in the mean value of $\lambda^0_{T\epsilon T}$ that we see is due to the fact that we fix $\Delta_\sigma$, which gives a slightly less accurate result for this coefficient. Nevertheless, the value we determine is still well within the error bar of the result in \citep{Poland:2023vpn}. Note that we use a convention where $\lambda_{\sigma\sigma T}<0$ and $\lambda_{\epsilon\epsilon T}<0$, which correlates with the signs of other OPE coefficients containing a single $T$ being negative. 

The coefficient $\lambda_{T\epsilon T}^0$ has been computed using the fuzzy sphere regularization approach~\cite{Hu:2023xak}, which gave the value $\lambda_{T\epsilon T}^0 \simeq 0.9162(73)$. It was also very precisely computed using the four-point bootstrap for the mixed bootstrap system containing $\sigma$, $\epsilon$, and $T_{\mu\nu}$ in \cite{Chang:2024whx}, which resulted in the value $\lambda_{T\epsilon T}^0 = 0.95331513(42)$.

Additionally, the following linear combination of the $\lambda_{C\epsilon C}^{n_{IJ}}$ OPE coefficients
\begin{equation}
f_{C\epsilon C}=\lambda_{C\epsilon C}^{4}-\frac{1}{3}\lambda_{C\epsilon C}^{3}+\frac{2}{15}\lambda_{C\epsilon C}^{2}-\frac{2}{35}\lambda_{C\epsilon C}^{1}+\frac{8}{315}\lambda_{C\epsilon C}^{0}
\end{equation}
has been recently roughly estimated in \cite{Lauchli:2025fii} to be around $0.5$. Here we find $f_{C\epsilon C}\simeq 0.57(5)$ when we treat $\lambda_{T\epsilon T}^0$ as an unknown (left, table~\ref{ssess}), or $f_{C\epsilon C}\simeq 0.59(6)$ with the $\lambda_{T\epsilon T}^0$ coefficient fixed to the central value from \cite{Chang:2024whx} (right, table~\ref{ssess}).

\subsection{$\langle \sigma \sigma \epsilon' \sigma \sigma \rangle$}\label{sec-ssepss}
Next we consider the $\langle \sigma \sigma \epsilon' \sigma \sigma \rangle$ correlator. To begin with, we truncate the $\sigma \times \sigma$ OPEs by including the following operators in the truncated crossing relation:
\begin{equation}\label{isingS-ep-one}
\mathcal{S}_{\rm Ising}^{\langle\sigma \sigma \epsilon' \sigma \sigma \rangle} =\{\mathbf{1}, \epsilon, \epsilon', T_{\mu\nu}, C_{\mu\nu\rho\sigma}| \,\, {\rm all \,\, pairs}\}.
\end{equation}
We also carry out the analogous calculation by adding some of the spin-6 contributions, as we did in our analysis of $\langle \sigma \sigma \epsilon \sigma \sigma \rangle$. In particular, we take
\begin{equation}\label{isingS-ep-two}
\begin{split}
\mathcal{S}_{\rm Ising}^{\langle\sigma \sigma \epsilon' \sigma \sigma \rangle} =\{&\mathbf{1}, \epsilon, \epsilon', T_{\mu\nu}, C_{\mu\nu\rho\sigma}| \,\, {\rm all \,\, pairs},\\
&(\epsilon, S_6), (S_6, \epsilon), (T_{\mu\nu}, S_6), (S_6, T_{\mu\nu})  \},
\end{split}
\end{equation}
and approximate the rest by the corresponding contributions in the disconnected correlator 
\begin{equation}
\langle \sigma(x_1)\sigma(x_2)\epsilon'(x_3)\sigma(x_4)\sigma(x_5) \rangle_{\rm d} = \langle \sigma(x_1)\sigma(x_2) \rangle \langle \epsilon'(x_3)\sigma(x_4)\sigma(x_5) \rangle + {\rm perm.}
\end{equation}
The pairs of operators that contribute to the disconnected correlator in the operator product expansion are $(\mathbf{1}, \epsilon')$, $(\epsilon',\mathbf{1})$, and $([\sigma,\sigma]_{n,\ell},[\sigma,\sigma]_{n',\ell'})$. 
In order to approximate the truncated operators in the critical Ising $\langle\sigma\sigma\epsilon' \sigma\sigma \rangle$ correlator, we need to subtract the following contributions from the disconnected correlator when we use \eqref{isingS-ep-one}:
\begin{equation}\label{disconnectedS-ep-one}
\begin{split}
\mathcal{S}_{\rm disc.}^{\langle\sigma \sigma \epsilon' \sigma \sigma \rangle} = \{&[\sigma,\sigma]_{0,0}, [\sigma,\sigma]_{1,0}, [\sigma,\sigma]_{0,2}, [\sigma,\sigma]_{0,4} |\,\, {\rm all \,\, pairs},\\
& (\mathbf{1}, \epsilon'), (\epsilon', \mathbf{1}) \},
\end{split}
\end{equation}
or when we use \eqref{isingS-ep-two} it is enlarged to
\begin{equation}\label{disconnectedS-ep-two}
\begin{split}
\mathcal{S}_{\rm disc.}^{\langle\sigma \sigma \epsilon' \sigma \sigma \rangle} = \{&[\sigma,\sigma]_{0,0}, [\sigma,\sigma]_{1,0}, [\sigma,\sigma]_{0,2}, [\sigma,\sigma]_{0,4} |\,\, {\rm all \,\, pairs},\\
& (\mathbf{1}, \epsilon'), (\epsilon', \mathbf{1}),\\
& ([\sigma,\sigma]_{0,0},[\sigma,\sigma]_{0,6}), ([\sigma,\sigma]_{0,6},[\sigma,\sigma]_{0,0}), ([\sigma,\sigma]_{0,2},[\sigma,\sigma]_{0,6}), ([\sigma,\sigma]_{0,6},[\sigma,\sigma]_{0,2}) \}.
\end{split}
\end{equation}

We take the unknown OPE coefficients to be $\lambda_{T\epsilon' T}^{0}$, $\lambda_{T\epsilon' C}^{0}$, $\lambda_{T\epsilon' S}^{0}$, $\lambda_{C\epsilon' C}^{n_{IJ}}$, $\lambda_{\epsilon \epsilon' C}$, $\lambda_{\epsilon \epsilon' S}$, and $\lambda_{\epsilon' \epsilon' C}$, which we rescale as
\begin{equation}\label{rescaling}
\begin{split}
&\lambda_{T\epsilon' T}^0 = \Lambda_{[\sigma, \sigma]_{0,2}\epsilon' [\sigma, \sigma]_{0,2}}^0 L_{T\epsilon' T}^0, \qquad \lambda_{T\epsilon' C}^0 = \Lambda_{[\sigma, \sigma]_{0,2}\epsilon' [\sigma, \sigma]_{0,4}}^{0} L_{T\epsilon' C}^0,\\
& \lambda_{C\epsilon' C}^{n_{IJ}} = \Lambda_{[\sigma, \sigma]_{0,4}\epsilon' [\sigma, \sigma]_{0,4}}^{n_{IJ}}  L_{C\epsilon' C}^{n_{IJ}}, \qquad \lambda_{\epsilon\epsilon' C} = \Lambda_{[\sigma, \sigma]_{0,0}\epsilon' [\sigma, \sigma]_{0,4}}^{0} L_{\epsilon\epsilon' C},\\
& \lambda_{\epsilon' \epsilon' C}^0 = \Lambda_{[\sigma, \sigma]_{1,0}\epsilon' [\sigma, \sigma]_{0,4}}^{0} L_{\epsilon' \epsilon' C}, \qquad \lambda_{T\epsilon' S}^0 = \Lambda_{[\sigma, \sigma]_{0,2}\epsilon' [\sigma, \sigma]_{0,6}}^0 L_{T\epsilon' S}^0\\
& \lambda_{\epsilon \epsilon' S} = \Lambda_{[\sigma, \sigma]_{0,0}\epsilon' [\sigma, \sigma]_{0,6}}^{0} L_{\epsilon \epsilon' S},
\end{split}
\end{equation}
where $\Lambda$ denotes the corresponding OPE coefficient in the disconnected correlator. As was done above, we consider the constraints obtained by taking up to three  derivatives in $\CD_i$.

\begin{figure}[t]
\centering
\includegraphics[width=0.315\textwidth]{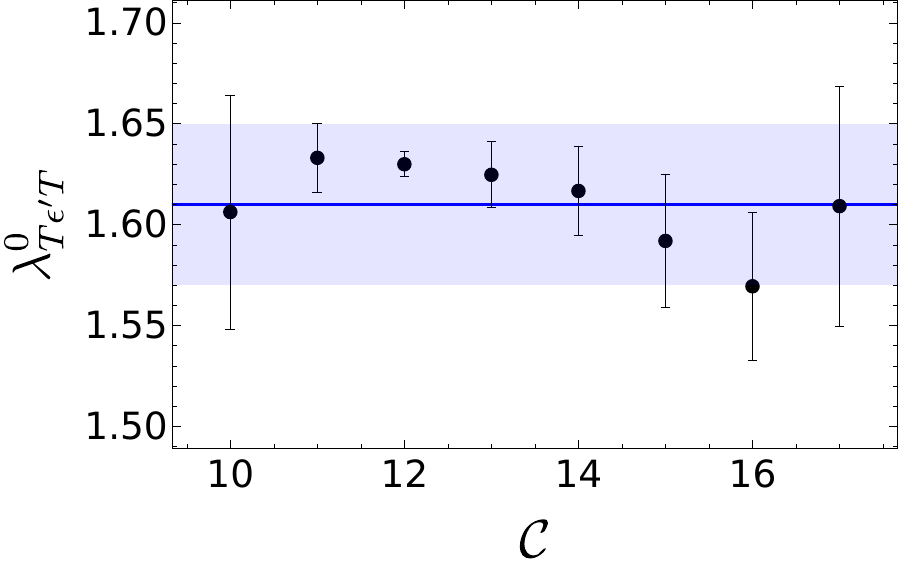}
\hspace{0.01\textwidth}
\includegraphics[width=0.315\textwidth]{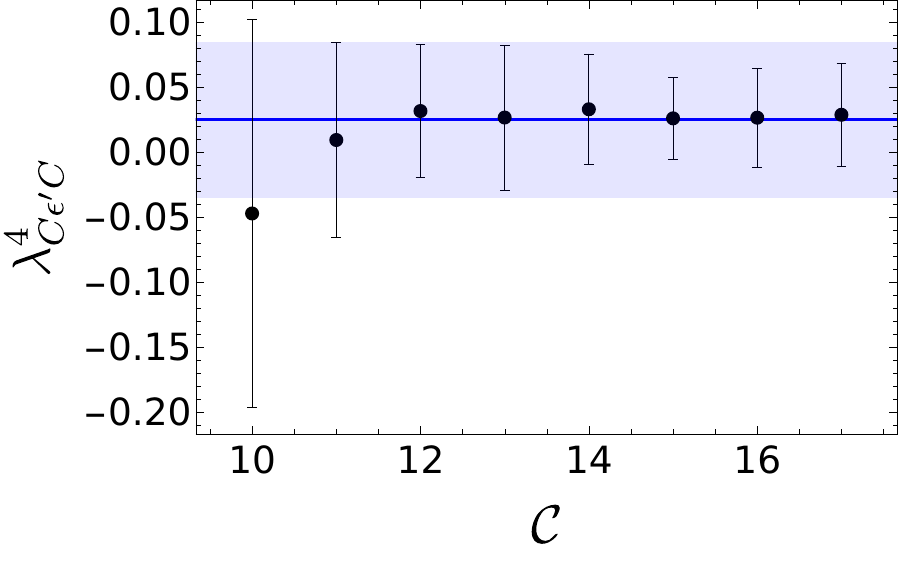}
\hspace{0.01\textwidth}
\includegraphics[width=0.315\textwidth]{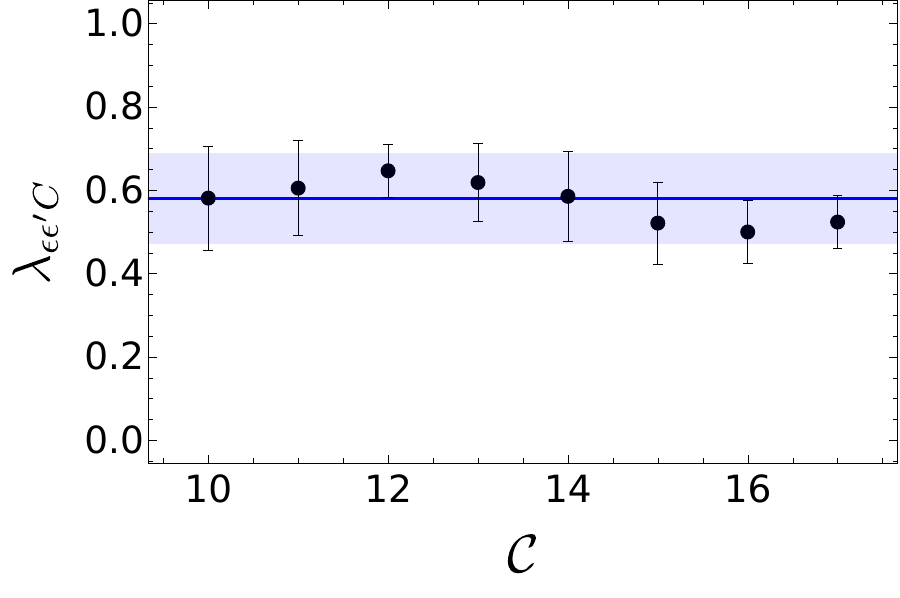}
\caption{{OPE coefficients computed from $\langle\sigma\sigma\epsilon'\sigma\sigma \rangle$ (without spin-6 contributions) by averaging over the positions of the minima of the cost function for 20 sets of $\CC$ constraints with the largest $\CI$. The horizontal blue lines and shaded regions correspond to the results given in table~\ref{ssepss}, left. These values are computed by averaging results with $11\leqslant \CC \leqslant 16$. One can again notice that the standard deviations for multiple OPE coefficients are large for low values of $\CC$, which is a consequence of the presence of flat directions of the cost function near the minima. Adding more constraints lifts these flat directions. Additionally, for large $\CC$ some the OPE coefficients have large standard deviations; in these cases, we have too many constraints in the cost function that we are unable to satisfy at this level of the OPE truncation. Other OPE coefficients show similar behavior with larger standard deviations.}}
\label{fig:ssepss-1}
\end{figure}

\begin{figure}[t]
\centering
\includegraphics[width=0.315\textwidth]{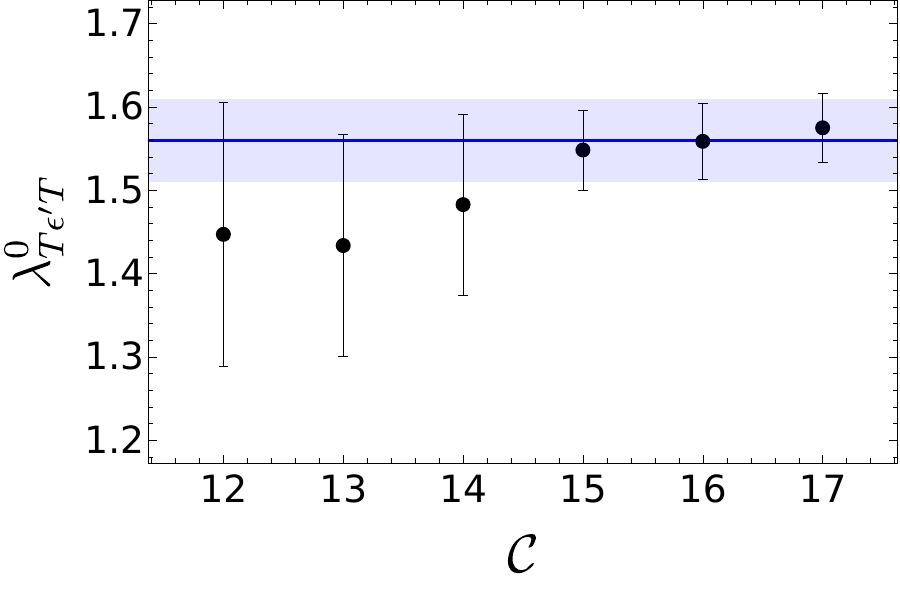}
\hspace{0.01\textwidth}
\includegraphics[width=0.315\textwidth]{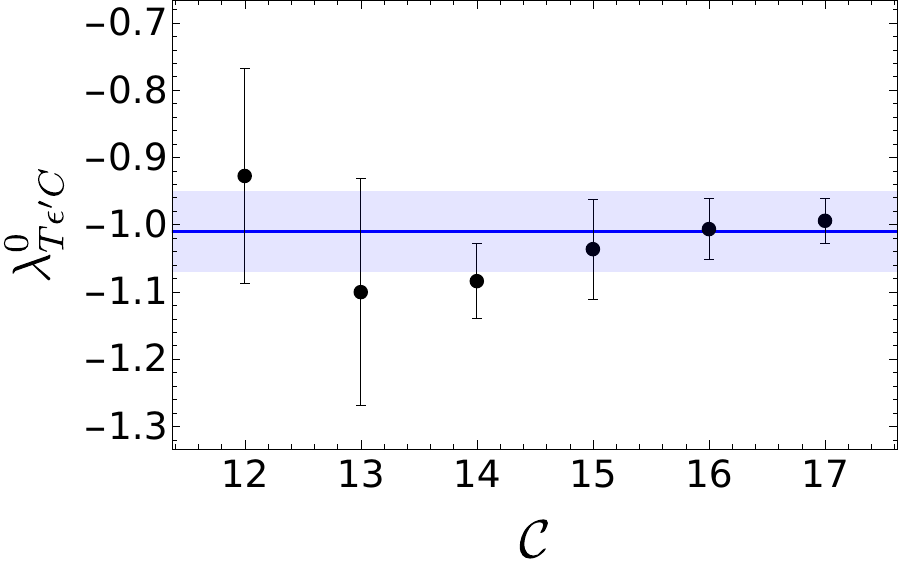}
\hspace{0.01\textwidth}
\includegraphics[width=0.315\textwidth]{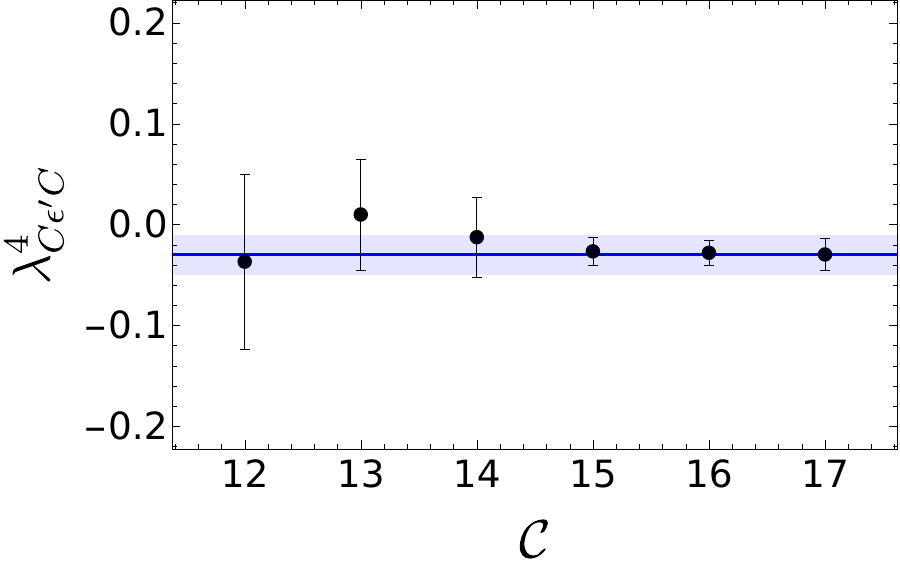}
\caption{{OPE coefficients computed from $\langle\sigma\sigma\epsilon'\sigma\sigma \rangle$ (with spin-6 contributions) by averaging over the positions of the minima of the cost function for 10 sets of $\CC$ constraints with the largest $\CI$. The horizontal blue lines and shaded regions correspond to the results given in table~\ref{ssepss}, right, computed by averaging the results for $15\leqslant \CC \leqslant 17$. One can again notice that the standard deviations for multiple OPE coefficients are large for low values of $\CC$, which is a consequence of the presence of flat directions of the cost function near the minima. Adding more constraints lifts these flat directions.  Other OPE coefficients show similar behavior with larger standard deviations.}}
\label{fig:ssepss-2}
\end{figure}

The results from these computations are shown in table~\ref{ssepss} and in figs.~\ref{fig:ssepss-1} and~\ref{fig:ssepss-2}. We observe that the error bars for the unknown OPE coefficients of the  form $\lambda_{X \epsilon' Y}$ are greater than the ones for $\lambda_{X \epsilon Y}$. This is a consequence of taking the irrelevant operator $\epsilon'$ as the external state as opposed to the relevant operator $\epsilon$. The presence of $\epsilon'$ seemingly changes the dominant contributions from the OPE to the correlator. One may notice that the estimates for $\lambda_{\epsilon \epsilon' C}$ here are consistent with those stemming from our analysis of $\langle\sigma\sigma\epsilon\sigma\sigma\rangle$ given in table~\ref{ssess}.

\begin{figure}[t]
\centering
\includegraphics[width=0.87\textwidth]{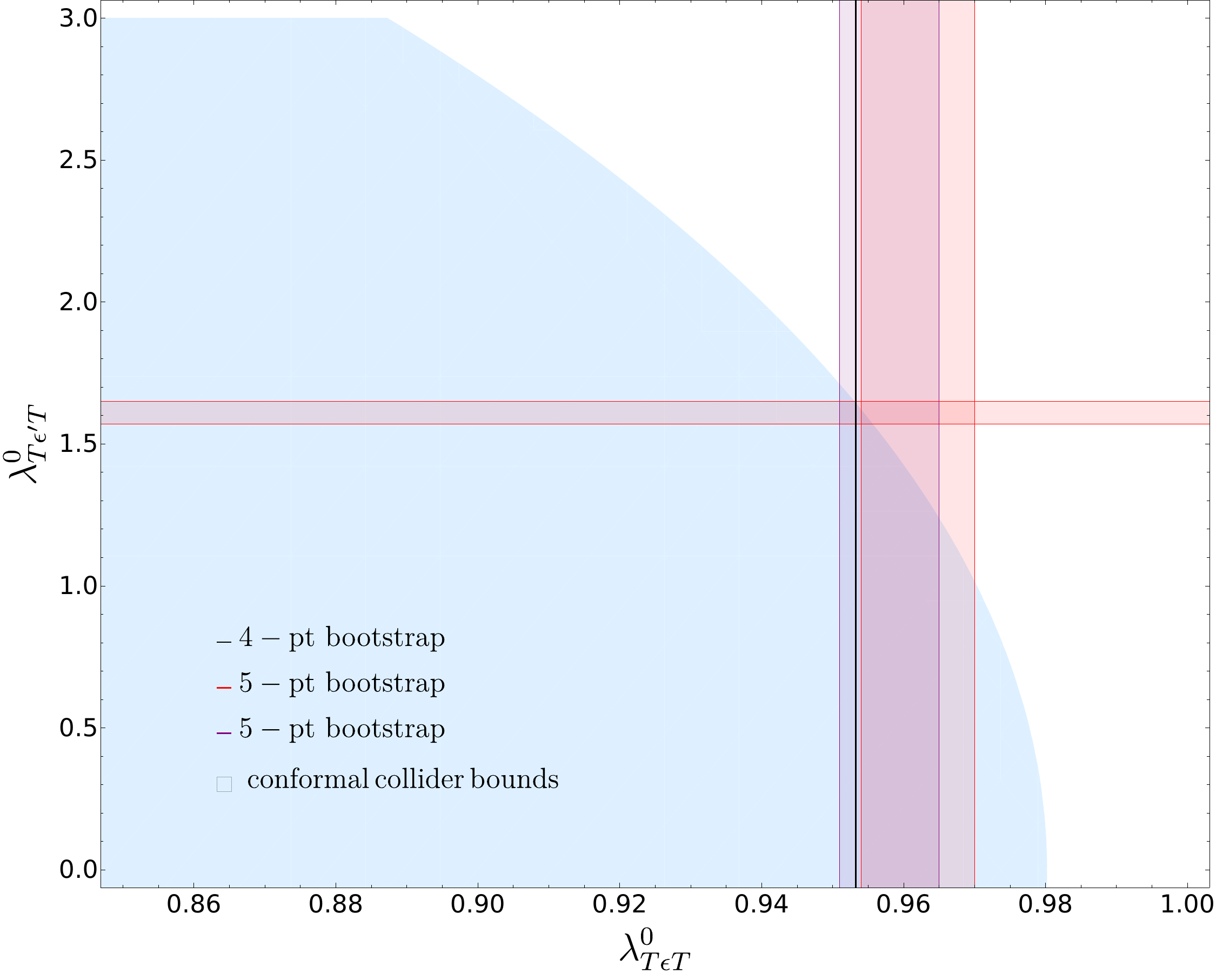}
\caption{{We show the conformal collider bounds on $\lambda_{T\epsilon T}^0$ and $\lambda_{T\epsilon' T}^0$ (blue region) \cite{Cordova:2017zej} in addition to the value of $\lambda_{T\epsilon T}^0$ computed from the four-point bootstrap  \cite{Chang:2024whx} (black line). These are compared with the approximate values computed using the five-point bootstrap, both from our earlier work~\cite{Poland:2023bny} (purple lines) and in the present work, in section~\ref{sec-ssepss} (red lines), for $\lambda_{T\epsilon T}^0$, given in table~\ref{ssess}, and $\lambda_{T\epsilon' T}^0$, given in table~\ref{ssepss}, left.}}
\label{isingv1-plots}
\end{figure}

It is interesting to compare our result $\lambda_{T\epsilon' T}^0 \simeq 1.61(4)$ with the bounds from interference effects in conformal collider physics~\cite{Cordova:2017zej}. In fig.~\ref{isingv1-plots} we show the region in $\{\lambda_{T\epsilon T}^0, \lambda_{T\epsilon' T}^0\}$ allowed by the conformal collider bounds as compared to the bootstrap determinations of these coefficients. Nontrivially, the intersection of these determinations is allowed but can be seen to sit very close to the boundary of the allowed region from~\cite{Cordova:2017zej}.

\begin{table}[t!]
\centering
\begin{minipage}{0.48\textwidth}
\centering
\begin{tabular}{|l|l|}
\hline
                                                   & est.\,values \\ \hline
$\lambda_{T\epsilon' T}^{0}$      				   & 1.61(4)         \\
$\lambda_{T\epsilon' C}^{0}$      				   & -1.2(3)         \\
$\lambda_{C\epsilon' C}^{4}$   					   & 0.02(6)         \\
$\lambda_{C\epsilon' C}^{3}$   					   & 2(2)           \\
$\lambda_{C\epsilon' C}^{2}$   					   & 3(5)           \\
$\lambda_{C\epsilon' C}^{1}$  				       & -4(10)          \\
$\lambda_{C\epsilon' C}^{0}$   					   & 3(8)          \\ 
$\lambda_{\epsilon \epsilon' C}$				   & 0.58(11)          \\
$\lambda_{\epsilon' \epsilon' C}$ 				   & 2.8(7)            \\ \hline
\end{tabular}
\end{minipage}
\hfill
\begin{minipage}{0.48\textwidth}
\centering
\begin{tabular}{|l|l|}
\hline
                                                   & est.\,values \\ \hline
$\lambda_{T\epsilon' T}^{0}$      				   & 1.56(5)         \\
$\lambda_{T\epsilon' C}^{0}$      				   & -1.01(6)         \\
$\lambda_{C\epsilon' C}^{4}$   					   & -0.03(2)         \\
$\lambda_{C\epsilon' C}^{3}$   					   & 1.7(4)           \\
$\lambda_{C\epsilon' C}^{2}$   					   & 4(2)           \\
$\lambda_{C\epsilon' C}^{1}$  				       & 6(4)          \\
$\lambda_{C\epsilon' C}^{0}$   					   & 5(3)          \\ 
$\lambda_{\epsilon \epsilon' C}$				   & 0.62(12)          \\
$\lambda_{\epsilon' \epsilon' C}$ 				   & 2.1(4)            \\
$\lambda_{\epsilon \epsilon' S}$ 				   & 1.1(3)            \\
$\lambda_{T \epsilon' S}^0$ 				       & -1.5(3)            \\ \hline
\end{tabular}
\end{minipage}
\caption{{Left: Numerical data for the unknown OPE coefficients in the truncated $\langle\sigma\sigma\epsilon'\sigma\sigma \rangle$ correlator, obtained from the 20 sets of constraints with the largest $\mathcal{I}$ for $11\leqslant \mathcal{C}\leqslant 16$. We used the operators \eqref{isingS-ep-one} and \eqref{disconnectedS-ep-one} in the Ising and disconnected correlator, respectively. The correlation matrix $\hat{\rho}_1$ for the $\lambda_{C\epsilon'C}^{n_{IJ}}$ OPE coefficients is given in Appendix~\ref{correlationmatrix}. Right: Numerical data for the unknown OPE coefficients in the truncated $\langle\sigma\sigma\epsilon'\sigma\sigma \rangle$ correlator, obtained from the 10 sets of constraints with the largest $\mathcal{I}$ for $15\leqslant \mathcal{C}\leqslant 17$. We used the operators \eqref{isingS-ep-two} and \eqref{disconnectedS-ep-two} in the Ising and disconnected correlator, respectively. The correlation matrix $\hat{\rho}_2$ for the $\lambda_{C\epsilon'C}^{n_{IJ}}$ OPE coefficients is given in Appendix~\ref{correlationmatrix}.}}\label{ssepss}
\end{table}

\section{$\sigma \times \epsilon$ OPE}\label{secfour}

In this section we study correlators expanded in the $\sigma \times \epsilon$ OPE. At the outset, we treat the four-point correlator $\langle \sigma \epsilon \sigma \epsilon \rangle$ to confirm the validity of the truncation technique and then proceed to study the five-point correlators $\langle \sigma \epsilon \epsilon \epsilon \sigma \rangle$, $\langle \sigma \epsilon \epsilon' \epsilon \sigma \rangle$, and $\langle \sigma \sigma \sigma \sigma \epsilon \rangle$, where we gain access to the $\mathbb{Z}_2$-odd states. Again, we keep a small number of exchanged operators in the OPE and approximate the rest with the double-trace contributions $[\sigma, \epsilon]_{n,\ell}$ from either the MFT correlator (in the case of $\langle\sigma \epsilon \sigma \epsilon\rangle$) or an appropriate disconnected correlator (in the case of a five-point function).

\subsection{$\langle \sigma \epsilon \sigma \epsilon \rangle$}

Let us consider the four-point correlator $\langle \sigma(x_1) \epsilon(x_2) \sigma(x_3) \epsilon(x_4) \rangle$. The associated crossing relation can be written as
\begin{equation}
\sum_{\Sigma} \lambda_{\sigma \epsilon \Sigma}^2 F_{\Sigma}(u, v)=0,
\end{equation}
where the sum runs over the primary states $\Sigma$ that appear in the $\sigma \times \epsilon$ OPE. Here $F_\Sigma(u, v)$ is given by 
\begin{equation}
F_{\Sigma}(u,v)=v^{\frac{\Delta_\sigma +\Delta_\epsilon}{2}}g_{\Sigma}(u,v)-u^{\frac{\Delta_\sigma +\Delta_\epsilon}{2}}g_{\Sigma}(v,u),
\end{equation}
where $g_\Sigma(u, v)$ are conformal blocks. We use the conformal block normalization given by eq.~(52) in~\cite{Poland:2018epd}. Here $u, v$ are the standard four-point cross ratios defined by
\begin{equation}
u=\frac{x_{12}^2 x_{34}^2}{x_{13}^2x_{24}^2},\qquad v=\frac{x_{14}^2 x_{23}^2}{x_{13}^2x_{24}^2}.
\end{equation}

We choose to keep the states with lowest conformal dimension up to spin $\ell_{\Sigma}\leqslant 7$ as well as the scalar $\sigma'$. We therefore have
\begin{equation}
S_{\rm Ising}^{\langle \sigma \epsilon \sigma \epsilon \rangle}=\{\sigma, \sigma', \Sigma_{\mu_1 \mu_2 \ldots \mu_{\ell}}| 2\leqslant \ell \leqslant 7\}.
\end{equation}
We subtract the contributions of the  following MFT operators from the MFT correlator $\langle \sigma \epsilon \sigma \epsilon \rangle_{\rm MFT} \equiv \langle \sigma  \sigma \rangle  \langle \epsilon \epsilon \rangle$:
\begin{equation}
\begin{split}
\mathcal{S}_{\rm MFT}^{\langle \sigma \epsilon \sigma \epsilon \rangle}=\{&[\sigma,\epsilon]_{0,0},  [\sigma,\epsilon]_{2,0},  [\sigma,\epsilon]_{0,2}, [\sigma,\epsilon]_{0,3}, [\sigma,\epsilon]_{0,4}, [\sigma,\epsilon]_{0,5}, [\sigma,\epsilon]_{0,6}, [\sigma,\epsilon]_{0,7},\\
&[\sigma,\epsilon]_{0,1}, [\sigma,\epsilon]_{1,0} \}.
\end{split}
\end{equation}
The logic here is that $\sigma'$ in the critical Ising spectrum of dimension $\Delta_{\sigma'} \simeq 5.29$ could be roughly (but poorly) approximated by the MFT operator $[\sigma,\epsilon]_{1,0}$ of dimension $\Delta_{\sigma} + \Delta_{\epsilon} + 2 \simeq 3.93$, or perhaps better approximated by the operator $[\sigma,\epsilon]_{2,0}$ of dimension $\Delta_{\sigma} + \Delta_{\epsilon} + 4 \simeq 5.93$. In practice, we remove both of these operators, replacing them with a single $\sigma'$ contribution. In addition, the critical Ising spectrum does not have an analog of $[\sigma,\epsilon]_{0,1}$, so we remove it as well. 

Now, the approximate crossing relation can be written as
\begin{equation}\label{4ptcrossrel}
\sum_{\Sigma \in \mathcal{S}_{\rm Ising}^{\langle \sigma \epsilon \sigma \epsilon \rangle}} \lambda_{\sigma \epsilon \Sigma}^2 F_{\Sigma}(u,v)- \sum_{[\sigma,\epsilon]_{n,\ell} \in \mathcal{S}_{\rm MFT}^{\langle \sigma \epsilon \sigma \epsilon \rangle}} \Lambda_{\sigma \epsilon [\sigma,\epsilon]_{n,\ell}}^2 F_{[\sigma,\epsilon]_{n,\ell}}(u,v)\simeq 0,
\end{equation}
where $\Lambda_{\sigma \epsilon [\sigma,\epsilon]_{n,\ell}}$ are the MFT OPE coefficients. 

Here, we fix all conformal dimensions of both the external and exchanged states, as well as the scalar OPE coefficients $\lambda_{\sigma \epsilon \sigma}$ and $\lambda_{\sigma \epsilon \sigma'}$, to the central values of their best determinations from~\cite{Simmons-Duffin:2016wlq, Chang:2024whx}. We treat the OPE coefficients of the spinning operators as unknown. To compute these, we generate additional constraints by taking derivatives of the crossing relation \eqref{4ptcrossrel},
\begin{equation}\label{constraints-sese}
e_i(\lambda)=\mathcal{D}_i \partial_{a}\left(\sum_{\Sigma \in \mathcal{S}_{\rm Ising}^{\langle \sigma \epsilon \sigma \epsilon \rangle}} \lambda_{\sigma \epsilon \Sigma}^2 F_{\Sigma}- \sum_{[\sigma,\epsilon]_{n,\ell} \in \mathcal{S}_{\rm MFT}^{\langle \sigma \epsilon \sigma \epsilon \rangle}} \Lambda_{\sigma \epsilon [\sigma,\epsilon]_{n,\ell}}^2 F_{[\sigma,\epsilon]_{n,\ell}}\right)\Bigg|_{a=1,\, b=0},
\end{equation}
where $\lambda$ collectively denotes the unknown OPE coefficients. Derivatives are taken with respect to the $(a, b)$ cross-ratios defined by
\begin{equation}
u=\frac{a^2-b}{4}, \qquad v=\left(1-\frac{a}{2}\right)^2-\frac{b}{4}.
\end{equation}
We consider the constraints obtained by taking up to 7 derivatives with respect to the $(a,b)$ cross-ratios in $\CD_i$. Here, only an even number of derivatives in $\CD_i$ with respect to $a$ give non-zero constraints. In order to get a nonzero constraint at the crossing symmetric point $u=v=\frac{1}{4}$ ($a=1, \, b=0$), one needs to take an extra derivative with respect to $a$. In view of this, we have explicitly separated $\partial_{a}$ from $\CD_i$. We obtain 20 constraints in this way. Note that we use the same definition of the cost function as in \eqref{costf}.

\begin{figure}[t]
\centering
\includegraphics[width=0.315\textwidth]{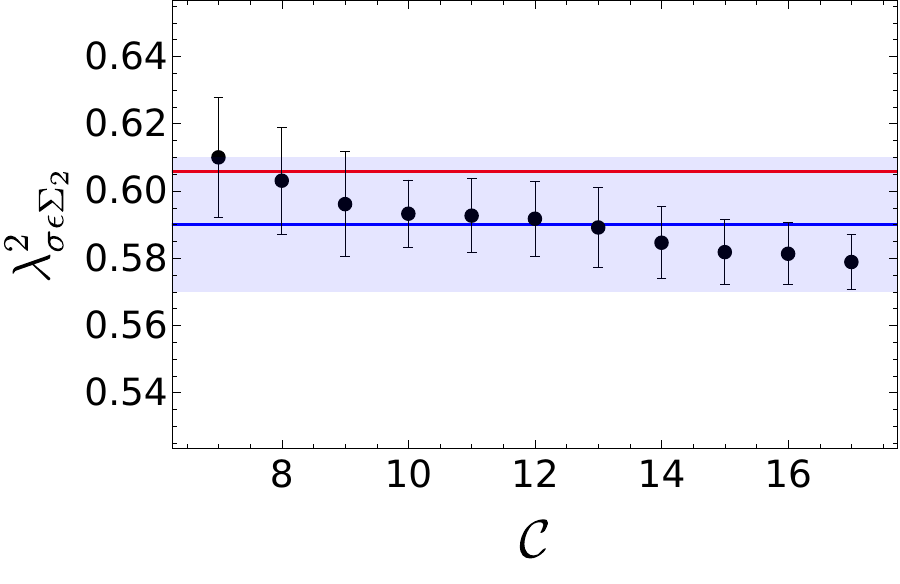}
\hspace{0.01\textwidth}
\includegraphics[width=0.315\textwidth]{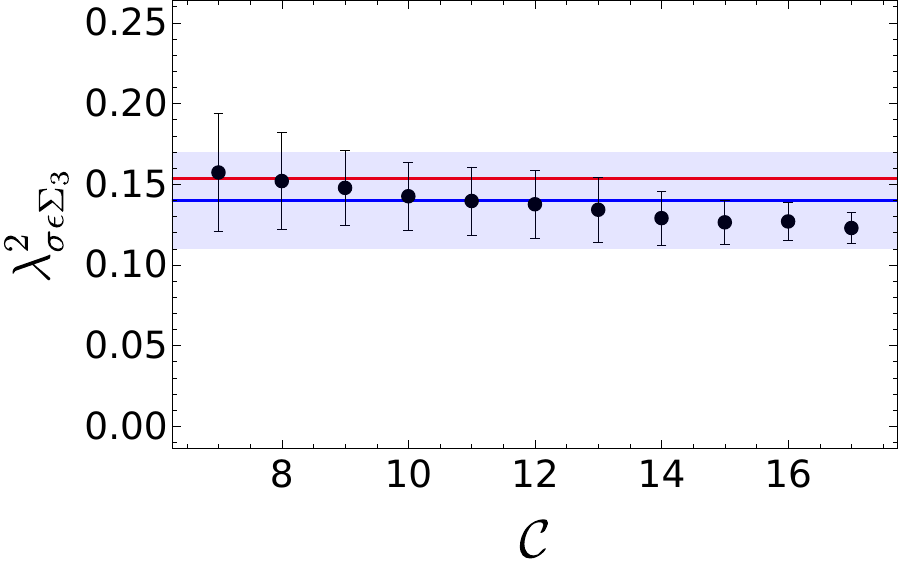}
\hspace{0.01\textwidth}
\includegraphics[width=0.315\textwidth]{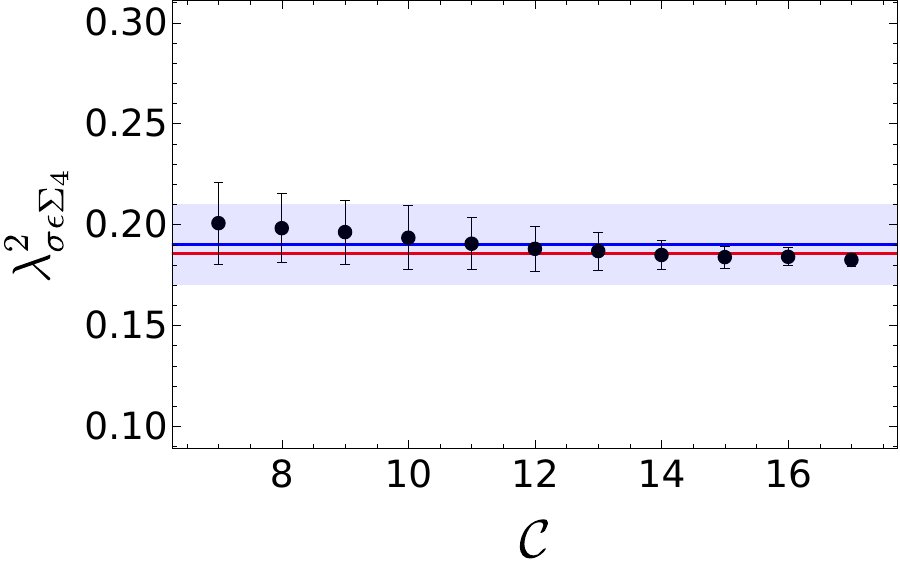}
\caption{{OPE coefficients computed from $\langle \sigma \epsilon \sigma \epsilon \rangle$ by averaging over the positions of the minima of the cost function for the 20 sets of $\CC$ constraints with the largest $\CI$, where the weights of individual constraints in each set are pseudo-randomly varied. The horizontal blue lines and shaded regions correspond to the results given in table~\ref{sese}. These values are computed by averaging the results with $7\leqslant \CC \leqslant 17$. The horizontal red lines represent the best known values from the four-point bootstrap~\cite{Simmons-Duffin:2016wlq}. Other OPE coefficients exhibit similar behavior.}}
\label{fig:sese}
\end{figure}

As was done above, we rescale the unknown OPE coefficients,
\begin{equation}
\lambda_{\sigma \epsilon \Sigma_{\mu_1 \ldots \mu_{\ell}}} = \Lambda_
{\sigma \epsilon [\sigma, \epsilon]_{0,\ell}} L_{\sigma \epsilon \Sigma_{\mu_1 \ldots \mu_{\ell}}},
\end{equation}
and use $L_{\sigma \epsilon \Sigma_{\mu_1 \ldots \mu_{\ell}}}$ as our unknowns. Again, in order to select the sets of constraints that we include in the cost function for each $\mathcal{C}$, we compute \eqref{dethessian} and use the 20 sets of constraints with the largest $\mathcal{I}$ for $7\leqslant \mathcal{C}\leqslant 12$. The results we obtain are shown in table~\ref{sese} and fig.~\ref{fig:sese}. From these, we conclude that our method has the capacity to roughly estimate the values of OPE coefficients of states on the leading Regge trajectory in the $\sigma \times \epsilon$ OPE.

\begin{table}[t!]
\centering
\begin{tabular}{|l|l|l|}
\hline
                                                   & est.\,values  & \cite{Simmons-Duffin:2016wlq}   \\ \hline
$\lambda_{\sigma\epsilon \Sigma_2}^{2}$    & 0.59(2)       & 0.60578	\\
$\lambda_{\sigma\epsilon \Sigma_3}^{2}$    & 0.14(3)        & 0.15346	\\
$\lambda_{\sigma\epsilon \Sigma_4}^{2}$	   & 0.19(2)       & 0.18561 	\\
$\lambda_{\sigma\epsilon \Sigma_5}^{2}$	   & 0.078(6)          & 0.05622	\\
$\lambda_{\sigma\epsilon \Sigma_6}^{2}$	   & 0.058(8)          & 0.05268	\\
$\lambda_{\sigma\epsilon \Sigma_7}^{2}$    & 0.008(6)        & 0.01726 	\\ \hline
\end{tabular}
\caption{Numerical data for the OPE coefficients of $\sigma, \epsilon$, and $\Sigma_{\mu_1 \mu_2 \ldots \mu_{\ell}}$ states with the lowest conformal dimensions  with spins $2\leqslant s_\Sigma \leqslant 7$ from the truncated $\langle \sigma \epsilon \sigma \epsilon \rangle$ correlator, obtained using 20 sets of constraints with the largest $\mathcal{I}$ for $7\leqslant \mathcal{C}\leqslant 17$, as compared with the best known values from~\cite{Simmons-Duffin:2016wlq}.}\label{sese}
\end{table}

\subsection{$\langle \sigma \epsilon \epsilon \epsilon \sigma \rangle$}

We next turn to the $\langle \sigma \epsilon \epsilon \epsilon \sigma \rangle$ correlator. For this case, we retain the following operators in the truncated crossing relation:
\begin{equation}\label{cont-ising-seees}
\begin{split}
\mathcal{S}_{\rm Ising}^{\langle\sigma \epsilon\epsilon\epsilon\sigma \rangle}=\{&\sigma, \sigma', \Sigma_{\mu\nu}, \Sigma_{\nu\mu\rho}| {\, \,\rm all \, \, pairs}, \\
& (\sigma, \Sigma_{\ell}), (\Sigma_{\ell}, \sigma)| \,\, 4\leqslant \ell \leqslant 7\},
\end{split}
\end{equation}
where $\Sigma_{\ell}$ denotes the spin-$\ell$ states with the lowest conformal dimension. We add the $(\sigma, \Sigma_{\ell})$ contributions for $4\leqslant \ell \leqslant 7$ in order to improve the disconnected approximation for the truncated states. We include operators up to spin 7 by analogy to our treatment of the $\langle\sigma \epsilon \sigma \epsilon \rangle$ correlator in the previous section.

The disconnected correlator in this case is given by
\begin{equation}\label{disconnected-seees}
\begin{split}
\langle \sigma(x_1) \epsilon(x_2) \epsilon(x_3) \epsilon(x_4) \sigma(x_5) \rangle_{\rm d} =\,& \langle \sigma(x_1) \sigma(x_5) \rangle \langle \epsilon(x_2) \epsilon(x_3) \epsilon(x_4) \rangle \\
&+ \langle \epsilon(x_2) \epsilon(x_3) \rangle \langle \sigma(x_1) \epsilon(x_4) \sigma(x_5) \rangle + {\rm perm.}
\end{split}
\end{equation}
We remark that there are two types of contributions to this correlator that arise here, namely $(\sigma, [\sigma, \epsilon]_{n,\ell})$ and $([\sigma, \epsilon]_{n,\ell}, [\sigma, \epsilon]_{n',\ell'})$. It follows that in order to approximate the truncated contributions in the critical Ising model, we need to subtract the contributions of the following operators from the crossing relation:
\begin{equation}\label{cont-disc-seees}
\begin{split}
\mathcal{S}_{\rm disc.}^{\langle\sigma \epsilon\epsilon\epsilon\sigma \rangle}=\{&\sigma, [\sigma,\epsilon]_{2,0}, [\sigma,\epsilon]_{0,2}, [\sigma,\epsilon]_{0,3}, \,|\,\, {\rm all \,\, pairs,}\\
& (\sigma, [\sigma,\epsilon]_{0,\ell}), ([\sigma,\epsilon]_{0,\ell}, \sigma)\,|\,\, 4\leqslant \ell \leqslant 7,\\
& (\sigma,[\sigma,\epsilon]_{0,0}), ([\sigma,\epsilon]_{0,0},\sigma), (\sigma,[\sigma,\epsilon]_{0,1}), ([\sigma,\epsilon]_{0,1},\sigma), (\sigma,[\sigma,\epsilon]_{1,0}), ([\sigma,\epsilon]_{1,0},\sigma), \\
& ([\sigma,\epsilon]_{0,0},[\sigma,\epsilon]_{n,\ell}), ([\sigma,\epsilon]_{n,\ell},[\sigma,\epsilon]_{0,0})\, | \, \, 2n+\ell\leqslant 7, \\
& ([\sigma,\epsilon]_{0,1},[\sigma,\epsilon]_{n,\ell}), ([\sigma,\epsilon]_{n,\ell},[\sigma,\epsilon]_{0,1})\, | \, \, 2n+\ell\leqslant 7, \\
& ([\sigma,\epsilon]_{1,0},[\sigma,\epsilon]_{n,\ell}), ([\sigma,\epsilon]_{n,\ell},[\sigma,\epsilon]_{1,0})\, | \, \, 2n+\ell\leqslant 7 \}.
\end{split}
\end{equation} 
The first two lines correspond to the contributions included in the Ising model \eqref{cont-ising-seees}. Then, we have to take into account that in the Ising model there are no contributions close to those of $[\sigma,\epsilon]_{0,0}$, $[\sigma,\epsilon]_{0,1}$, and $[\sigma,\epsilon]_{1,0}$, so we need to explicitly subtract these. In particular, we subtract off all pairs that contain these states and operators with $2n+\ell\leqslant 7$.

\begin{figure}[t]
\centering
\includegraphics[width=0.315\textwidth]{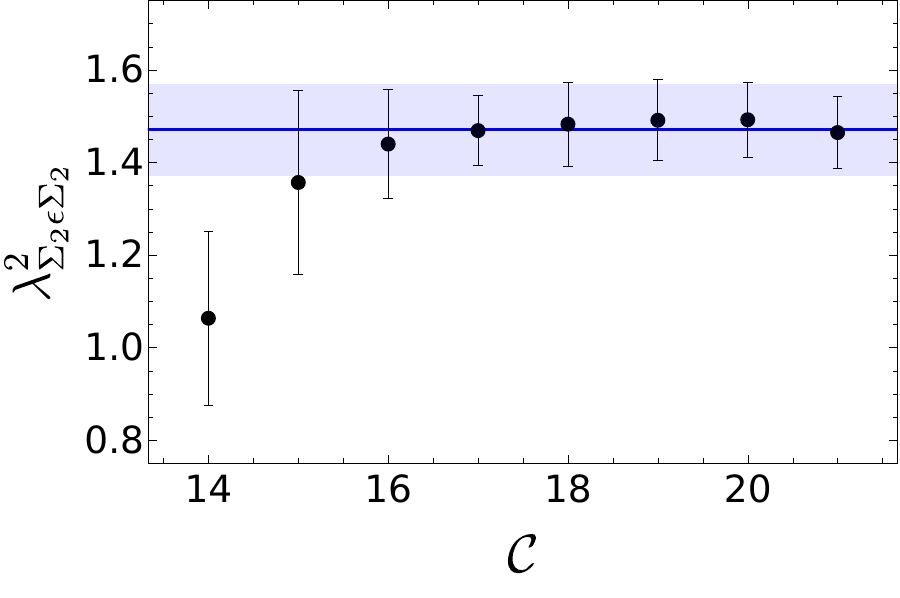}
\hspace{0.01\textwidth}
\includegraphics[width=0.315\textwidth]{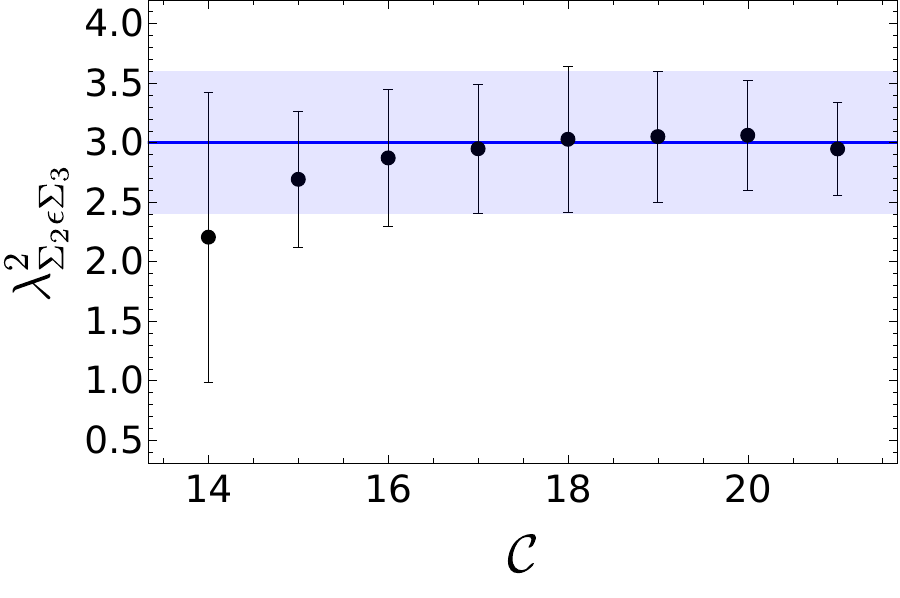}
\hspace{0.01\textwidth}
\includegraphics[width=0.315\textwidth]{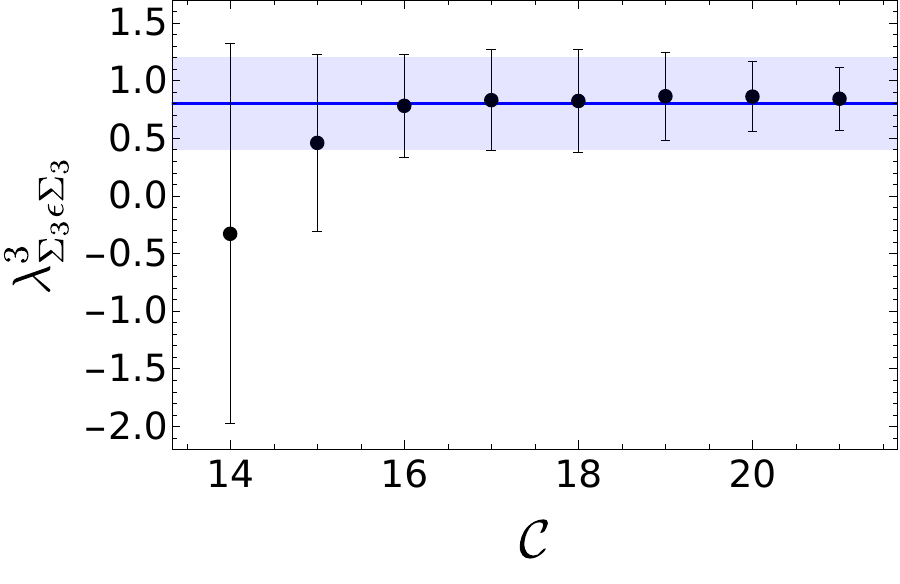}
\caption{{OPE coefficients computed from $\langle\sigma\epsilon\epsilon\epsilon\sigma \rangle$ by averaging over the positions of the minima of the cost function for the 20 sets of $\CC$ constraints with the largest $\CI$, where the weights of individual constraints in each set are pseudo-randomly varied. The horizontal blue lines and shaded regions correspond to the results given in table~\ref{seees}. These values are obtained by averaging the results with $16\leqslant \CC \leqslant 21$. One can again notice that the standard deviations for multiple OPE coefficients are large for low values of $\CC$, which is a consequence of the presence of flat directions of the cost function near its minima. Adding more constraints lifts these flat directions. Other OPE coefficients show similar behavior with larger standard deviations.}}
\label{fig:seees}
\end{figure}

In this case, we have more unknown OPE coefficients than in the case of the $\langle \sigma \sigma \epsilon' \sigma \sigma \rangle$ correlator. Hence, we also include some of the constraints obtained by taking four derivatives in $\CD_i$ of the crossing relation in \eqref{constraints}. We pick the derivatives that yield the best-satisfied constraints when used for the disconnected correlator upon including the contributions in \eqref{cont-disc-seees}. The fourth-order derivatives we choose are
\begin{equation}
\CD_i = \{\partial_{b^+}\partial_{w}^3, \quad \partial_{b^+}^4, \quad \partial_w^4, \quad \partial_{b^+}^3\partial_{w} \}.
\end{equation}
Here, we rescale the unknown OPE coefficients as
\begin{equation}
\begin{split}
&\lambda_{\Sigma_i \epsilon \Sigma_j}^{n_{IJ}}=\Lambda_{[\sigma,\epsilon]_{0,i} \epsilon [\sigma,\epsilon]_{0,j}}^{n_{IJ}} L^{n_{IJ}}_{\Sigma_i \epsilon \Sigma_j}, \quad 
\lambda_{\sigma' \epsilon \Sigma_j}= \Lambda_{[\sigma,\epsilon]_{2,0} \epsilon [\sigma,\epsilon]_{0,j}}^{0} L_{\sigma' \epsilon \Sigma_j},\\
&\lambda_{\sigma' \epsilon \sigma'}= \Lambda_{[\sigma,\epsilon]_{2,0} \epsilon [\sigma,\epsilon]_{2,0}}^{0} L_{\sigma' \epsilon \sigma'},
\end{split}
\end{equation}
where $\Lambda$ denotes the OPE coefficients from the disconnected correlator \eqref{disconnected-seees}. We treat the $L$'s as the unknowns when evaluating $\mathcal{I}$ in \eqref{dethessian}. Upon defining the cost function as in \eqref{costf}, we scan over all sets of $\mathcal{C}$ constraints for $16 \leqslant\CC\leqslant 21$. We select more than the minimal number of constraints (14 here) in order to lift the flat directions in the cost function. For each $\CC$, we use the 20 sets of constraints that yield the largest $\mathcal{I}$ in \eqref{dethessian}. Further, for each of these sets, we randomly vary the weights of the constraints. We then average over all solutions for the 20 sets for different values of $\CC$. The results we obtain are shown in table~\ref{seees} and fig.~\ref{fig:seees}. Changing the number of sets for each $\CC$  would shift the mean values within the error bars reported here.

\begin{table}[t!]
\centering
\begin{minipage}{0.48\textwidth}
\centering
\begin{tabular}{|l|l|}
\hline
                                                   & est.\,values \\ \hline
$\lambda_{\Sigma_2\epsilon \Sigma_2}^{2}$      				   & 1.47(10)         \\
$\lambda_{\Sigma_2\epsilon \Sigma_2}^{1}$      				   & -0.2(8)      \\
$\lambda_{\Sigma_2\epsilon \Sigma_2}^{0}$   					   & 1.5(8)         \\ \hline
$\lambda_{\Sigma_3\epsilon \Sigma_3}^{3}$   					   & 0.8(4)           \\
$\lambda_{\Sigma_3\epsilon \Sigma_3}^{2}$   					   & -7(3)           \\
$\lambda_{\Sigma_3\epsilon \Sigma_3}^{1}$  				       & 7(3)          \\
$\lambda_{\Sigma_3\epsilon \Sigma_3}^{0}$   					   & -5(2)          \\ \hline
\end{tabular}
\end{minipage}
\hfill
\begin{minipage}{0.48\textwidth}
\centering
\begin{tabular}{|l|l|}
\hline
                                                   & est.\,values \\ \hline
$\lambda_{\Sigma_2\epsilon \Sigma_3}^{2}$				   & 3.0(6)          \\
$\lambda_{\Sigma_2\epsilon \Sigma_3}^{1}$				   & 3(2)          \\
$\lambda_{\Sigma_2\epsilon \Sigma_3}^{0}$				   & -3(2)          \\ \hline
$\lambda_{\sigma' \epsilon \Sigma_2}$ 				   & -1(2)            \\
$\lambda_{\sigma' \epsilon \Sigma_3}$ 				   & -16(3)            \\ 
$\lambda_{\sigma' \epsilon \sigma'}$ 				   & 26(6)            \\ \hline
\end{tabular}
\end{minipage}
\caption{Numerical data for the unknown OPE coefficients in the truncated $\langle\sigma\epsilon\epsilon\epsilon\sigma \rangle$ correlator, obtained using the 20 sets of constraints with the largest $\mathcal{I}$ for $16\leqslant \mathcal{C}\leqslant 21$. The correlation matrices for the OPE coefficients of the same operators with different tensor structures are given in Appendix~\ref{correlationmatrix}.}
\label{seees}
\end{table}

We observe large error bars for the OPE coefficients involving the $\sigma'$ operator. This suggests that our method is not sensitive to these contributions, and we take this to mean that they cannot be accurately estimated by the truncation method at this level. Generally, it seems difficult to reliably predict the OPE coefficients of $\sigma'$ operator, which is likely a consequence of its large twist. It appears that we are typically able to extract the most accurate results for the states on the leading, lowest-twist, Regge trajectory. For example, the OPE coefficient $\lambda_{\sigma' \epsilon \sigma'}$ comes with a large standard deviation, which indicates the sensitivity of the predicted value to the number of constraints in the cost function as well as to the particular choice of constraints. We  expect, therefore, that our estimate for this coefficient here is not accurate. We note that the fuzzy sphere regularization method~\cite{Hu:2023xak} gives $|\lambda_{\sigma' \epsilon \sigma'}| \simeq 2.98(13)$ for this coefficient.

\subsection{$\langle \sigma \epsilon \epsilon' \epsilon \sigma \rangle$}

Just as in the $\langle \sigma \sigma \epsilon' \sigma \sigma \rangle$ case, we can take an irrelevant scalar to be the middle operator without significantly affecting the convergence of the OPE. We will then study the correlator $\langle \sigma \epsilon \epsilon' \epsilon \sigma \rangle$ next. 

In the critical Ising correlator we include the following exchanged operators:
\begin{equation}\label{cont-ising-seepes}
\begin{split}
\mathcal{S}_{\rm Ising}^{\langle\sigma \epsilon\epsilon'\epsilon\sigma \rangle}=\{&\sigma, \sigma', \Sigma_{\mu\nu}, \Sigma_{\nu\mu\rho}| {\, \,\rm all \, \, pairs}, \\
& (\sigma, \Sigma_{\ell}), (\Sigma_{\ell}, \sigma)| \,\, 4\leqslant \ell \leqslant 7\}.
\end{split}
\end{equation}
We have the disconnected correlator 
\begin{equation}\label{disconnected-seepes}
\begin{split}
\langle \sigma(x_1) \epsilon(x_2) \epsilon'(x_3) \epsilon(x_4) \sigma(x_5) \rangle_{\rm d} =\,& \langle \sigma(x_1) \sigma(x_5) \rangle \langle \epsilon(x_2) \epsilon'(x_3) \epsilon(x_4) \rangle \\
&+ \langle \epsilon(x_2) \epsilon(x_4) \rangle \langle \sigma(x_1) \epsilon'(x_3) \sigma(x_5) \rangle. 
\end{split}
\end{equation}
The only type of contribution to this correlator in the operator product expansion is due to pairs of double-trace operators $([\sigma,\epsilon]_{n,\ell},[\sigma,\epsilon]_{n',\ell'})$. We can roughly approximate the truncated contributions in the critical Ising $\langle \sigma \epsilon \epsilon' \epsilon \sigma \rangle$ correlator if we subtract the following contributions  from the disconnected correlator:
\begin{equation}\label{cont-disc-seepes}
\begin{split}
\mathcal{S}_{\rm disc.}^{\langle\sigma \epsilon\epsilon'\epsilon\sigma \rangle}=\{&    [\sigma,\epsilon]_{0,2},  [\sigma,\epsilon]_{0,3}, [\sigma,\epsilon]_{2,0}\,|\,\, {\rm all \,\, pairs,}\\
& ([\sigma,\epsilon]_{0,0},[\sigma,\epsilon]_{n,\ell}), ([\sigma,\epsilon]_{n,\ell},[\sigma,\epsilon]_{0,0})\, | \, \, 2n+\ell\leqslant 7, \\
& ([\sigma,\epsilon]_{0,1},[\sigma,\epsilon]_{n,\ell}), ([\sigma,\epsilon]_{n,\ell},[\sigma,\epsilon]_{0,1})\, | \, \, 2n+\ell\leqslant 7, \\
& ([\sigma,\epsilon]_{1,0},[\sigma,\epsilon]_{n,\ell}), ([\sigma,\epsilon]_{n,\ell},[\sigma,\epsilon]_{1,0})\, | \, \, 2n+\ell\leqslant 7 \}.
\end{split}
\end{equation}
In this case, there are no individual $\sigma$-contributions in the disconnected correlator. This is the reason why we include additional $\sigma$-contributions in the critical Ising correlator in the second line of \eqref{cont-ising-seepes}. By analogy to our analysis of $\langle\sigma \epsilon\epsilon\epsilon \sigma\rangle$, we maximally remove the contributions of $[\sigma,\epsilon]_{0,0}$, $[\sigma,\epsilon]_{0,1}$, and $[\sigma,\epsilon]_{1,0}$, since states close to these do not exist in the critical 3d Ising model. 

The unknown OPE coefficients here are $\lambda_{\Sigma_i \epsilon' \Sigma_j}^{n_{IJ}}$ for $i,j=2,3$, $\lambda_{\sigma' \epsilon' \Sigma_j}$ for $j=2,3$,  $\lambda_{\sigma \epsilon' \sigma'}$, $\lambda_{\sigma' \epsilon' \sigma'}$, and $\lambda_{\sigma \epsilon' \Sigma_j}$ for $2\leqslant j\leqslant 7$. Some of these are rescaled as
\begin{equation}
\begin{split}
\lambda_{\Sigma_i \epsilon' \Sigma_j}^{n_{IJ}}&=\Lambda_{[\sigma,\epsilon]_{0,i} \epsilon' [\sigma,\epsilon]_{0,j}}^{n_{IJ}} L^{n_{IJ}}_{\Sigma_i \epsilon' \Sigma_j}, \qquad 
\lambda_{\sigma' \epsilon' \Sigma_j}=\Lambda_{[\sigma,\epsilon]_{2,0} \epsilon' [\sigma,\epsilon]_{0,j}}^{0}L_{\sigma' \epsilon' \Sigma_j},\\
\lambda_{\sigma' \epsilon' \sigma'}&=\Lambda_{[\sigma,\epsilon]_{2,0} \epsilon' [\sigma,\epsilon]_{2,0}}^{0}L_{\sigma' \epsilon' \sigma'},
\end{split}
\end{equation}
where $\Lambda$ are the OPE coefficients from the disconnected correlator \eqref{disconnected-seepes}. We do not rescale $\lambda_{\sigma \epsilon' \sigma'}$ and $\lambda_{\sigma \epsilon' \Sigma_j}$, since their analogs do not exist in the disconnected correlator. In addition to the constraints obtained by taking up to three derivatives $\CD_i$ of the crossing relation, we include the following subset of fourth-order derivatives:
\begin{equation}
\CD_i = \{ \partial_{w}\partial_{b^+}\partial_{a^-}^2, \,\, \partial_{b^+}^3\partial_{w}, \,\, \partial_{w}^4, \,\, \partial_{b^+}^2\partial_{a^-}^2, \,\, \partial_{a^+}^2\partial_{a^-}^2, \,\, \partial_{a^+}^2\partial_{b^+}\partial_{w}, \,\, \partial_{a^+}^2 \partial_{b^-}^2, \,\, \partial_{a^-}^2 \partial_{b^-}^2 \}.
\end{equation}
We have chosen these such that the constraints that arise from this subset are the ones that are best-satisfied for the disconnected $\langle \sigma \epsilon \epsilon' \epsilon \sigma \rangle_{\rm d}$ correlator.

\begin{table}[t!]
\centering
\begin{minipage}{0.48\textwidth}
\centering
\begin{tabular}{|l|l|}
\hline
                                                   & est.\,values \\ \hline
$\lambda_{\sigma\epsilon' \Sigma_2}$      				   & -0.28(8)         \\
$\lambda_{\sigma\epsilon' \Sigma_3}$      				   & -0.11(10)         \\
$\lambda_{\sigma\epsilon' \Sigma_4}$      				   & -0.11(3)      \\
$\lambda_{\sigma\epsilon' \Sigma_5}$   					   & -0.03(8)         \\ 
$\lambda_{\sigma\epsilon' \Sigma_6}$   					   & -0.06(3)           \\
$\lambda_{\sigma\epsilon' \Sigma_7}$   					   & 0.01(5)           \\
$\lambda_{\sigma \epsilon' \sigma'}$				       & 0.4(4)          \\ \hline
$\lambda_{\Sigma_2\epsilon' \Sigma_2}^{2}$  				       & 0.0(5)          \\
$\lambda_{\Sigma_2\epsilon' \Sigma_2}^{1}$   					   & -7(2)          \\ 
$\lambda_{\Sigma_2\epsilon' \Sigma_2}^{0}$				   & 1(5)          \\ \hline
\end{tabular}
\end{minipage}
\hfill
\begin{minipage}{0.48\textwidth}
\centering
\begin{tabular}{|l|l|}
\hline
                                                   & est.\,values \\ \hline
$\lambda_{\Sigma_3\epsilon' \Sigma_3}^{3}$				   & -0.7(7)          \\
$\lambda_{\Sigma_3\epsilon' \Sigma_3}^{2}$				   & -7(7)          \\ 
$\lambda_{\Sigma_3\epsilon' \Sigma_3}^{1}$ 				   & 9(10)            \\
$\lambda_{\Sigma_3\epsilon' \Sigma_3}^{0}$ 				   & -10(10)            \\ \hline
$\lambda_{\Sigma_2\epsilon' \Sigma_3}^{2}$  				       & 0.7(7)          \\
$\lambda_{\Sigma_2\epsilon' \Sigma_3}^{1}$   					   & -5(6)          \\
$\lambda_{\Sigma_2\epsilon' \Sigma_3}^{0}$   					   & 3(7)          \\ \hline
$\lambda_{\sigma'\epsilon' \Sigma_2}$				   & 9(3)          \\ 
$\lambda_{\sigma'\epsilon' \Sigma_3}$				   & 3(8)          \\ 
$\lambda_{\sigma'\epsilon' \sigma'}$				   & 12(11)          \\ \hline
\end{tabular}
\end{minipage}
\caption{Numerical data for the unknown OPE coefficients in the truncated $\langle\sigma\epsilon\epsilon'\epsilon\sigma \rangle$ correlator, obtained with the 20 sets of constraints with the largest $\mathcal{I}$ for $22\leqslant \mathcal{C}\leqslant 25$. The correlation matrices for the OPE coefficients of the same operators for different tensor structures are given in Appendix~\ref{correlationmatrix}.}
\label{seepes}
\end{table}

\begin{figure}[t]
\centering
\includegraphics[width=0.315\textwidth]{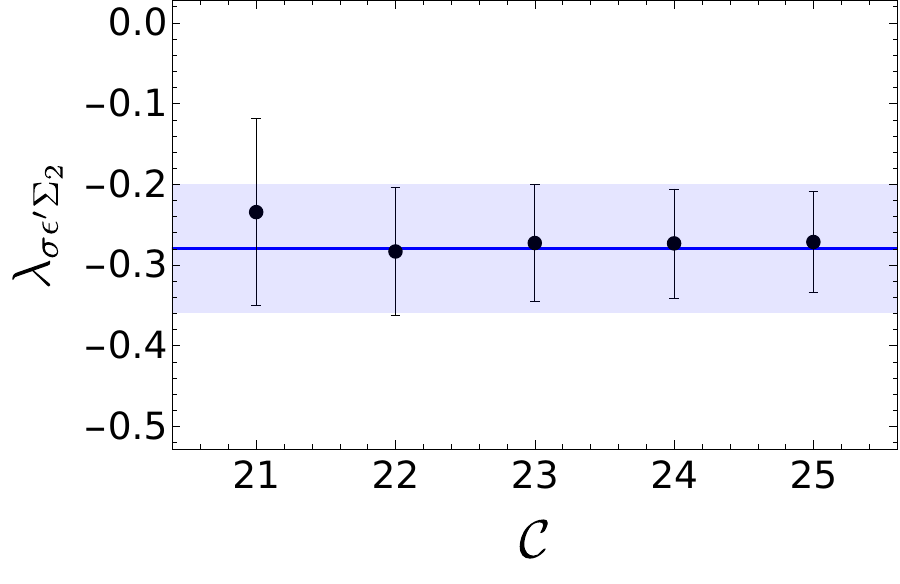}
\hspace{0.01\textwidth}
\includegraphics[width=0.315\textwidth]{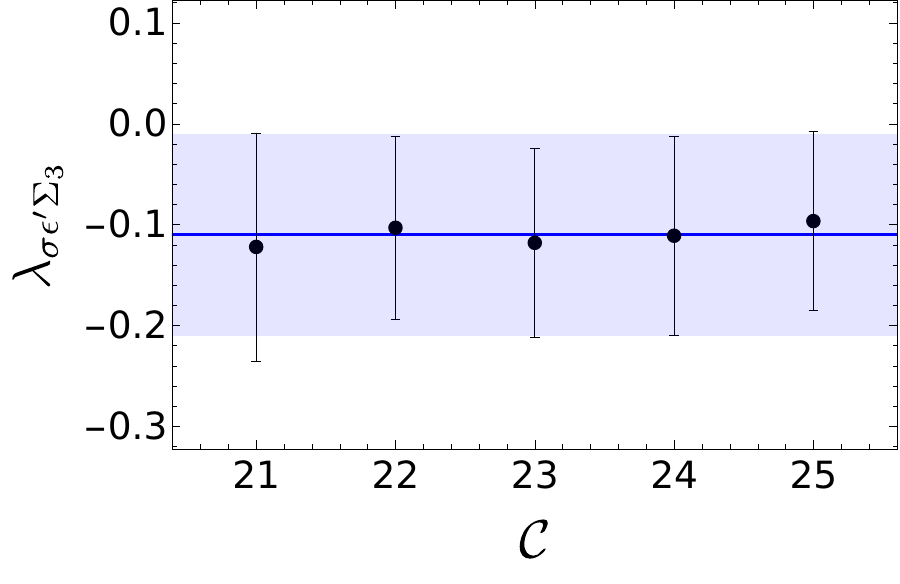}
\hspace{0.01\textwidth}
\includegraphics[width=0.315\textwidth]{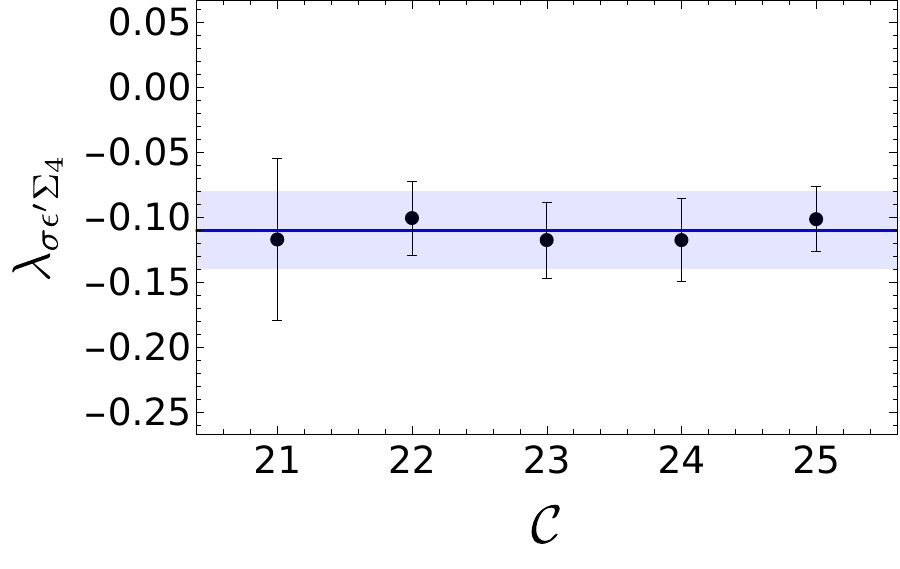}
\caption{{OPE coefficients computed from $\langle\sigma\epsilon\epsilon'\epsilon\sigma \rangle$ by averaging over the positions of the minima of the cost function for 20 sets of $\CC$ constraints with the largest $\CI$, where the weights of individual constraints in each set are pseudo-randomly varied. The horizontal blue lines and shaded regions correspond to the results given in table~\ref{seepes}. These values are obtained by averaging the results with $22\leqslant \CC \leqslant 25$. Other OPE coefficients show similar behavior with larger standard deviations.}}
\label{fig:seepes}
\end{figure}

Our results are presented in table~\ref{seepes} and fig.~\ref{fig:seepes}. Again, we observe that the relative errors of the unknown OPE coefficients in the $\langle\sigma \epsilon \epsilon' \epsilon \sigma \rangle$ correlator are generally larger than in $\langle\sigma \epsilon \epsilon \epsilon \sigma \rangle$. We interpret this as a consequence of the large conformal dimension of the external operator,  which presumably changes the dominant contributions from the OPE to the correlator.

Here, we can compare our prediction $\lambda_{\sigma\epsilon' \Sigma_2} \simeq -0.28(8)$ to the one obtained via the fuzzy sphere regularization approach \cite{Hu:2023xak}. According to our conventions, the corresponding result in \cite{Hu:2023xak} is given by $\lambda_{\sigma\epsilon' \Sigma_2} \simeq -0.365(2)$, which falls within the error bar of our result. 

\subsection{$\langle \sigma \sigma \sigma \sigma \epsilon \rangle$}

Lastly, we study the $\langle \sigma(x_1) \sigma(x_2) \sigma(x_3) \sigma(x_4) \epsilon(x_5) \rangle$ correlator. For this correlator, we consider the $\sigma \times \sigma$ and $\sigma \times \epsilon$ operator product expansions. 

We include the following exchanged operators in the critical Ising correlator:
\begin{equation}\label{cont-ising-sssse}
\begin{split}
\mathcal{S}_{\rm Ising}^{\langle\sigma \sigma\sigma\sigma\epsilon \rangle}=\{& (\mathcal{O},\mathcal{O}')|\, \mathcal{O}\in \{ \mathbf{1}, \epsilon, \epsilon', T_{\mu\nu}, C_{\mu\nu\rho\sigma}\}, \, \mathcal{O}' \in \{ \sigma, \sigma', \Sigma_2, \Sigma_3 \},\\
& (\epsilon,\Sigma_i), (\epsilon', \Sigma_{\ell}), (T_{\mu\nu}, \Sigma_{\ell})\, |\, 4\leqslant \ell \leqslant 7 \}.
\end{split}
\end{equation}
In this case, the disconnected correlator is defined as
\begin{equation}\label{disconnected-seepes}
\begin{split}
\langle \sigma(x_1) \sigma(x_2) \sigma(x_3) \sigma(x_4) \epsilon(x_5) \rangle_{\rm d} = \langle \sigma(x_1) \sigma(x_2) \rangle \langle \sigma(x_3) \sigma(x_4) \epsilon(x_5) \rangle + {\rm perm.} 
\end{split}
\end{equation}
There are four types of contributions to this correlator in the operator product expansion that appear, namely $(\mathbf{1},\sigma)$, $(\epsilon,[\sigma,\epsilon]_{n,\ell})$, $([\sigma,\sigma]_{n,\ell}, \sigma)$, and $([\sigma,\sigma]_{n,\ell},[\sigma,\epsilon]_{n',\ell'})$. The contributions that we have to subtract from the disconnected correlator in order to roughly approximate the truncated contributions in this correlator are given by
\begin{equation}\label{cont-disc-sssse}
\begin{split}
\mathcal{S}_{\rm disc.}^{\langle\sigma \sigma\sigma\sigma\epsilon \rangle}=\{&([\sigma,\sigma]_{0,0}, [\sigma, \epsilon]_{n,\ell})\, |\, 2n+\ell \leqslant 7,\\
& ([\sigma,\sigma]_{n,\ell}, [\sigma, \epsilon]_{0,0})\, |\, 0< 2n+\ell  \leqslant 10, \, \ell-{\rm even},\\
& ([\sigma,\sigma]_{n,\ell}, [\sigma, \epsilon]_{1,0})\, |\, 0< 2n+\ell  \leqslant 10, \, \ell-{\rm even},\\
& ([\sigma,\sigma]_{n,\ell}, [\sigma, \epsilon]_{0,1})\, |\, 0< 2n+\ell  \leqslant 10, \, \ell-{\rm even},\\
& (\epsilon, [\sigma, \epsilon]_{0,\ell})\, |\, 0\leqslant \ell \leqslant 7, \\
&(\epsilon, [\sigma, \epsilon]_{n,0})\, |\, 1\leqslant n \leqslant 2, \\
&(\mathcal{O}, \sigma)\, | \, \mathcal{O}\in \{\mathbf{1}, [\sigma,\sigma]_{0,0}, [\sigma,\sigma]_{1,0}, [\sigma,\sigma]_{0,2}, [\sigma,\sigma]_{0,4}\},\\
&(\mathcal{O},\mathcal{O}')|\, \mathcal{O}\in \{[\sigma,\sigma]_{1,0}, [\sigma,\sigma]_{0,2}, [\sigma,\sigma]_{0,4}\}, \, \mathcal{O}' \in \{ [\sigma,\epsilon]_{2,0}, [\sigma,\epsilon]_{0,2}, [\sigma,\epsilon]_{0,3} \},\\
& ([\sigma,\sigma]_{1,0}, [\sigma, \epsilon]_{0,\ell})\, |\, 4\leqslant \ell \leqslant 7,\\
& ([\sigma,\sigma]_{0,2}, [\sigma, \epsilon]_{0,\ell})\, |\, 4\leqslant \ell \leqslant 7 
\}.
\end{split}
\end{equation}
We employ similar reasoning here as in the above analysis. In particular, the logic is that the  disconnected correlator analog of the $\epsilon$ operator, namely $[\sigma,\sigma]_{0,0}$, is redundant, since we have $\epsilon$ explicitly included. We therefore need to maximally remove the contributions that contain $[\sigma,\sigma]_{0,0}$. In the same spirit, we try to remove $[\sigma,\epsilon]_{0,0}$, $[\sigma,\epsilon]_{0,1}$, $[\sigma,\epsilon]_{1,0}$. We rescale the unknown OPE coefficients as
\begin{equation}
\begin{split}
&\lambda_{C \sigma \Sigma_{\ell}}^{n_{IJ}}=\Lambda_{[\sigma,\sigma]_{0,4} \sigma [\sigma,\epsilon]_{0,\ell}}^{n_{IJ}} L^{n_{IJ}}_{C \sigma \Sigma_{\ell}}, \qquad \ell=2,3,
\\
&\lambda_{T \sigma \Sigma_{\ell}}^{0}=\Lambda_{[\sigma,\sigma]_{0,2} \sigma [\sigma,\epsilon]_{0,\ell}}^{0} L^{0}_{T \sigma \Sigma_{\ell}}, \qquad \ell=2,3, \ldots, 7,\\
&\lambda_{\epsilon' \sigma \sigma'}=\Lambda_{[\sigma,\sigma]_{1,0} \sigma [\sigma,\epsilon]_{2,0}}^{0} L_{\epsilon' \sigma \sigma'},\\
&\lambda_{C \sigma \sigma'}=\Lambda_{[\sigma,\sigma]_{0,4} \sigma [\sigma,\epsilon]_{2,0}}^{0} L_{C \sigma \sigma'}.
\end{split}
\end{equation}
We use the values of $\lambda_{\epsilon' \sigma \Sigma_{\ell}}$ given in table~\ref{seepes}, since we are unable to accurately and precisely estimate the values of these coefficients from this correlator. Along with the constraints obtained by taking up to three derivatives $\CD_i$ of the crossing relation, we include the following subset of fourth order derivatives:
\begin{equation}
\CD_i = \{ \partial_{a^+}^4, \,\, \partial_{w}^4, \,\, \partial_{w}^3\partial_{b^+}, \,\, \partial_{w}^2\partial_{b^-}^2, \,\, \partial_{b^-}^4 \}.
\end{equation}
We have chosen these such that the constraints we obtain from this subset are the best-satisfied ones for the disconnected $\langle \sigma \sigma \sigma \sigma \epsilon \rangle_{\rm d}$ correlator.

\begin{figure}[t]
\centering
\includegraphics[width=0.315\textwidth]{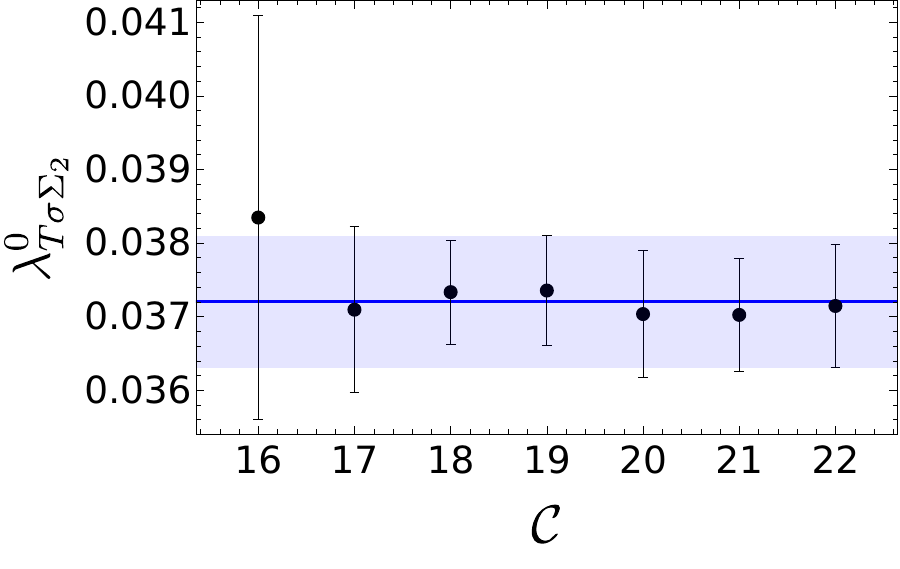}
\hspace{0.01\textwidth}
\includegraphics[width=0.315\textwidth]{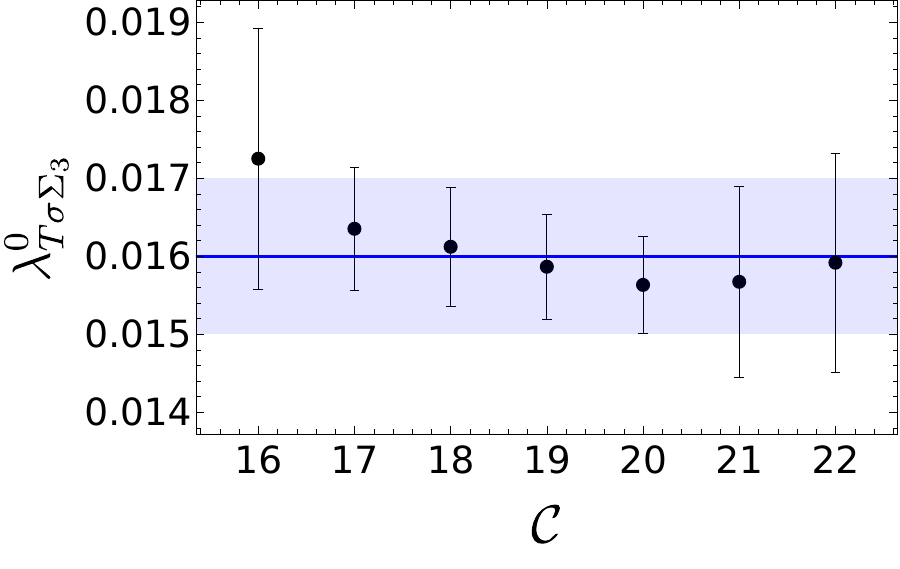}
\hspace{0.01\textwidth}
\includegraphics[width=0.315\textwidth]{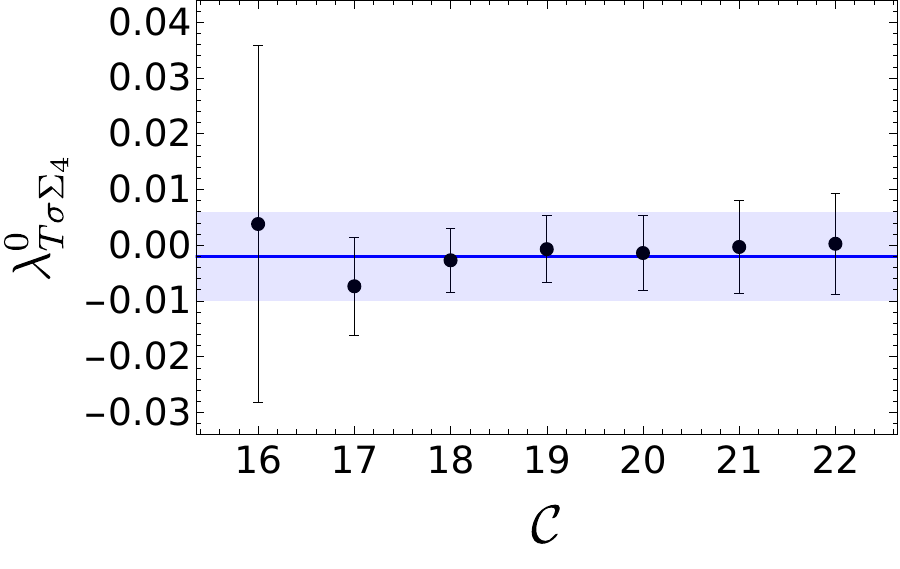}\\
\includegraphics[width=0.315\textwidth]{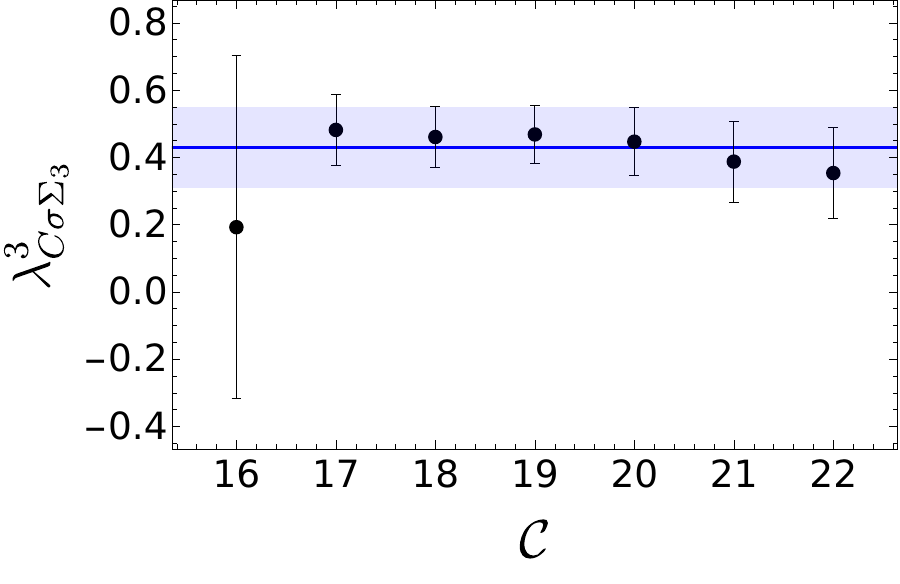}
\hspace{0.01\textwidth}
\includegraphics[width=0.315\textwidth]{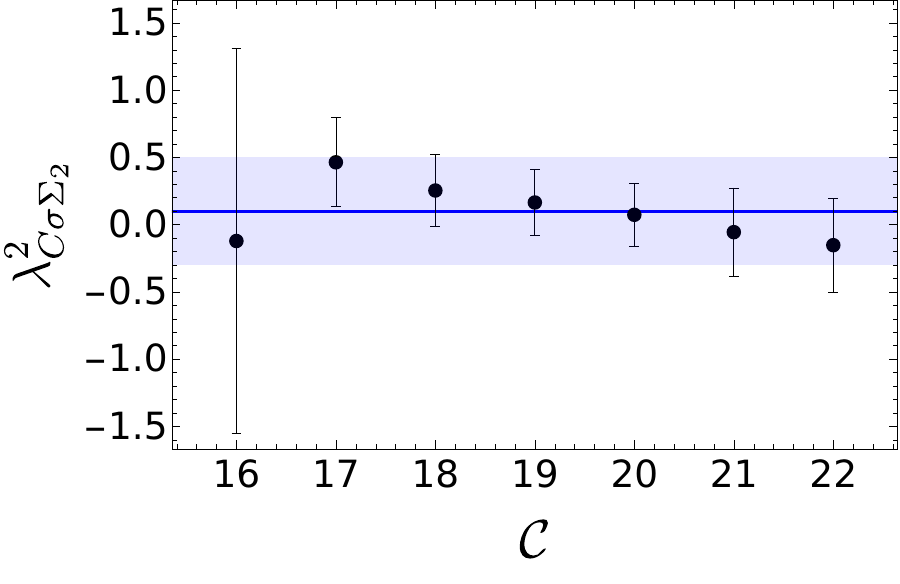}
\hspace{0.01\textwidth}
\includegraphics[width=0.315\textwidth]{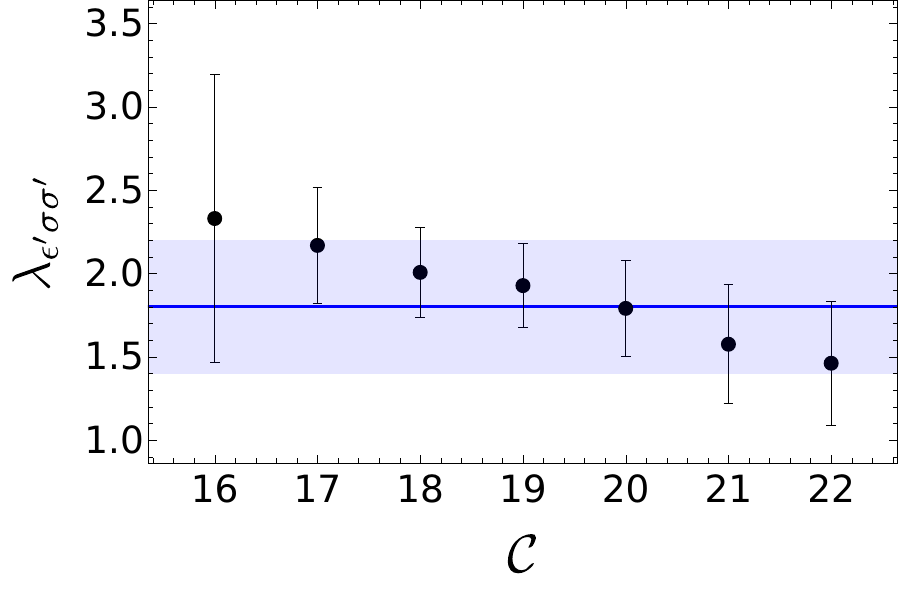}
\caption{{OPE coefficients computed from $\langle\sigma\sigma\sigma\sigma\epsilon \rangle$ by averaging over the positions of the minima of the cost function for 20 sets of $\CC$ constraints with the largest $\CI$, where the weights of individual constraints in each set are pseudo-randomly varied. The horizontal blue lines and shaded regions correspond to the results given in table~\ref{sssse}. These values are obtained by averaging results with $17\leqslant \CC \leqslant 22$. One can again notice that the standard deviations for multiple OPE coefficients are large for low values of $\CC$, which is a consequence of the presence of flat directions of the cost function near the minima. Adding more constraints lifts these flat directions. Other OPE coefficients exhibit similar behavior with larger standard deviations.}}
\label{fig:sssse}
\end{figure}

We display the results for the estimated values of the unknown OPE coefficients in table~\ref{sssse} and fig.~\ref{fig:sssse}. We note that the fuzzy sphere regularization method~\cite{Hu:2023xak} gives $\lambda_{\sigma' \sigma \epsilon'}\simeq 1.294(51)$ in our conventions, while our rough estimate is $\lambda_{\sigma' \sigma \epsilon'}\simeq 1.8(4)$, which is almost within the error bar. 

\begin{table}[t!]
\centering
\begin{minipage}{0.48\textwidth}
\centering
\begin{tabular}{|l|l|}
\hline
                                                   & est.\,values \\ \hline
$\lambda_{T\sigma \Sigma_2}^0$      				   & 0.0372(9)         \\
$\lambda_{T\sigma \Sigma_3}^0$      				   & 0.0160(10)         \\
$\lambda_{T\sigma \Sigma_4}^0$      				   & -0.002(8)      \\
$\lambda_{T\sigma \Sigma_5}^0$   					   & -0.0010(7)         \\ 
$\lambda_{T\sigma \Sigma_6}^0$   					   & -0.008(10)           \\
$\lambda_{T\sigma \Sigma_7}^0$   					   & -0.018(7)           \\
$\lambda_{\epsilon' \sigma \sigma'}$				       & 1.8(4)  \\     
$\lambda_{C \sigma \sigma'}^{0}$  				       & -1.2(9)    \\  \hline
\end{tabular}
\end{minipage}
\hfill
\begin{minipage}{0.48\textwidth}
\centering
\begin{tabular}{|l|l|}
\hline
                                                   & est.\,values \\ \hline
                                                   $\lambda_{C \sigma \Sigma_2}^{2}$  				       & 0.1(4)          \\
$\lambda_{C \sigma \Sigma_2}^{1}$   					   & 1(1)          \\ 
$\lambda_{C \sigma \Sigma_2}^{0}$				   & 1(1)          \\
$\lambda_{C \sigma \Sigma_3}^{3}$				   & 0.43(12)          \\
$\lambda_{C \sigma \Sigma_3}^{2}$				   & 1(2)          \\ 
$\lambda_{C \sigma \Sigma_3}^{1}$ 				   & 3(2)            \\
$\lambda_{C \sigma \Sigma_3}^{0}$ 				   & 2(2)            \\ \hline
\end{tabular}
\end{minipage}
\caption{Numerical data for the unknown OPE coefficients in truncated $\langle\sigma\sigma\sigma\sigma\epsilon \rangle$ correlator, obtained with 20 sets of constraints with the largest $\mathcal{I}$ for $17\leqslant \mathcal{C}\leqslant 22$. Correlation matrices for OPE coefficients of the same operators for different tensor structures are provided in Appendix~\ref{correlationmatrix}.}
\label{sssse}
\end{table}

\section{OPE coefficients from the fuzzy sphere}\label{fuzzysphere}

In this section we compute some of the unknown OPE coefficients in critical 3d Ising model using the fuzzy sphere regularization method. We compare the fuzzy sphere predictions to some of our computations in section~\ref{secfour} and observe agreement for those OPE coefficients which we consider to be accurately determined, i.e.~with relatively small error bars. We directly use the algorithm for OPE computation developed in \cite{Hu:2023xak} and the code for fuzzy sphere regularization written in \cite{Zhou:2025liv}.

We fix the parameters in the Hamiltonian at the critical point to the values found in \cite{Zhu:2022gjc}: $V_1/V_0 = 4.75$ and $h/V_0 = 3.16$. To fix the sign of the OPE coefficients, we invoke the fact that $\lambda_{\sigma\sigma \epsilon}>0$, $\lambda_{\epsilon\epsilon \epsilon'}>0$, $\lambda_{\sigma\sigma T}<0$, $\lambda_{\sigma\epsilon \sigma'}>0$, $\lambda_{\sigma\epsilon \Sigma_2}>0$, and $\lambda_{\sigma\epsilon \Sigma_3}>0$ in our conventions. We then compute OPE coefficients for various values of the number of electrons on the fuzzy sphere $N$, in particular, for $8 \leqslant N\leqslant 18$, and then extrapolate the results in the limit $N\to \infty$. The extrapolation is done by considering the subleading contributions from descendants and new primaries in the $N\to \infty$ limit, as described in \cite{Hu:2023xak}.

All OPE coefficients we compute by means of the fuzzy sphere regularization technique contain $\sigma$ or $\epsilon$ operators. These operators are represented on the fuzzy sphere by the density operator $n^z(\Omega)$, in the case of $\sigma$, and by $\CO_\epsilon(\Omega)$ or the density operator $n^x(\Omega)$, in the case of $\epsilon$. The density operators $n^i(\Omega)$ and $\CO_\epsilon(\Omega)$ are defined in eqs.~(S18) and (S21) of \cite{Hu:2023xak}, respectively. 

\begin{figure}[t]
\centering
\includegraphics[width=0.8\textwidth]{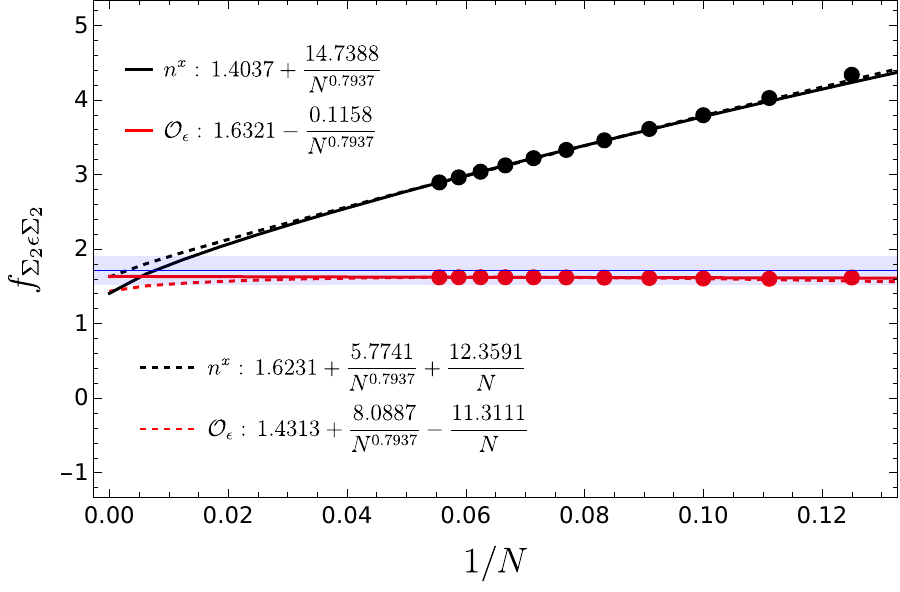}
\caption{{We show the fuzzy sphere results for the OPE coefficient $f_{\Sigma_2\epsilon\Sigma_2}$. The black and red dots and fits represent different ways in which the $\epsilon$ operator can be represented on the fuzzy sphere. The fitting was performed using the 6 points with the largest $N$. The horizontal blue line and shaded region show the results from the five-point bootstrap given in table~\ref{seees}.}}
\label{S2eS2-f}
\end{figure}
On the fuzzy sphere, $f_{\Sigma_2\epsilon\Sigma_2}$ is computed as
\begin{equation}
\begin{split}
&\frac{\langle \Sigma_2, m=0| \int d\Omega Y_{0,0}(\Omega) n^x(\Omega)|\Sigma_2, m=0\rangle - \langle 0| \int d\Omega Y_{0,0}(\Omega) n^x(\Omega)|0\rangle}{\langle \epsilon| \int d\Omega Y_{0,0}(\Omega) n^x(\Omega)|0\rangle} \\ 
&\qquad\qquad= f_{\Sigma_2\epsilon\Sigma_2}+\frac{c_1}{N^\frac{\Delta_T -\Delta_\epsilon}{2}} +\frac{c_2}{N}+\CO\left(N^{-\frac{\Delta_{\epsilon'} -\Delta_\epsilon}{2}}\right),
\end{split}
\end{equation}
where $Y_{\ell,n}$ denote spherical harmonics. A similar computation yields $f_{\Sigma_2\epsilon\Sigma_2}$ when the $n^x$ operator is replaced by $\CO_\epsilon$. The subleading corrections arise from the stress tensor and descendants of $\epsilon$, while the $c$-coefficients are determined by fitting the data. In the standard box basis, $f_{\Sigma_2\epsilon\Sigma_2}$ is given by
\begin{equation}
f_{\Sigma_2\epsilon\Sigma_2}=\lambda_{\Sigma_2\epsilon\Sigma_2}^2-\frac{1}{3}\lambda_{\Sigma_2\epsilon\Sigma_2}^1+\frac{2}{15}\lambda_{\Sigma_2\epsilon\Sigma_2}^0.
\end{equation}
Upon inserting our computed values from table~\ref{seees}, we obtain   $f_{\Sigma_2\epsilon\Sigma_2}\simeq 1.7(2)$, which agrees with the fuzzy sphere result shown in fig.~\ref{S2eS2-f}. We note that in \cite{Lauchli:2025fii}, $f_{\Sigma_2\epsilon\Sigma_2}$ was reported as $f_{\Sigma_2\epsilon\Sigma_2}\simeq 1.46$.

\begin{figure}[t]
\centering
\includegraphics[width=0.48\textwidth]{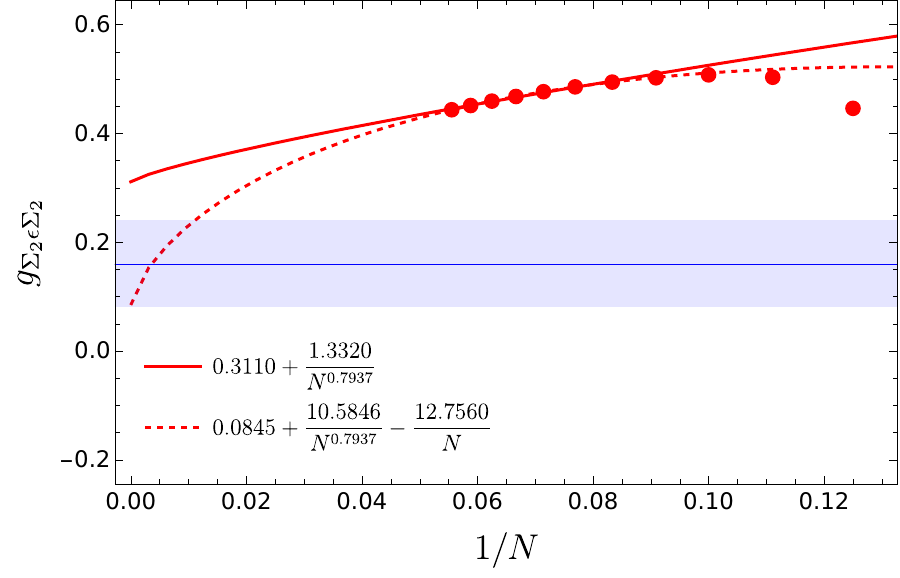}
\hspace{0.01\textwidth}
\includegraphics[width=0.48\textwidth]{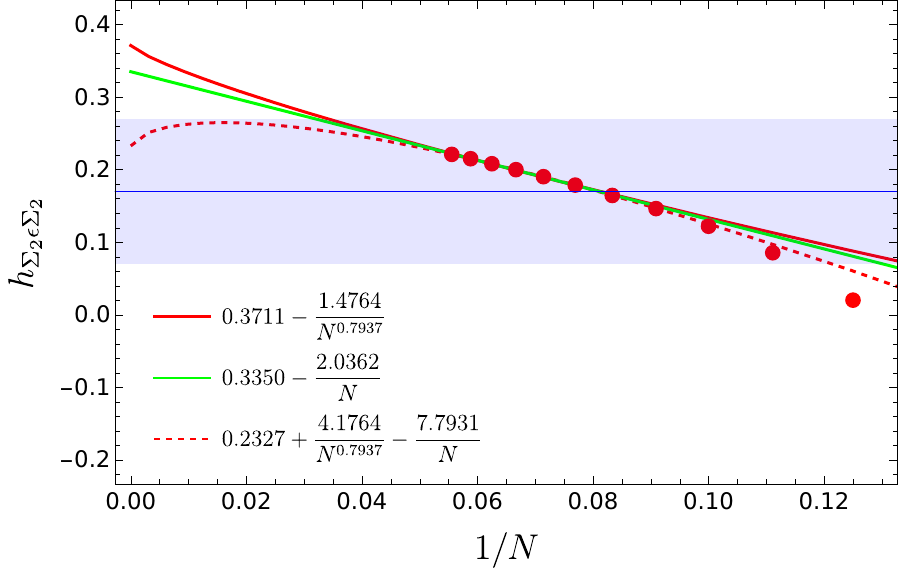}
\caption{{We show the fuzzy sphere results for the OPE coefficients $g_{\Sigma_2\epsilon\Sigma_2}$ and $h_{\Sigma_2\epsilon\Sigma_2}$ (see eq.~(\ref{ghS2S2}) for their relation to standard box basis OPE coefficients). Here, the $\epsilon$ operator is represented by the $n^{x}$ operator on the fuzzy sphere. The fitting was performed using the 6 points with the largest $N$. The horizontal blue lines and shaded regions show the results from the five-point bootstrap given in table~\ref{seees}.}}
\label{S2eS2-gh}
\end{figure}

On the fuzzy sphere, $g_{\Sigma_2\epsilon\Sigma_2}$ and $h_{\Sigma_2\epsilon\Sigma_2}$ are computed as
\begin{equation}
\begin{split}
&\frac{\langle \Sigma_2, m=0| \int d\Omega Y_{2,0}(\Omega) n^x(\Omega)|\Sigma_2, m=0\rangle}{\langle \epsilon| \int d\Omega Y_{0,0}(\Omega) n^x(\Omega)|0\rangle} =g_{\Sigma_2\epsilon\Sigma_2}+\frac{c_1}{N^\frac{\Delta_T -\Delta_\epsilon}{2}} +\frac{c_2}{N}+\CO\left(N^{-\frac{\Delta_{\epsilon'} -\Delta_\epsilon}{2}}\right),\\
&\frac{\langle \Sigma_2, m=0| \int d\Omega Y_{4,0}(\Omega) n^x(\Omega)|\Sigma_2, m=0\rangle}{\langle \epsilon| \int d\Omega Y_{0,0}(\Omega) n^x(\Omega)|0\rangle} =h_{\Sigma_2\epsilon\Sigma_2}+\frac{c_1}{N^\frac{\Delta_T -\Delta_\epsilon}{2}} +\frac{c_2}{N}+\CO\left(N^{-\frac{\Delta_{\epsilon'} -\Delta_\epsilon}{2}}\right),
\end{split}
\end{equation}
where $|\Sigma_2, m=0\rangle$ is defined analogously to $|T_{\mu\nu}, m=0\rangle$ in \cite{Hu:2023xak} (see Appendix~\ref{fuzzydetails} for details on how CFT operators relate to the states on the fuzzy sphere), to be specific,
\begin{equation}\label{spin2st}
|\Sigma_2, m=0\rangle =\lim_{{\bf r}\to 0} \sqrt{\frac{2}{27}}\left(2\Sigma_{33}({\bf r})-\Sigma_{11}({\bf r})-\Sigma_{22}({\bf r})\right)|0\rangle.
\end{equation}
In the standard box basis, $g_{\Sigma_2\epsilon\Sigma_2}$ and $h_{\Sigma_2\epsilon\Sigma_2}$ are given by
\begin{equation}\label{ghS2S2}
\begin{split}
&g_{\Sigma_2\epsilon\Sigma_2}=-\frac{1}{3\sqrt{5}}\lambda_{\Sigma_2\epsilon\Sigma_2}^1+\frac{4}{21\sqrt{5}}\lambda_{\Sigma_2\epsilon\Sigma_2}^0,\\
&h_{\Sigma_2\epsilon\Sigma_2}=\frac{4}{35}\lambda_{\Sigma_2\epsilon\Sigma_2}^0,
\end{split}
\end{equation}
and, upon inserting our values from table~\ref{seees}, we find $g_{\Sigma_2\epsilon\Sigma_2}\simeq 0.16(8)$ and $h_{\Sigma_2\epsilon\Sigma_2}\simeq 0.17(10)$, which are roughly consistent with the fuzzy sphere results shown in fig.~\ref{S2eS2-gh}.

\begin{figure}[t]
\centering
\includegraphics[width=0.315\textwidth]{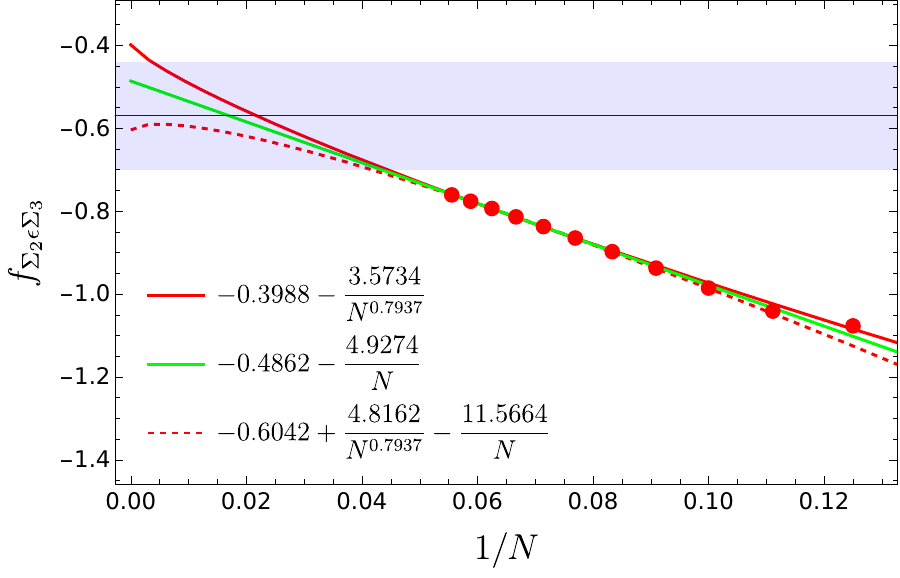}
\hspace{0.01\textwidth}
\includegraphics[width=0.315\textwidth]{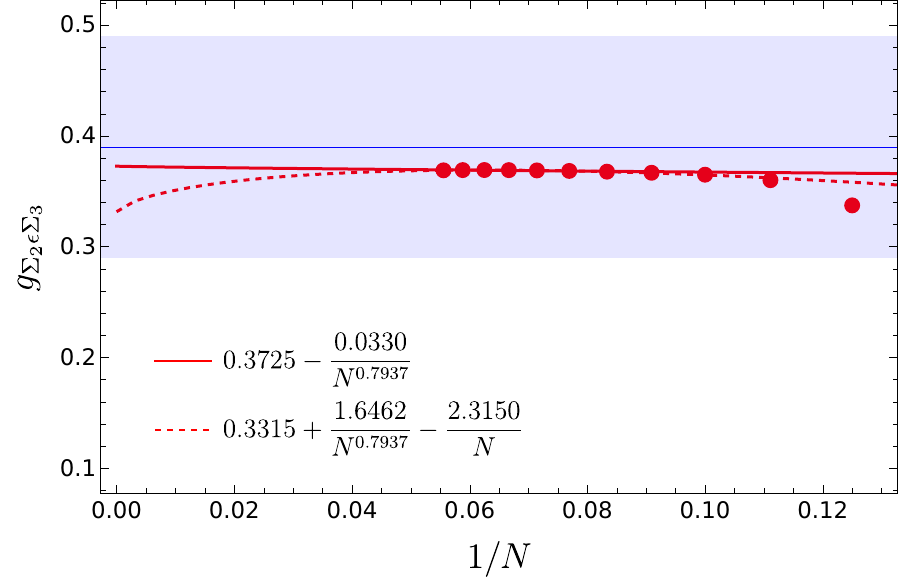}
\hspace{0.01\textwidth}
\includegraphics[width=0.315\textwidth]{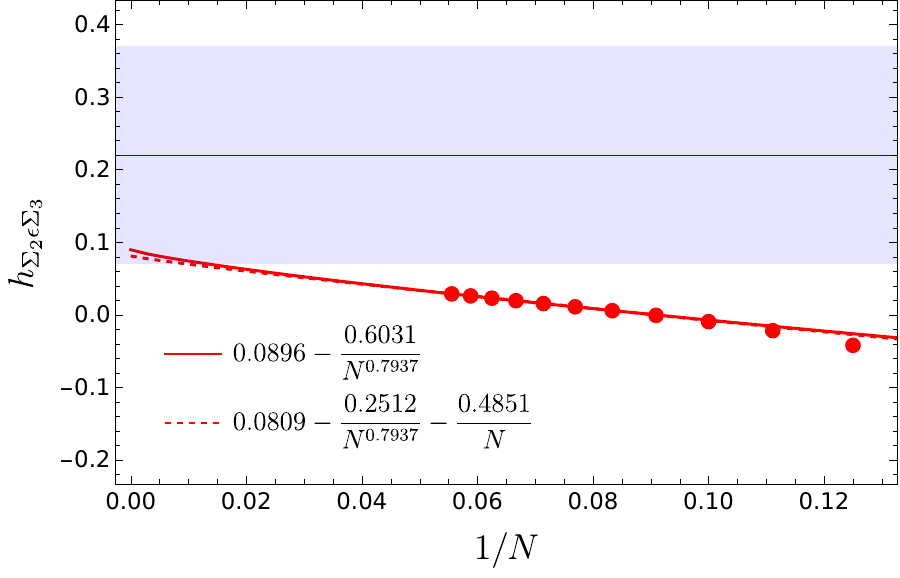}
\caption{{We show the fuzzy sphere results for the OPE coefficients $f_{\Sigma_2\epsilon\Sigma_3}$, $g_{\Sigma_2\epsilon\Sigma_3}$, and $h_{\Sigma_2\epsilon\Sigma_3}$ (see eq.~(\ref{ghS2S3}) for their relation to the standard box basis OPE coefficients). Here, the $\epsilon$ operator is represented by the $n^{x}$ operator on the fuzzy sphere. The fitting was performed using the 6 points with the largest $N$. The horizontal blue lines and shaded regions show the results from the five-point bootstrap given in table~\ref{seees}.}}
\label{S2eS3-fgh}
\end{figure}
Further, on the fuzzy sphere, $f_{\Sigma_2\epsilon\Sigma_3}$, $g_{\Sigma_2\epsilon\Sigma_3}$, and $h_{\Sigma_2\epsilon\Sigma_3}$ are computed as
\begin{equation}
\begin{split}
&\frac{\langle \Sigma_2, m=0| \int d\Omega Y_{1,0}(\Omega) n^x(\Omega)|\Sigma_3, m=0\rangle}{\langle \epsilon| \int d\Omega Y_{0,0}(\Omega) n^x(\Omega)|0\rangle} =f_{\Sigma_2\epsilon\Sigma_3}+\frac{c_1}{N^\frac{\Delta_T -\Delta_\epsilon}{2}} +\frac{c_2}{N}+\CO\left(N^{-\frac{\Delta_{\epsilon'} -\Delta_\epsilon}{2}}\right),\\
&\frac{\langle \Sigma_2, m=0| \int d\Omega Y_{3,0}(\Omega) n^x(\Omega)|\Sigma_3, m=0\rangle}{\langle \epsilon| \int d\Omega Y_{0,0}(\Omega) n^x(\Omega)|0\rangle} =g_{\Sigma_2\epsilon\Sigma_3}+\frac{c_1}{N^\frac{\Delta_T -\Delta_\epsilon}{2}} +\frac{c_2}{N}+\CO\left(N^{-\frac{\Delta_{\epsilon'} -\Delta_\epsilon}{2}}\right),\\
&\frac{\langle \Sigma_2, m=0| \int d\Omega Y_{5,0}(\Omega) n^x(\Omega)|\Sigma_3, m=0\rangle}{\langle \epsilon| \int d\Omega Y_{0,0}(\Omega) n^x(\Omega)|0\rangle} =h_{\Sigma_2\epsilon\Sigma_3}+\frac{c_1}{N^\frac{\Delta_T -\Delta_\epsilon}{2}} +\frac{c_2}{N}+\CO\left(N^{-\frac{\Delta_{\epsilon'} -\Delta_\epsilon}{2}}\right),
\end{split}
\end{equation}
where
\begin{equation}\label{spin3st}
|\Sigma_3, m=0\rangle =\lim_{{\bf r}\to 0} \sqrt{\frac{8}{10125}}\left(2\Sigma_{333}({\bf r})-3\Sigma_{113}({\bf r})-3\Sigma_{223}({\bf r})\right)|0\rangle.
\end{equation}

In the standard box basis, $f_{\Sigma_2\epsilon\Sigma_3}$,  $g_{\Sigma_2\epsilon\Sigma_3}$, and  $h_{\Sigma_2\epsilon\Sigma_3}$ are given by
\begin{equation}\label{ghS2S3}
\begin{split}
&f_{\Sigma_2\epsilon\Sigma_3}=-\frac{6}{35 \sqrt{5}}\lambda_{\Sigma_2\epsilon\Sigma_3}^0+\frac{2}{5 \sqrt{5}}\lambda_{\Sigma_2\epsilon\Sigma_3}^1-\frac{1}{\sqrt{5}}\lambda_{\Sigma_2\epsilon\Sigma_3}^2,\\
&g_{\Sigma_2\epsilon\Sigma_3}=-\frac{8}{15 \sqrt{105}}\lambda_{\Sigma_2\epsilon\Sigma_3}^0+\frac{4}{5 \sqrt{105}}\lambda_{\Sigma_2\epsilon\Sigma_3}^1,\\
&h_{\Sigma_2\epsilon\Sigma_3}=-\frac{4}{21} \sqrt{\frac{5}{33}} \lambda_{\Sigma_2\epsilon\Sigma_3}^0.
\end{split}
\end{equation}
After inserting our values from table~\ref{seees}, we obtain $f_{\Sigma_2\epsilon\Sigma_3}\simeq -0.57(13)$, $g_{\Sigma_2\epsilon\Sigma_3}\simeq 0.39(10)$ and $h_{\Sigma_2\epsilon\Sigma_3}\simeq 0.22(15)$,  which all roughly agree with the fuzzy sphere results shown in fig.~\ref{S2eS3-fgh}. 

\begin{figure}[t]
\centering
\includegraphics[width=0.48\textwidth]{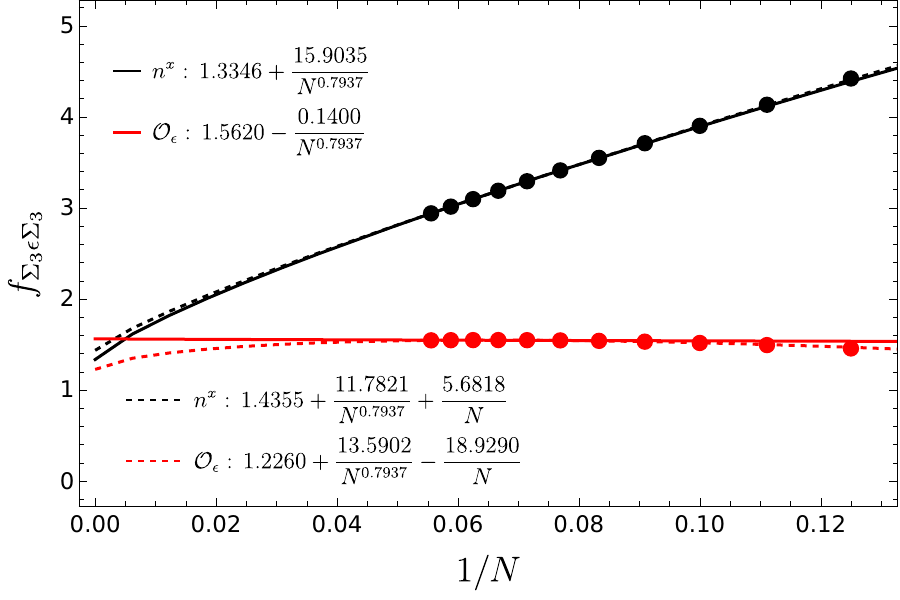}
\hspace{0.01\textwidth}
\includegraphics[width=0.48\textwidth]{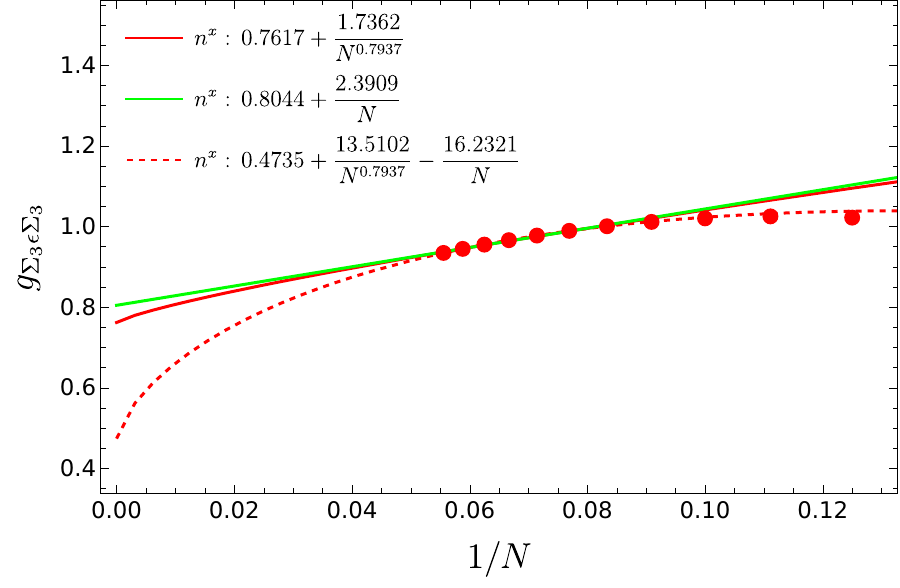}\\
\includegraphics[width=0.48\textwidth]{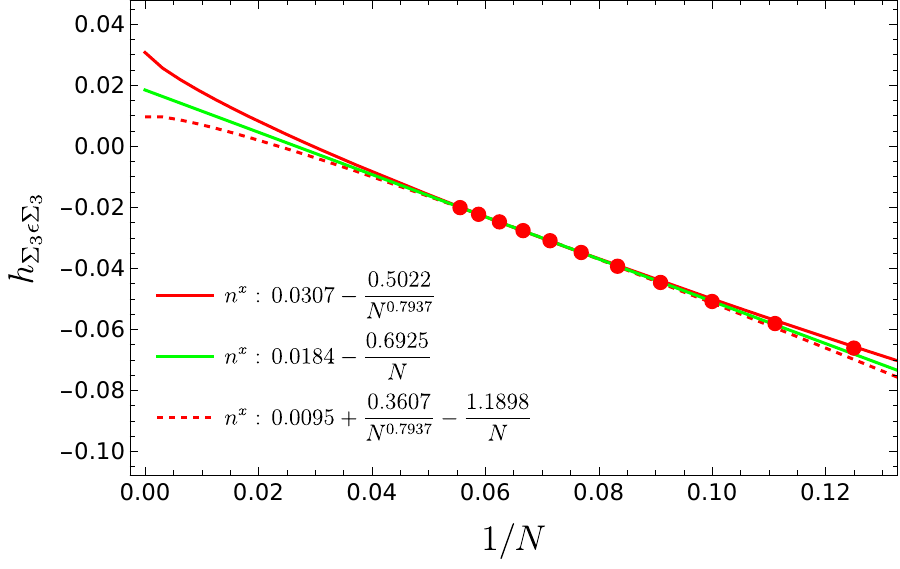}
\hspace{0.01\textwidth}
\includegraphics[width=0.48\textwidth]{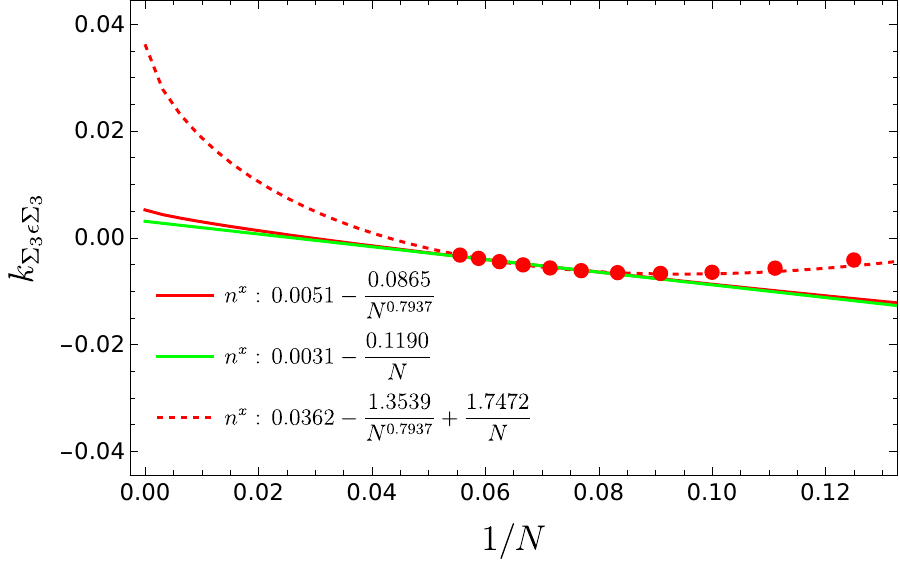}
\caption{{We show the fuzzy sphere results for the OPE coefficients $f_{\Sigma_3\epsilon\Sigma_3}$, $g_{\Sigma_3\epsilon\Sigma_3}$, $h_{\Sigma_3\epsilon\Sigma_3}$, and $k_{\Sigma_3\epsilon\Sigma_3}$ (see eq.~(\ref{fghkS3S3}) for their relation to the standard box basis OPE coefficients). Here, the $\epsilon$ operator is represented by the $\CO_\epsilon$ or $n^{x}$ operator on the fuzzy sphere. All fits are computed using the 6 points with the largest $N$.}}
\label{S3eS3-fghk}
\end{figure}
On the fuzzy sphere the OPE coefficients for $\langle \Sigma_3\epsilon\Sigma_3\rangle$ are calculated analogously to those for $\langle \Sigma_2\epsilon\Sigma_2\rangle$. We use the $Y_{0,0}$, $Y_{2,0}$, $Y_{4,0}$, and $Y_{6,0}$ spherical harmonics for the $f$-, the $g$-, the $h$-, and the $k$- coefficient, respectively. In the standard box basis, $f_{\Sigma_3\epsilon\Sigma_3}$,  $g_{\Sigma_3\epsilon\Sigma_3}$,  $h_{\Sigma_3\epsilon\Sigma_3}$, and $k_{\Sigma_3\epsilon\Sigma_3}$ are given by
\begin{equation}\label{fghkS3S3}
\begin{split}
&f_{\Sigma_3\epsilon\Sigma_3}=-\frac{2}{35}\lambda_{\Sigma_3\epsilon\Sigma_3}^0+\frac{2}{15}\lambda_{\Sigma_3\epsilon\Sigma_3}^1-\frac{1}{3}\lambda_{\Sigma_3\epsilon\Sigma_3}^2+\lambda_{\Sigma_3\epsilon\Sigma_3}^3,\\
&g_{\Sigma_3\epsilon\Sigma_3}=-\frac{8}{105 \sqrt{5}}\lambda_{\Sigma_3\epsilon\Sigma_3}^0+\frac{16}{105 \sqrt{5}}\lambda_{\Sigma_3\epsilon\Sigma_3}^1-\frac{4}{15 \sqrt{5}}\lambda_{\Sigma_3\epsilon\Sigma_3}^2,\\
&h_{\Sigma_3\epsilon\Sigma_3}=-\frac{12}{385}\lambda_{\Sigma_3\epsilon\Sigma_3}^0+\frac{4}{105}\lambda_{\Sigma_3\epsilon\Sigma_3}^1,\\
&k_{\Sigma_3\epsilon\Sigma_3}=-\frac{40}{231 \sqrt{13}}\lambda_{\Sigma_3\epsilon\Sigma_3}^0.
\end{split}
\end{equation}
After inserting our values from table~\ref{seees}, we arrive at $f_{\Sigma_3\epsilon\Sigma_3}\simeq 4(1)$, $g_{\Sigma_3\epsilon\Sigma_3}\simeq 1.5(4)$, $h_{\Sigma_3\epsilon\Sigma_3}\simeq 0.4(2)$, and $k_{\Sigma_3\epsilon\Sigma_3}\simeq 0.2(1)$. These values do not quite agree with the fuzzy sphere results shown in fig.~\ref{S3eS3-fghk}, but this is not surprising, since the systematic error due to truncation in the five-point bootstrap increases for higher-spin states. This may be inferred from the observation that these OPE coefficients have larger error bars in table~\ref{seees}. Also, we note that the convergence in the fuzzy sphere calculation is not ideal; to improve it, one would presumably need to go to higher $N$, which is computationally challenging. In \cite{Lauchli:2025fii}, $f_{\Sigma_3\epsilon\Sigma_3}$ was calculated as $f_{\Sigma_3\epsilon\Sigma_3}\simeq 1.38$.

\begin{figure}[t]
\centering
\includegraphics[width=0.48\textwidth]{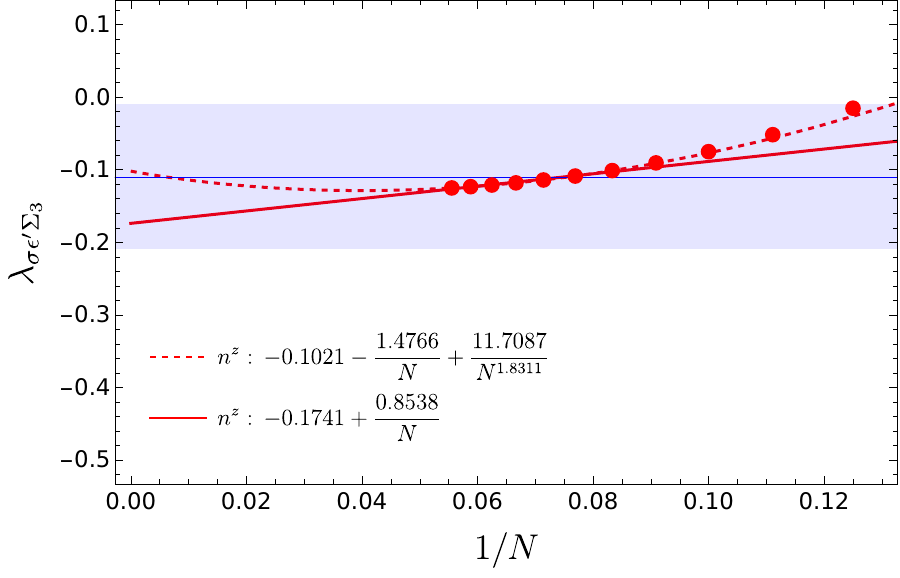}
\hspace{0.01\textwidth}
\includegraphics[width=0.48\textwidth]{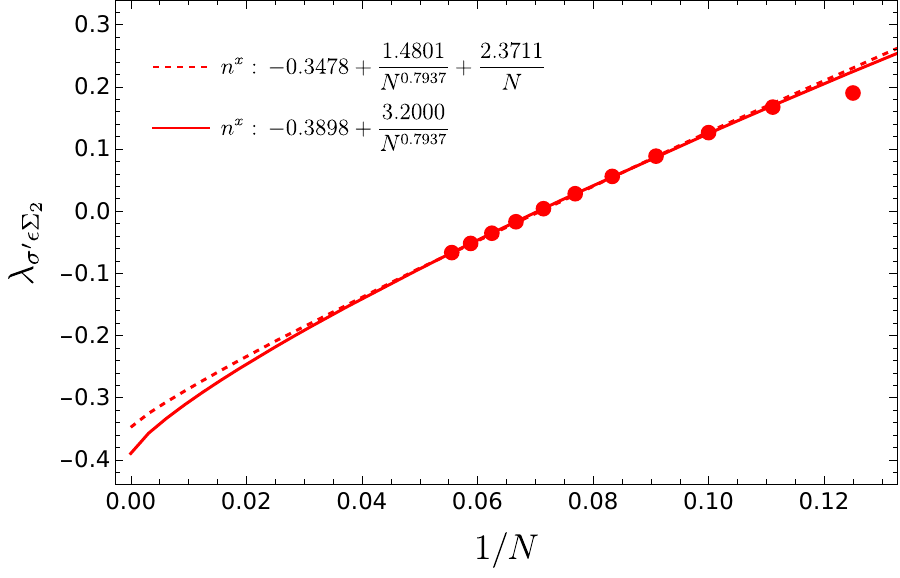}
\caption{{\small We show the fuzzy sphere results for the OPE coefficients $\lambda_{\sigma\epsilon'\Sigma_3}$ and $\lambda_{\sigma'\epsilon\Sigma_2}$. Here, the $\sigma$ and $\epsilon$ states on the fuzzy sphere are represented by the $n^z$ and  $n^x$ operators, respectively. Both fits are computed using the 6 points with the largest $N$. The horizontal blue line and shaded region show the results from the five-point bootstrap given in table~\ref{seepes}.}}
\label{sepS3-f}
\end{figure}

Further, we can compute $\lambda_{\sigma\epsilon'\Sigma_3}$ and $\lambda_{\sigma'\epsilon\Sigma_2}$ on the fuzzy sphere using 
\begin{equation}
\begin{split}
&\frac{\langle \epsilon'    | \int d\Omega Y_{3,0}(\Omega) n^z(\Omega)|\Sigma_3, m=0\rangle}{\langle \sigma  | \int d\Omega Y_{0,0}(\Omega) n^z(\Omega)|0\rangle} =\sqrt{\frac{2}{35}}\lambda_{\sigma\epsilon'\Sigma_3}+\frac{c_1}{N} +\frac{c_2}{N^{\frac{\Delta_{\Sigma_2}-\Delta_\sigma}{2}}}+\CO\left(N^{-2}\right),\\
&\frac{\langle \sigma'      | \int d\Omega Y_{2,0}(\Omega) n^x(\Omega)|\Sigma_2, m=0\rangle}{\langle \epsilon| \int d\Omega Y_{0,0}(\Omega) n^x(\Omega)|0\rangle} =\sqrt{\frac{2}{15}}\lambda_{\sigma'\epsilon\Sigma_2}+\frac{c_1}{N^\frac{\Delta_T -\Delta_\epsilon}{2}} +\frac{c_2}{N}+\CO\left(N^{-\frac{\Delta_{\epsilon'} -\Delta_\epsilon}{2}}\right).
\end{split}
\end{equation}
We can compare the fuzzy sphere prediction for $\lambda_{\sigma\epsilon'\Sigma_3}$ with our result in table~\ref{seepes}, where we obtained $\lambda_{\sigma\epsilon' \Sigma_3}\simeq -0.11(10)$.  We remark that the five-point bootstrap does not give a precise prediction for $\lambda_{\sigma'\epsilon\Sigma_2}$, since in table~\ref{seees} we have $\lambda_{\sigma'\epsilon\Sigma_2}\simeq -1(2)$. The results for these coefficients are presented in fig.~\ref{sepS3-f}. However, the result for $\lambda_{\sigma\epsilon'\Sigma_3}$ seems to be good in agreement with the corresponding bootstrap result. 

\begin{figure}[t]
\centering
\includegraphics[width=0.48\textwidth]{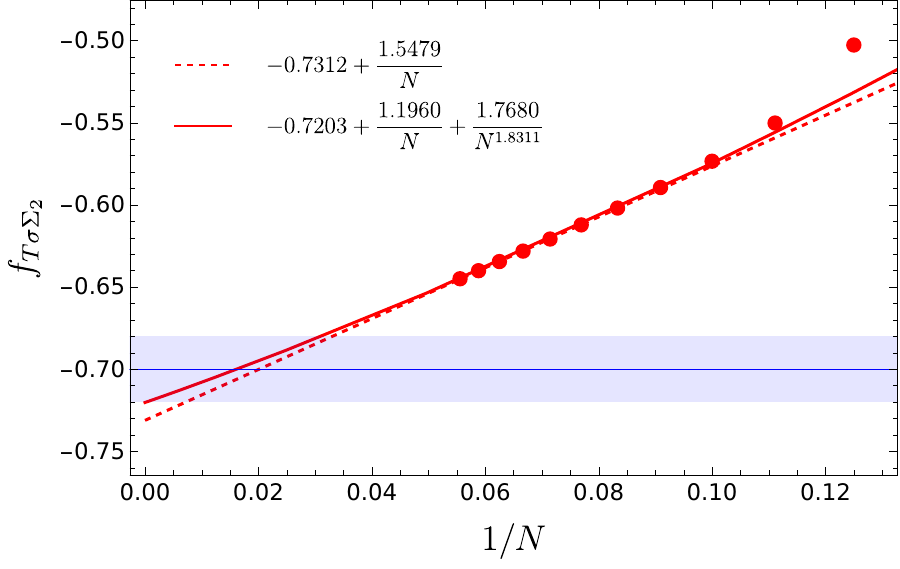}
\hspace{0.01\textwidth}
\includegraphics[width=0.48\textwidth]{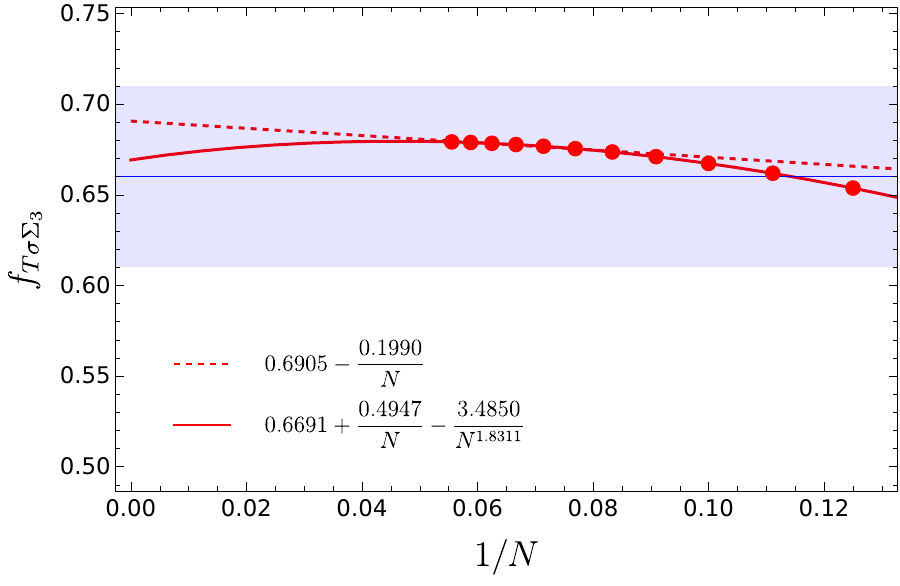}
\caption{{\small We show the fuzzy sphere results for the OPE coefficients $f_{T \sigma \Sigma_2}$ and $f_{T \sigma \Sigma_3}$. Here, the $\sigma$ state on the fuzzy sphere is represented by the $n^z$ operator. The horizontal blue lines and shaded regions show the results from the five-point bootstrap given in table~\ref{sssse}.}}
\label{TsS23-f}
\end{figure}
Finally, the $f_{T \sigma \Sigma_2}$ and $f_{T \sigma \Sigma_3}$ OPE coefficients can be computed on the fuzzy sphere in a similar way as  $f_{\Sigma_2 \epsilon \Sigma_2}$ (without subtracting the identity contribution) and $f_{\Sigma_2 \epsilon \Sigma_3}$, respectively. But, here, the subleading contributions are different, as they come from descendants of $\sigma$ and $\Sigma_2$. 

In particular, these coefficients are related to the OPE coefficients in the standard box basis by
\begin{equation}
\begin{split}
&f_{T \sigma \Sigma_2}=\lambda_{T \sigma \Sigma_2}^2-\frac{1}{3}\lambda_{T \sigma \Sigma_2}^1+\frac{2}{15}\lambda_{T \sigma \Sigma_2}^0,\\
&f_{T \sigma \Sigma_3}=-\frac{6}{35 \sqrt{5}}\lambda_{T \sigma \Sigma_3}^0+\frac{2}{5 \sqrt{5}}\lambda_{T \sigma \Sigma_3}^1-\frac{1}{\sqrt{5}}\lambda_{T \sigma \Sigma_3}^2.
\end{split}
\end{equation}
After enforcing Ward identities (since the stress tensor is a conserved current), we can relate these coefficients to $\lambda_{T\sigma \Sigma_2}^0$ and $\lambda_{T\sigma \Sigma_3}^0$, respectively. Upon applying our results from table~\ref{sssse}, we find
\begin{equation}
\begin{split}
&f_{T \sigma \Sigma_2}=\left(\frac{4}{5}-\frac{14}{\Delta _{\sigma }-\Delta _{\Sigma _2}+5}+\frac{6}{\Delta _{\sigma }-\Delta _{\Sigma
   _2}+3}\right)\lambda_{T\sigma \Sigma_2}^0 \simeq -0.70(2),\\
& f_{T \sigma \Sigma_3}=  -\frac{4 \left(\Delta _{\sigma }-\Delta _{\Sigma _3}-3\right) \left(\Delta _{\sigma
   }-\Delta _{\Sigma _3}-1\right)}{7 \sqrt{5} \left(\Delta _{\sigma }-\Delta _{\Sigma _3}+4\right)
   \left(\Delta _{\sigma }-\Delta _{\Sigma _3}+6\right)}\lambda_{T\sigma \Sigma_3}^0\simeq 0.66(5).
\end{split}
\end{equation}
We note that these appear to be consistent with the fuzzy sphere result shown in fig.~\ref{TsS23-f}.

\section{Discussion}\label{secfive}

In this work we studied several five-point correlators containing the $\sigma$, $\epsilon$, and $\epsilon'$ operators in the critical 3d Ising model. We evaluated these correlators by considering the $\sigma \times \sigma$ and $\sigma \times \epsilon$ operator product expansions. We used the truncation method~\cite{Gliozzi:2013ysa, Gliozzi:2014jsa} in the OPE and  approximated the truncated contributions by the corresponding contributions from an appropriate disconnected five-point correlator, as was done in~\cite{Poland:2023bny}. Using this approach we estimated several previously unknown OPE coefficients, determining some with relatively small error bars. 

To rigorously bound all the OPE coefficients we had access to with five-point correlators using the four-point bootstrap, one would need to study the system of four-point correlators of the following operators: $\sigma$, $\epsilon$, $\sigma'$, $\epsilon'$, $T_{\mu\nu}$, $C_{\mu\nu\sigma\rho}$, $\Sigma_{\mu\nu}$, and $\Sigma_{\mu\nu\rho}$. The size of this system as well as the high spin of the external operators (up to spin-4) make this calculation extremely difficult. Of course, it is possible to study a subset of these correlators (as was done in~\cite{Chang:2024whx}) and to get rigorous bounds on a subset of the OPE coefficients we computed. We leave this for future work.

We observed that choosing the irrelevant scalar $\epsilon'$ as the middle external operator with which no OPEs are taken makes the relative error of the unknown OPE coefficients somewhat larger than in the case when the fifth operator is the relevant scalar $\epsilon$. This is the reason why we did not consider the $\langle \sigma \sigma \sigma' \sigma \epsilon \rangle$ correlator, for example, since we expect that the presence of $\sigma'$ would increase the error bars even further. 

We compared our results to those obtained using the fuzzy sphere regularization approach, and we observed good agreement for those OPE coefficients which we were able to determine with small error bars. For many of the new OPE coefficients we computed using the fuzzy sphere regularization method, the error bars, which we roughly estimated using different ways of extrapolating the data, seem to be comparable to those obtained by means of the five-point bootstrap. It would be interesting to compute these OPE coefficients using larger values of $N$, e.g.~using the DMRG algorithm, and subsequently compare with the predictions derived here.

The bootstrap method presented here is orthogonal to the typical four-point bootstrap approach, as it relies on all the data obtained from the four-point bootstrap as input parameters. It would be interesting to apply this method to other theories, where much  conformal data is already available through the four-point bootstrap or other methods. For example, one could study models with global $O(2)$~\cite{Chester:2019ifh, Liu:2020tpf} and $O(3)$~\cite{Chester:2020iyt} symmetry (see e.g.~\cite{Henriksson:2022rnm} for an overview of available CFT data). In addition, one could study five-point correlators in non-unitary CFTs, such as Lee-Yang models, where CFT data has been computed both using truncation in the four-point correlators \cite{Gliozzi:2014jsa} and the fuzzy sphere regularization approach \cite{ArguelloCruz:2025zuq, Fan:2025bhc, Miro:2025jnz}. Moreover, it would be interesting to explore how machine learning techniques, such as those developed in \cite{Kantor:2021kbx, Kantor:2021jpz} for four-point correlators, could further improve the efficiency of truncated OPE conformal bootstrap methods for five-point correlators.

\subsection*{Acknowledgments}
We thank Slava Rychkov, Matthew Walters, Yuan Xin, and Zheng Zhou for discussions. 
The work of D.P. is supported by U.S. DOE mgrant DE-SC00-17660 and Simons Foundation grant 488651 (Simons Collaboration on the Nonperturbative Bootstrap). 
The work of P.T. is supported by the Royal Society.
Computations in this work were performed on the Yale Grace computing cluster, supported by the facilities and staff of the Yale University Faculty of Sciences High Performance Computing Center.

\appendix

\section{Cross-ratios for derivatives of the crossing relation}\label{parametrization}

We use the following parametrization of the five-point conformal cross-ratios:
\begin{equation}\label{new-coordinates}
\begin{split}
u_1'=\,& \frac{1}{4} \left((a^{-}+a^{+})^2-b^{-}-b^{+}\right) \,,\\
v_1'=\,& \frac{1}{4} \left((a^{-}+a^{+}-2)^2-b^{-}-b^{+}\right)\,,\\
u_2'=\,& \frac{1}{4} \left((a^{+}-a^{-})^2+b^{-}-b^{+}\right) \,,\\
v_2'=\,& \frac{1}{4} \left((a^{+}-a^{-}-2)^2+b^{-}-b^{+}\right)\,,\\
w' =\,& \frac{1}{4} \Big((a^{-}+a^{+})^2+2 (a^{+}-a^{-}-2) (a^{-}+a^{+})+(a^{+}-a^{-}-4) (a^{+}-a^{-})\\
&+2 (2 w-1) \sqrt{b^{+}-b^{-}} \sqrt{b^{-}+b^{+}}-2 b^{+}+4\Big)\,.
\end{split}
\end{equation}
We take derivatives $\mathcal{D}_i$ of the truncated crossing relation with respect to $(a^+, b^+, a^-, b^-, w)$ and then evaluate the derivatives at the configuration \eqref{configurationab}. One should note that only even numbers of derivatives with respect to $a^+, a^-, b^-$ in $\CD_i$ give non-zero constraints. When selecting which constraints to include, we impose a hierarchical condition. In particular, we only retain sets of constraints where, for each cross-ratio, all lower-order derivatives are selected before any higher-order ones.

\section{Correlation matrices}\label{correlationmatrix}

Here we present correlation matrices for OPE coefficients for different tensor structures $n_{IJ}$. We define the correlation matrices as
\begin{equation}
\left(\hat{\rho}[\lambda]\right)_{i,j}=\frac{{\rm Cov}(\lambda^{i+1},\lambda^{j+1})}{\sigma_{\lambda^{i+1}}\sigma_{\lambda^{j+1}}},
\end{equation}
where the indices $i,j$ denote the tensor structures $n_{IJ}$, $\sigma_\lambda$ is the standard deviation, and ${\rm Cov}(\ldots)$ is the covariance defined by ${\rm Cov}(X,Y)=\langle(X-\langle X \rangle)(Y-\langle Y \rangle) \rangle$. Using these, one can compute the OPE coefficients in any basis of tensor structures with the appropriate error bars.\footnote{We thank Slava Rychkov for this suggestion.} 

\begin{equation}
\hat{\rho}_1[\lambda_{C\epsilon C}]=\left(
\begin{array}{ccccc}
 1 & 0.95904 & 0.76955 & 0.88110 & 0.79929 \\
 0.95904 & 1 & 0.90765 & 0.94986 & 0.80317 \\
 0.76955 & 0.90765 & 1 & 0.92308 & 0.69645 \\
 0.88110 & 0.94986 & 0.92308 & 1 & 0.77840 \\
 0.79929 & 0.80317 & 0.69645 & 0.77840 & 1 \\
\end{array}
\right),
\end{equation}

\begin{equation}
\hat{\rho}_2[\lambda_{C\epsilon C}]=\left(
\begin{array}{ccccc}
 1 & 0.97488 & 0.86319 & 0.84245 & 0.64988 \\
 0.97488 & 1 & 0.94265 & 0.87291 & 0.66530 \\
 0.86319 & 0.94265 & 1 & 0.89833 & 0.71925 \\
 0.84245 & 0.87291 & 0.89833 & 1 & 0.58819 \\
 0.64988 & 0.66530 & 0.71925 & 0.58819 & 1 \\
\end{array}
\right),
\end{equation}

\begin{equation}
\hat{\rho}_1[\lambda_{C\epsilon' C}]=\left(
\begin{array}{ccccc}
 1 & 0.95690 & 0.97259 & 0.97948 & 0.40758 \\
 0.95690 & 1 & 0.99675 & 0.92067 & 0.21396 \\
 0.97259 & 0.99675 & 1 & 0.94726 & 0.27486 \\
 0.97948 & 0.92067 & 0.94726 & 1 & 0.44812 \\
 0.40758 & 0.21396 & 0.27486 & 0.44812 & 1 \\
\end{array}
\right),
\end{equation}

\begin{equation}
\hat{\rho}_2[\lambda_{C\epsilon' C}]=\left(
\begin{array}{ccccc}
 1 & 0.92752 & 0.94330 & 0.58841 & -0.22235 \\
 0.92752 & 1 & 0.82440 & 0.59296 & -0.02774 \\
 0.94330 & 0.82440 & 1 & 0.76523 & -0.40407 \\
 0.58841 & 0.59296 & 0.76523 & 1 & -0.46256 \\
 -0.22235 & -0.02774 & -0.40407 & -0.46256 & 1 \\
\end{array}
\right),
\end{equation}

\begin{equation}
\hat{\rho}[\lambda_{\Sigma_2\epsilon\Sigma_2}]=\left(
\begin{array}{ccc}
 1 & 0.77363 & 0.70601 \\
 0.77363 & 1 & 0.69674 \\
 0.70601 & 0.69674 & 1 \\
\end{array}
\right),
\end{equation}

\begin{equation}
\hat{\rho}[\lambda_{\Sigma_2\epsilon\Sigma_3}]=\left(
\begin{array}{ccc}
 1 & 0.82153 & 0.78834 \\
 0.82153 & 1 & 0.95401 \\
 0.78834 & 0.95401 & 1 \\
\end{array}
\right),
\end{equation}

\begin{equation}
\hat{\rho}[\lambda_{\Sigma_3\epsilon\Sigma_3}]=\left(
\begin{array}{cccc}
 1 & 0.12587 & -0.00657 & -0.40997 \\
 0.12587 & 1 & 0.54080 & 0.52940 \\
 -0.00657 & 0.54080 & 1 & 0.20140 \\
 -0.40997 & 0.52940 & 0.20140 & 1 \\
\end{array}
\right),
\end{equation}

\begin{equation}
\hat{\rho}[\lambda_{\Sigma_2\epsilon'\Sigma_2}]=\left(
\begin{array}{ccc}
 1 & 0.38622 & 0.94536 \\
 0.38622 & 1 & 0.43314 \\
 0.94536 & 0.43314 & 1 \\
\end{array}
\right),
\end{equation}

\begin{equation}
\hat{\rho}[\lambda_{\Sigma_2\epsilon'\Sigma_3}]=\left(
\begin{array}{ccc}
 1 & 0.40129 & -0.42856 \\
 0.40129 & 1 & 0.46503 \\
 -0.42856 & 0.46503 & 1 \\
\end{array}
\right),
\end{equation}

\begin{equation}
\hat{\rho}[\lambda_{\Sigma_3\epsilon'\Sigma_3}]=\left(
\begin{array}{cccc}
 1 & 0.60931 & -0.75727 & 0.84671 \\
 0.60931 & 1 & -0.06887 & 0.56090 \\
 -0.75727 & -0.06887 & 1 & -0.77847 \\
 0.84671 & 0.56090 & -0.77847 & 1 \\
\end{array}
\right),
\end{equation}

\begin{equation}
\hat{\rho}[\lambda_{C\sigma\Sigma_2}]=\left(
\begin{array}{ccc}
 1 & 0.99125 & 0.96651 \\
 0.99125 & 1 & 0.98502 \\
 0.96651 & 0.98502 & 1 \\
\end{array}
\right),
\end{equation}

\begin{equation}
\hat{\rho}[\lambda_{C\sigma\Sigma_3}]=\left(
\begin{array}{cccc}
 1 & 0.99030 & 0.93059 & 0.47955 \\
 0.99030 & 1 & 0.96139 & 0.46544 \\
 0.93059 & 0.96139 & 1 & 0.44595 \\
 0.47955 & 0.46544 & 0.44595 & 1 \\
\end{array}
\right).
\end{equation}

\section{Numerical data}\label{a:data}

We fix the conformal data in our calculations to the values given in tables~\ref{z2even} and~\ref{z2odd}. These were extracted from \cite{Simmons-Duffin:2016wlq, Chang:2024whx, Su:unpublised}. 
\begin{table}[h]
\centering
\begin{tabular}{|l|l|l|l|l|}
\hline
$\CO$                  & $\Delta_{\CO}$ & $\lambda_{\sigma \sigma \CO}$ & $\lambda_{\epsilon \epsilon \CO}$ & $\lambda_{\epsilon' \epsilon' \CO}$ \\ \hline
$\epsilon$             & 1.412625       & 1.0518537                     & 1.532435                          & 2.3955808                           \\
$\epsilon'$            & 3.82968        & 0.053012                      & 1.536                             & 7.6771                              \\
$T_{\mu\nu}$           & 3              & -0.65227552                   & -1.7782942                        & -4.8210194                                    \\
$C_{\mu\nu\rho\sigma}$ & 5.022665       & 0.276304                      & 0.99168                           &   --                                  \\
$S_6$                  & 7.028488       & 0.1259328                     & 0.529088                          &       --                              \\
$X_8$                  & 9.031023       & 0.05896                       & 0.277088                          &  --                                   \\
$X_{10}$               & 11.0324141     & 0.02801984                    & 0.1433952                         &    --                                 \\ \hline
\end{tabular}
\caption{CFT data for the $\mathbb{Z}_2$-even part of the spectrum in the 3d critical Ising model.}
\label{z2even}
\end{table}

\begin{table}[h]
\centering
\begin{tabular}{|l|l|l|}
\hline
$\CO$      & $\Delta_{\CO}$ & $\lambda_{\sigma\epsilon\CO}$ \\ \hline
$\sigma$   & 0.5181489      & 1.0518537                     \\
$\sigma'$  & 5.2906         & 0.057235                      \\
$\Sigma_2$ & 4.180305       & 0.77831882                    \\
$\Sigma_3$ & 4.63804        & 0.39173716                    \\
$\Sigma_4$ & 6.112674       & 0.4308208                     \\
$\Sigma_5$ & 6.709778       & 0.2371098                     \\
$\Sigma_6$ & 8.08097        & 0.2295216                     \\
$\Sigma_7$ & 8.747293       & 0.1313810                     \\ \hline
\end{tabular}
\caption{CFT data for the $\mathbb{Z}_2$-odd part of the spectrum in the 3d critical Ising model.}
\label{z2odd}
\end{table}

\section{Fuzzy sphere correlators}\label{fuzzydetails}

Here we provide more detail on the computations of OPE coefficients on the fuzzy sphere. The two-point function of a spin-$s$ primary operator is given by
\begin{equation}
\langle \CO_s(x_1,z_1)\CO_s(x_2,z_2) \rangle = \frac{1}{x_{12}^{2\Delta_{\CO}}}\left(z_1\cdot z_2 -\frac{(z_1 \cdot x_{12})(z_2\cdot x_{12})}{x_{12}^2}\right)^s,
\end{equation}
while the three-point functions in the standard box basis \cite{Costa:2011mg} are given by
\begin{equation}
\langle \CO_{s_1}(x_1,z_1)\phi(x_2)\CO_{s_2}(x_3,z_3) \rangle = \sum_{n_{IJ}=0}^{{\rm min}(s_1,s_3)}\lambda^{n_{IJ}}_{\CO_{s_1}\phi\CO_{s_3}}\frac{H_{13}^{n_{IJ}} V_{1,23}^{s_1-n_{IJ}}V_{3,12}^{s_3-n_{IJ}}}{x_{12}^{\tau_1+\tau_2-\tau_3}x_{13}^{\tau_1+\tau_3-\tau_2}x_{23}^{\tau_2+\tau_3-\tau_1}},
\end{equation}
where
\begin{equation}
\begin{split}
&H_{13}=(x_1-x_3)^2 z_1\cdot z_3-2(x_1-x_3)\cdot z_1 (x_1-x_3)\cdot z_3,\\
&V_{i,jk}=\frac{(x_i-x_j)^2 (x_i-x_k)\cdot z_i - (x_i-x_k)^2 (x_i-x_j)\cdot z_i}{(x_j-x_k)^2},
\end{split}
\end{equation}
and $\tau_i = \Delta_i + s_i$. To express the correlation functions explicitly in terms of free indices, one may act on the given correlators using derivatives with respect to the polarization vectors $z$
\begin{equation}
D_{z,\mu}=\frac{1}{2}\frac{\partial}{\partial z^{\mu}}+z\cdot \frac{\partial}{\partial z} \frac{\partial}{\partial z^{\mu}}-\frac{1}{2}z_{\mu} \frac{\partial^2}{\partial z \cdot \partial_z}.
\end{equation}
Let us define the indexed operators
\begin{equation}
\CO_{\mu_1\mu_2\ldots \mu_s}(x)\equiv D_{z,\mu_1}D_{z,\mu_2}\ldots D_{z,\mu_s}\CO(x,z).
\end{equation}
Then states $|\CO_{l,m}\rangle$ on the fuzzy sphere are related to these CFT operators by
\begin{equation}
|\CO_{l,m}\rangle = \lim_{x\to 0}\CP_{l,m}^{\mu_1\mu_2\ldots \mu_l}\CO_{\mu_1\mu_2\ldots \mu_l}(x)|0\rangle,
\end{equation}
where the $\CP_{l,m}^{\mu_1\mu_2\ldots \mu_l}$ are projectors to the $(l,m)$ representation of SO(3). As explained in \cite{ArguelloCruz:2025zuq}, these can be calculated recursively by starting from the spin-1 projectors,
\begin{equation}
\CP_{1,0} = \left( 
\begin{array}{ccc}
 0  \\
 0  \\
 1  \\
\end{array}
\right),\quad
\CP_{1,1} = -\frac{1}{\sqrt{2}}\left( 
\begin{array}{ccc}
 1  \\
 i  \\
 0  \\
\end{array}
\right),
\quad
\CP_{1,-1} = \frac{1}{\sqrt{2}}\left( 
\begin{array}{ccc}
 1  \\
 -i  \\
 0  \\
\end{array}
\right),
\end{equation}
and multiplying them by applying the Clebsch-Gordan rule
\begin{equation}
\CP_{l,m} =\frac{1}{l} \sum_{m'}\left(\sum_{k=0}^{l-1} \CR^{k}\left(\CP_{1,m'}\otimes\CP_{l-1,m-m'} \right)\right) \langle 1,m'; l-1, m-m'|l,m \rangle,
\end{equation}
where $\langle 1,m'; l-1, m-m'|l,m \rangle$ are Clebsch-Gordan coefficients and the function $\CR$ cyclically permutes the indices of the tensor it acts on, thus ensuring that we get symmetric $\CP$ tensors. 

Once we compute $\CP_{l,m}$, we can determine the $|\CO_{l,m}\rangle$ states on the fuzzy sphere explicitly in terms of indices in 3-dimensional spacetime. Further, we need to ensure that the fuzzy sphere states are unit normalized, that is,
 $\langle \CO_{l,m}|\CO_{l,m}\rangle = 1$. This gives the normalization factors $\sqrt{\frac{2}{27}}$ and $\sqrt{\frac{8}{10125}}$ in \eqref{spin2st} and \eqref{spin3st}.
Ket states on the fuzzy sphere are related to the CFT operators by
\begin{equation}
\langle \CO_{l,m}| = \CP^{\mu_1\ldots\mu_l}_{l,m}\lim_{x\to 0}x^{-2\Delta_{\CO}}{\CI_{\mu_1}}^{\nu_1}{\CI_{\mu_2}}^{\nu_2}\ldots {\CI_{\mu_l}}^{\nu_l}\CO_{\nu_1\nu_2\ldots \nu_l}\left(\frac{x}{x^2}\right)|0\rangle,
\end{equation}
where ${\CI_{\mu}}^{\nu} = \delta^{\mu}_{\nu}-2\frac{x_{\mu}x^{\nu}}{x^2}$. Note that we only consider states with $m=0$ quantum numbers on the fuzzy sphere. With the above, it is straightforward to relate the fuzzy sphere OPE coefficients to the ones in the standard box basis.

\bibliographystyle{JHEP}
\bibliography{refs.bib}{}
\end{document}